\documentclass[a4paper,11pt]{article}
\pdfoutput=1 % if your are submitting a pdflatex (i.e. if you have
\usepackage{soul}

\usepackage{jcappub} % for details on the use of the package, please
\usepackage[T1]{fontenc} % if needed
\usepackage[normalem]{ulem}

\usepackage{cancel}
\usepackage{multirow}
\usepackage{mathtools}
\newcommand\Ccancel[2][black]{\renewcommand\CancelColor{\color{#1}}\cancel{#2}}

\title{\boldmath  %Integrated Matter trispectrum: \\ 
Matter trispectrum:\\
theoretical modelling and comparison to N-body simulations}
\author[a,b]{Davide Gualdi,}
\author[a]{Sergi Novell,}
\author[a,b]{H\'ector Gil-Mar\'in}
\author[a,c]{and Licia Verde}

\affiliation[a]{Institut de Ci\`encies del Cosmos, University of Barcelona, ICCUB, Barcelona 08028, Spain}
\affiliation[b]{Institute of Space Studies of Catalonia (IEEC), E-08034 Barcelona, Spain}
\affiliation[c]{Instituci\'o Catalana de Recerca i Estudis Avan\c{c}ats, Passeig Llu\'is Companys 23, Barcelona 08010, Spain}

\emailAdd{dgualdi@icc.ub.edu}
\emailAdd{sergi.novell@icc.ub.edu}
\emailAdd{hectorgil@icc.ub.edu}
\emailAdd{liciaverde@icc.ub.edu}

\abstract{The power spectrum has long been the workhorse summary statistics for large-scale structure cosmological analyses. However, gravitational non-linear evolution moves precious cosmological information from the two-point statistics (such as the power spectrum) to higher-order correlations. Moreover, information about the primordial non-Gaussian signal lies also in higher-order correlations. Without tapping into these, that information remains hidden. While the three-point function (or the bispectrum), even if not extensively,  has been studied and applied to data, there has been only limited discussion about the four point/trispectrum. This is because the high-dimensionality of the statistics (in real space a skew-quadrilateral has 6 degrees of freedom), and the high number of skew-quadrilaterals, make the trispectrum numerically and algorithmically very challenging. Here we address this challenge by studying the i-trispectrum, an integrated trispectrum that only depends on four $k$-modes moduli.
We model and measure the matter i-trispectrum from a set of 5000 \textsc{Quijote} N-body simulations both in real and redshift space, finding good agreement between simulations outputs and model up to mildly non-linear scales. Using the power spectrum, bispectrum and i-trispectrum joint data-vector covariance matrix estimated from the simulations, we begin to quantify the added-value provided by the i-trispectrum. In particular, we forecast the i-trispectrum improvements on constraints on the local primordial non-Gaussianity amplitude parameters $f_\mathrm{nl}$ and $g_\mathrm{nl}$. For example, using the full joint data-vector,  we forecast $f_\mathrm{nl}$  constraints up to two times ($\sim32\%$) smaller in real (redshift) space than those obtained  without i-trispectrum.}

\begin{document}
\maketitle
\flushbottom

%%%%%%%%%%%%%%%%%%%%%%%%%%%%%%%%%%%%%%%%%%%%%
%%%%%%%%%%%%%%%%%%%%%%%%%%%%%%%%%%%%%%%%%%%%%
%%%%%%%%%%%%%%%%%%%%%%%%%%%%%%%%%%%%%%%%%%%%%
\section{Introduction}
% surveys and the future

The clustering of large-scale structure (LSS) encloses key information about both the primordial Universe, including  the physics of inflation,  and the growth of cosmological perturbations which are driven by gravity and the physics of the late-time Universe. 
Future and forthcoming large-scale structure surveys (e.g.,  DESI\footnote{\url{http://desi.lbl.gov}} \citep{Levi:2013gra}; Euclid \footnote{\url{http://sci.esa.int/euclid/}} \citep{Laureijs:2011gra}; PFS \footnote{\url{http://pfs.ipmu.jp}} \citep{Ellis:2012rn}; SKA\footnote{\url{https://www.skatelescope.org}} \citep{Bacon:2018dui}; LSST\footnote{\url{https://www.lsst.org/}} \citep{Abell:2009aa} and WFIRST\footnote{\url{https://www.cosmos.esa.int/web/wfirst}} \cite{Green:2012mj}) will cover unprecedented  effective volumes, giving in principle access to this wealth of  cosmological information.  

While Bayesian  hierarchical modelling, likelihood-free or forward modelling approaches (e.g.,  \cite{Tayloretal19,Ramanah2019,AlsingCFW19} and Refs. therein) are extremely promising, to date the workhorse approach still consists in considering and analysing  summary statistics. The most popular is, of course, the power spectrum --the Fourier space counterpart of the two-point correlation function--  which for a Gaussian random  field encloses all the information.  However, deviations from Gaussianity induced, e.g., by  non-linear gravitational evolution but also of primordial origin, generate higher-order statistics.
The bispectrum --the Fourier space counterpart of the three-point correlation function-- is the next-to-leading-order statistic and it has been studied somewhat extensively. However, bispectrum measurements from  galaxy surveys and their cosmological interpretation is still to a certain degree limited (\cite{Scoccimarro01,2dFbispectrum,Gil-Marin:2016wya,Sugiyama:2018yzo}). %
This is due to the fact that it is much more challenging and computationally intensive to measure the bispectrum signal, estimate its covariance matrix, account for its selection effects, than it is for the power spectrum.

The trispectrum  --the  connected part of the four-point correlation function in Fourier space-- is even more off the beaten track. The trispectrum is non-zero when four $k$-vectors make a closed \textit{skew-quadrilateral} (i.e.,
it is embedded in a 3D space). For simplicity we will loosely use the term {\it quadrilateral} from now on, but one should keep in mind that we refer to sets of four $k$-vectors that do not necessarily lie on the same plane. Note that the trispectrum is the lowest-order correlation where the connected part needs to be  explicitly disentangled from the unconnected part. In other words, the ensemble average of four Fourier modes whose $k$-vectors make a closed quadrilateral would be, in general, non-zero even for a Gaussian random field. But what carries additional information not enclosed in the power spectrum is the connected part of that statistic. 

While for the cosmic microwave background (CMB) fluctuations the trispectrum has been studied in detail and has been measured \cite{kunzetal2001,Komatsu:2002db,deTroia2003,Munshi:2009wy,Kamionkowski:2010me,Izumi:2011di,Regan_2015, Feng:2015pva,Fergusson:2010gn,Smith:2015uia,Namikawa:2017uke,PlanckTrispectrum18, Akrami:2019izv},  in the late-time large-scale structure context it has received limited  attention \cite{Verde:2001pf,Cooray:2008eb,Lazeyras:2017hxw,Bellomo:2018lew} mainly focused on its relation with the power spectrum covariance matrix \cite{2010A&A...514A..79P,Mohammed:2016sre,Taruya:2020qoy}.
An effective field theory model of the trispectrum was derived in \cite{Bertolini:2016bmt} while a formalism in angular space was recently proposed by \cite{Lee:2020ebj}.
In the presence of a primordial trispectrum, the correction to the non-Gaussian linear bias was derived in \cite{Lazeyras:2015giz}
Applications of the trispectrum  to data are at an even more embryonic stage.  The four-point correlation function in configuration space, was originally measured by \cite{Fry1978} from the Lick and Zwicky catalogs, from simulations by \cite{Suto:1993ua} and recently at small scales from the BOSS NGC CMASS galaxy catalog by \cite{Sabiu:2019kbh}.

CMB studies on the trispectrum have proved its additional constraining power in particular regarding possibly  primordial non-Gaussianities, which could confirm or rule out models of the very early Universe (e.g., single/multi field inflation). Even if more noisy and difficult to model due to non-linear gravitational evolution, the late-time 3D matter field trispectrum contains by definition many more modes than the primordial 2D CMB counterpart. It is reasonable to expect that, if measurable, it will provide additional information regarding the same primordial non-Gaussianity parameters. This was also the initial motivation stressed in \cite{Verde:2001pf} to look at this statistic.

Therefore the ultimate question we want to address is: "Is there additional information in the matter/halo/galaxy  fields which is not captured by the power spectrum and bispectrum but that could be extracted by considering  also the trispectrum?". In this paper, as a first step we focus on the matter field. Only if this idealised case shows that there is  additional  useful information in the trispectrum,  then it would provide motivation to extend the treatment to more complex and realistic cases. 

Studying the LSS trispectrum is massively more challenging for several reasons. On the theoretical and modelling side, several physical processes contribute to the trispectrum signal: not only a possible primordial signature and systematics/foreground effects as in the CMB, but also the mode-coupling arising from gravity and (non-linear) gravitational clustering. Moreover, the trispectrum of galaxies or other dark matter tracers is also affected by real-world effects such as redshift space distortions and galaxy bias, which  need to be modelled consistently (i.e.,  at least up to third order in perturbation theory).

On the more practical side of measuring the signal from (real or simulated) data, in the CMB only co-planar quadrilaterals  are considered, but in the three-dimensional LSS space, the number of quadrilateral configurations 
increases very rapidly. This has been to-date the showstopper for considering the trispectrum of LSS. 

In this paper we address this challenge by  considering a specific type of  integrated  trispectrum of the late-time dark matter overdensity field, which we will call for simplicity i-trispectrum and that depends only on the modulus of the  four $k$-modes $(k_1,k_2,k_3,k_4)$. A first attempt to consider such an estimator was done in \cite{Sefusatti:2004xz}. Here we pay particular attention to  the numerical implementation and the connection to the theoretical model. In particular in the approach presented here the theoretical model is significantly improved, by reducing drastically the adopted approximations and  by performing the  full multidimensional integration over the volume in Fourier space.

As it will become apparent later, the i-trispectrum is an integrated trispectrum over (skew) quadrilaterals with all possible folding angles around a diagonal and all possible lengths of  the diagonal.
This reduces greatly the computational and algorithmic challenge and makes the trispectrum signal accessible. 

The analysis is presented in both real and redshift space including possible primordial non-Gaussianities. The theoretical model can be easily extended to the galaxy field in redshift space using an appropriate bias expansion up to third order. The estimator can be directly applied to the galaxy field in redshift space, but this is left for future work.

Modelling the dark matter quantities is a necessarily unavoidable and non-trivial
first step (and may in principle be useful for weak lensing applications).
It enables us to assess how well the adopted theoretical model reproduces the measured statistics of the simulated dark matter field, before applying any kind of galaxy or halo bias prescription, or real-to-redshift space conversion.

The goal of this paper is to present a theoretical modelling of the  dark matter (connected) i-trispectrum signal observed in N-body simulations of  structure formation, and to highlight the challenges of characterising correctly its covariance matrix.
We also quantify the added value of the i-trispectrum, when combined with power spectrum and bispectrum, especially for constraints on primordial non-Gaussianities.  
The rest of the paper is organised as follows. In section \ref{sec:methodology}
the methodology of our analysis is presented, including the theoretical modelling of statistics (section \ref{sec:theo_models}), the estimators applied on the simulated data (section \ref{sec:estim}) and both local primordial non-Gaussianity theory and the formalism necessary for the Fisher forecasts (respectively sections \ref{sec:png_methodology} and \ref{sec:fisher_method}).
The results are presented in the second part of the paper, section \ref{sec:results}. In particular, the real space analysis comparing theory with measurements is described in section \ref{sec:meas_vs_model} while the Fisher forecast for the  primordial non-Gaussianity parameters is reported in section \ref{sec:png_forecasts}. Finally, the same analysis and forecasts performed in redshift space are reported in section \ref{sec:rsd_results}. We conclude in section \ref{sec:conclusions} where we also discuss possible further extension of this work.

%%%%%%%%%%%%%%%%%%%%%%%%%%%%%%%%%%%%%%%%%%%%
%%%%%%%%%%%%%%%%%%%%%%%%%%%%%%%%%%%%%%%%%%%%
%%%%%%%%%%%%%%%%%%%%%%%%%%%%%%%%%%%%%%%%%%%%

\begin{figure}[tbp]
\centering 
\includegraphics[width=0.99\textwidth]%,trim=0 380 0 200,clip]
{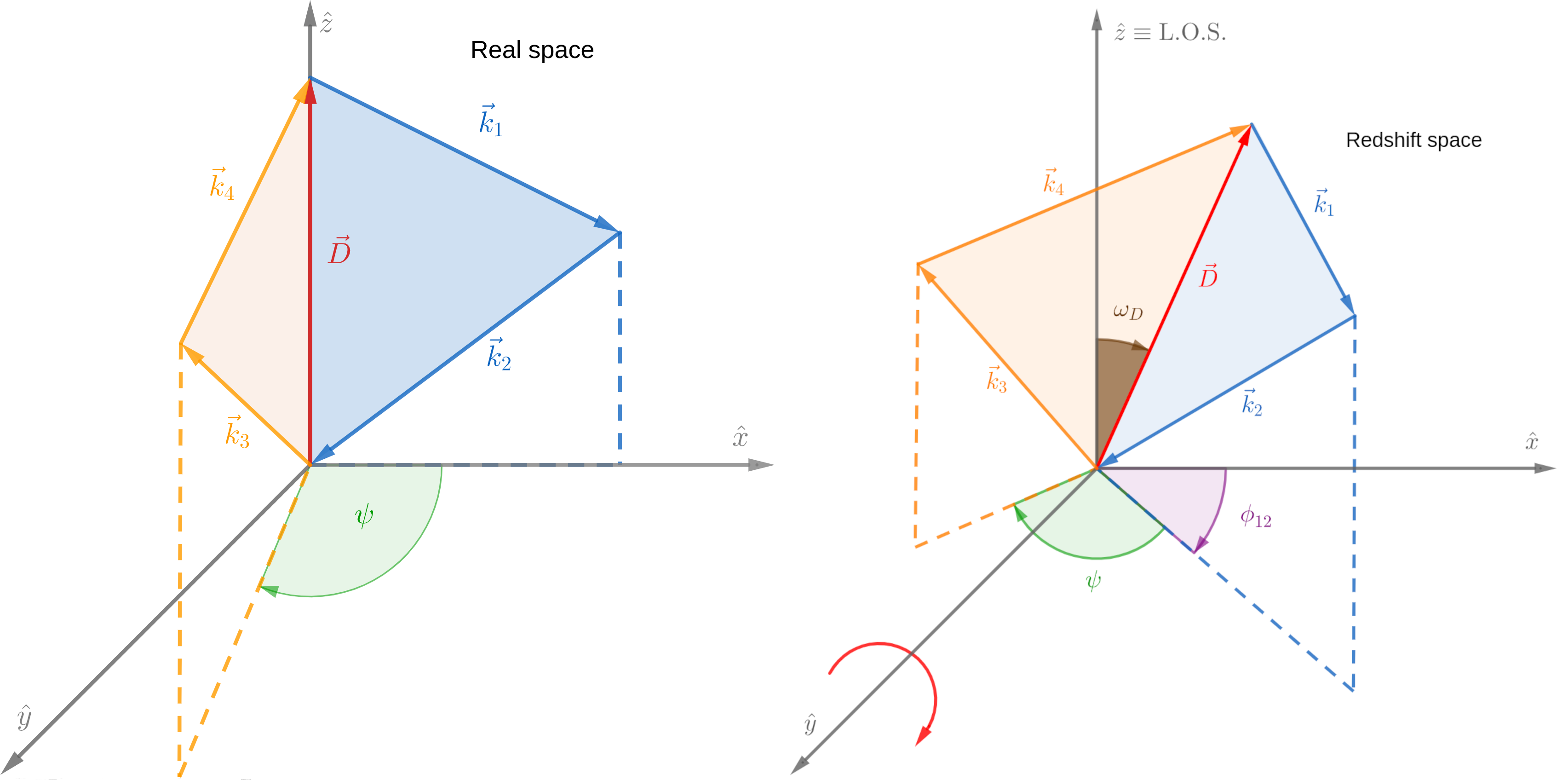}
\caption{\label{fig:coordinates_tk} Trispectrum coordinates used to describe each configuration in the analytical model for $T$. On the left hand side the real space coordinate system. Without loss of generality the diagonal ${\bf D}$ can be aligned with the $z$-axis and the triangle ${\bf k}_1+{\bf k}_2+{\bf D}=0$ can be on the $xz$-plane. $\psi$ denotes the folding angle of the quadrilateral. On the right hand side is the redshift space coordinate system. Now the line of sight is aligned with the $z$-axis and therefore another two angles are needed to fully describe the quadrilateral and its $T$: $\omega_D=\arccos{\mu_D}$ (the angle between ${\bf D}$ and the line of sight., and $\phi_{12}$ (the angle between the $xz$-plane and the triangle ${\bf k}_1+{\bf k}_2+{\bf D}=0$).} 
\end{figure}

%%%%%%%%%%%%%%%%%%%%%%%%%%%%%%%%%%%%%%%%%%%%%
%%%%%%%%%%%%%%%%%%%%%%%%%%%%%%%%%%%%%%%%%%%%%
%%%%%%%%%%%%%%%%%%%%%%%%%%%%%%%%%%%%%%%%%%%%%
\section{Methodology, theory modelling and estimators}
\label{sec:methodology}
In view of our goal to investigate whether and how the i-trispectrum could be a powerful tool for cosmology, especially  to further tighten the constraints on local primordial non-Gaussianity, we have to lay out a suitable methodology.
We set out to  model and measure the i-trispectrum signal of the dark matter particles field contained in a set of simulations with regular geometry. We use 5000 realisations of the \textsc{Quijote} N-body suite \cite{Villaescusa-Navarro:2019bje}. 

A theoretical model for the i-trispectrum and an estimator for its measurement are introduced and validated. For completeness  we also review the  adopted theory model and the estimators for the power spectrum and bispectrum. Power spectrum, bispectrum and trispectrum are the Fourier counterparts of the 2pt, 3pt and 4pt correlation functions in configuration space, respectively. In terms of averaged correlator of the Fourier transformed density field these are defined as

\begin{eqnarray}
\langle \delta(\mathbf{k}_1)\delta(\mathbf{k}_2)\rangle_\mathrm{c}
&=& (2\pi)^3\delta^D(\mathbf{k}_1+\mathbf{k}_2)\,P(k_1)\notag \\
\langle \delta(\mathbf{k}_1)\delta(\mathbf{k}_2)\delta(\mathbf{k}_3)\rangle_\mathrm{c}
&=& 
(2\pi)^3\delta^D(\mathbf{k}_1+\mathbf{k}_2+\mathbf{k}_3)\,
B(\mathbf{k}_1,\mathbf{k}_2,\mathbf{k}_3)
\notag \\
\langle \delta(\mathbf{k}_1)\delta(\mathbf{k}_2)\delta(\mathbf{k}_3)\delta(\mathbf{k}_4)\rangle_\mathrm{c} 
&=&
(2\pi)^3\delta^D(\mathbf{k}_1+\mathbf{k}_2+\mathbf{k}_3+\mathbf{k}_4)\,
T(\mathbf{k}_1,\mathbf{k}_2,\mathbf{k}_3,\mathbf{k}_4)\,,
\notag \\
\end{eqnarray}

 \noindent where the subscript "c" stands for connected. In the following sub-sections, after introducing  the modeling for the data-vectors, the i-trispectrum covariance (both auto and cross-covariance with the power spectrum and bispectrum) is presented and estimated. With this in hand, we perform a Fisher forecast using a theoretical model for the primordial non-Gaussianity signal.

The methodology is described here and the results  are presented in section \ref{sec:results}.

\subsection{Theoretical models of the signals}
\label{sec:theo_models}
 
Our theory model for the matter power spectrum is provided by the \textsc{class} code \cite{Lesgourgues:2011re}, which returns both the linear and non-linear power spectrum via the \textsc{halofit} \cite{Smith:2002dz,Bird:2011rb,Takahashi:2012em} fitting formula\footnote{A more recent fitting formula for the matter power spectrum also implemented in \textsc{class} is \textsc{HMcode} \cite{Mead:2016zqy}. We tested that for the chosen $k$-range the difference between \textsc{halofit} and \textsc{HMcode} is negligible for our purposes}.
We model the tree-level matter bispectrum in two ways: using the second-order standard perturbation theory kernel $F^{(2)}_\mathrm{SPT}\left[\mathbf{k}_1,\mathbf{k}_2\right]$ \cite{Peebles1980,Fry:1983cj}, which is derived from first principles, and with its effective version $F^{(2)}_\mathrm{EFF}\left[\mathbf{k}_1,\mathbf{k}_2\right]$ \cite{GilMarin:2011ik,Gil-Marin:2014pva}, which was calibrated on simulations using the measured non-linear matter power spectrum ($P_\mathrm{nl}$) as input. 
In what follows, unless otherwise stated, the linear power spectrum is indicated by $P$. Therefore for the bispectrum in real space we have
\begin{eqnarray}
B_\mathrm{SPT}(\mathbf{k}_1, \mathbf{k}_2, \mathbf{k}_3) &&=\, 2 F^{(2)}_\mathrm{SPT}\left[\mathbf{k}_1,\mathbf{k}_2\right]\,P\left(k_1\right)P\left(k_2\right)\quad + \quad 2 \,\,\mathrm{permutations}\,,
\notag\\
B_\mathrm{EFF}(\mathbf{k}_1, \mathbf{k}_2, \mathbf{k}_3) &&=\, 2 F^{(2)}_\mathrm{EFF}\left[\mathbf{k}_1,\mathbf{k}_2\right]\,P_\mathrm{nl}\left(k_1\right)P_\mathrm{nl}\left(k_2\right)\quad + \quad 2\,\mathrm{p.}\,\,,
\label{eq:BSPT}
\end{eqnarray}
\noindent where we indicate the cyclical terms by "permutations" or "p".
In the $B_\mathrm{EFF}$ case the non-linear power spectrum can either be provided by the measured quantity from the  N-body simulations (this is the way how the kernel was originally calibrated), or alternatively, by \textsc{halofit}.  
We also consider the ``educated guess" bispectrum model which uses the standard perturbation theory tree-level $F_2^{\rm SPT}$ kernel with  the \textsc{halofit} non-linear power spectrum rather than the linear one 
\begin{equation}
B_{\rm SPT-NL}(\mathbf{k}_1, \mathbf{k}_2, \mathbf{k}_3)=2F^{(2)}_\mathrm{SPT}\left[\mathbf{k}_1,\mathbf{k}_2\right]\,P_\mathrm{nl}\left(k_1\right)P_\mathrm{nl}\left(k_2\right)\quad + \quad 2\,\mathrm{p}.
\end{equation}
The tree-level trispectrum model $T_\mathrm{SPT}\left(\mathbf{k}_1,\mathbf{k}_2,\mathbf{k}_3,\mathbf{k}_4\right)$ has been already derived and presented in several works e.g., \cite{Fry:1983cj,Sefusatti:2009qh}; here we use the standard perturbation theory expression reported in the appendix B3 of \cite{Gualdi:2017iey} which we briefly summarise also in appendix \ref{app:theo_models_rsd}.
Appendix  \ref{app:theo_models_rsd}  also reports the full expressions for  the power spectrum,  bispectrum and trispectrum theory  models in redshift space for a biased tracer as a proxy for the galaxy field.

It is important to specify that by trispectrum, $T$, and thus by  trispectrum model $T_\mathrm{SPT}$,  we mean the signal corresponding to a  quadrilateral configuration of a given shape defined by the choice of the four $k$-vectors $\left(\mathbf{k}_1,\mathbf{k}_2,\mathbf{k}_3,\mathbf{k}_4\right)$. 
In particular, for the matter trispectrum in real space we use \cite{Fry:1983cj,2010A&A...514A..79P}
\begin{eqnarray}
\label{eq:tk_3d_matter_model}
T_\mathrm{SPT}(\mathbf{k}_1,\mathbf{k}_2,\mathbf{k}_3,\mathbf{k}_4)
&=&
4
P(k_1)P(k_2)
%\notag \\
%&\times&
\times \Big\{
F^{(2)}_{\rm SPT}\left[\mathbf{k}_1,-\mathbf{k}_{13}\right]
F^{(2)}_{\rm SPT}\left[\mathbf{k}_2, \mathbf{k}_{13}\right]
P(k_{13})
\notag \\
&+&
F^{(2)}_{\rm SPT}\left[\mathbf{k}_1,-\mathbf{k}_{14}\right]
F^{(2)}_{\rm SPT}\left[\mathbf{k}_2, \mathbf{k}_{14}\right]
P(k_{14})
\Big\}
\,+\,5\,\mathrm{p}.
\notag \\
&+&
6
\,\,F^{(3)}_{\rm SPT}\left[\mathbf{k}_{1},\mathbf{k}_2, \mathbf{k}_{3}\right]
P(k_1)P(k_2)P(k_3)
\,+\,3\,\mathrm{p}. ,
\label{eq:TSPT}
\end{eqnarray}
\noindent where $k_{ij}=|{\bf k}_i+{\bf k}_j|$.
The expression for the second- and third-order standard perturbation theory kernels $F^{(2,3)}_{\mathrm{SPT}}$ can be found in Ref. \cite{Gualdi:2017iey}.
As for the case of the bispectrum, in equation \ref{eq:TSPT} we might use the linear matter power spectrum for $P$ yielding $T_{\rm SPT}$ or a similar educated guess combining the non-linear matter power spectrum with standard perturbation theory kernels, yielding $T_{\rm SPT-NL}$.

From equation \ref{eq:TSPT}, the expression for the  theoretical model of the trispectrum, we can easily identify the  two terms obtained by expanding 
the four-point correlator up to sixth order in $\delta$ \cite{Fry:1983cj}:
$T_\mathrm{SPT}(\mathbf{k}_1,\mathbf{k}_2,\mathbf{k}_3,\mathbf{k}_4)
= T_{(1122)} + T_{(1113)}$ (see also section ~\ref{sec:png_methodology}, Appendices \ref{App:estimators} and \ref{sec:app_png}).

The matter trispectrum  (in real space) is a statistic measured on an isotropic field. This reduces the number of degrees of freedom necessary to describe a quadrilateral configuration from eight (needed when the signal depends on the orientation with respect to the line of sight) down to six. 

Here we use the convention illustrated in figure \ref{fig:coordinates_tk}. We use as diagonal, $\mathbf{D}$,  the one defined by $\mathbf{D} + \mathbf{k}_1 + \mathbf{k}_2 = 0$.
The module of this diagonal can then vary between a minimum and a maximum value defined by
\begin{eqnarray}
\label{eq:diag_min_max}
    D_{\mathrm{min}} = \mathrm{max}\,\left[|k_1-k_2|,|k_3-k_4|\right]
\quad \mathrm{and}\quad
 D_{\mathrm{max}} = \mathrm{min}\,\left[|k_1+k_2|,|k_3+k_4|\right]\,,
\end{eqnarray}
\noindent where the above conditions are imposed by the requirement that the triangles formed by the diagonal $\mathbf{D}$  and the two $k$-vectors ($\{ k_1$, $k_2\}$ and $\{k_3$, $k_4\}$) are closed.
In real space, the orientation with respect to the line of sight does not matter and therefore, for each quadrilateral, one could choose to set the diagonal vector $\mathbf{D}$ parallel to the line of sight direction (namely, $z$-axis). With this choice, the two triangles defined by $\mathbf{D}+\mathbf{k}_1+\mathbf{k}_2=0$ and $\mathbf{D}-\mathbf{k}_3-\mathbf{k}_4=0$ will be orthogonal to the $xy$-plane.

Finally, the angle $\psi$  describes the "folding" of the quadrilateral around the diagonal $\mathbf{D}$, in other words, the angle between the two planes defined by the two triangles $\mathbf{D}+\mathbf{k}_1+\mathbf{k}_2=0$ and $\mathbf{D}-\mathbf{k}_3-\mathbf{k}_4=0$.
It should be then clear that $T(\mathbf{k}_1,\mathbf{k}_2,\mathbf{k}_3,\mathbf{k}_4)\equiv  T(k_1, k_2, k_3, k_4, D,$ $\psi,[\omega, \phi_{12}])$ where  $\omega, \phi_{12}$  are varying only in the redshift space case (see section~\ref{sec:Tcoordredshiftspace}), because each $\mathbf{k}_1,\mathbf{k}_2,\mathbf{k}_3,\mathbf{k}_4$ set corresponds univocally to a $\{k_1, k_2, k_3, k_4, D, \psi,[\omega, \phi_{12}]\}$ set and vice versa.
We choose to study an averaged version of the trispectrum signal which we call i-trispectrum, with each configuration described only by the modulus of the four quadrilateral sides, $\mathcal{T}\left(k_1,k_2,k_3,k_4\right)$. In other words, we integrate out, by averaging over all their possible values, the additional two  coordinates needed to fully describe a quadrilateral configuration (in real  space). 
According to our adopted convention (figure~\ref{fig:coordinates_tk}) this means integrating over $D$ and $\psi$. 
The redshift space i-trispectrum (see section~\ref{sec:Tcoordredshiftspace}) monopole is further averaged over $\omega$ and $\phi_{12}$.

Therefore, differently from what happens for the power spectrum (line) and the bispectrum (triangle), the i-trispectrum signal as a function of just $(k_1,k_2,k_3,k_4)$  is an average over many different shapes characterised by different values of one diagonal $D$ and the folding angle around it $\psi$. The second diagonal that could be used as coordinate and its folding angle is fixed by choosing the first coordinates pair $\{D,\psi\}$.

Following these conventions, we define matter i-trispectrum in real space as a function of the module of the four $k$-vectors  $k_1,k_2,k_3,k_4$ as %
\begin{eqnarray}
\label{eq:tk_int_real}
\mathrm{{\cal T}}\left(k_1,k_2,k_3,k_4\right)
&&=\,
\dfrac{1}{3}\sum_{\substack{k_1,k_2,k_3,k_4 \\ k_1,k_3,k_2,k_4 \\ k_1,k_2,k_4,k_3}}
\dfrac{1}{2\pi \Delta D}
\int^{D_\mathrm{max}}_{D_\mathrm{min}}dD\int^{2\pi}_0 d\psi\,
T\left(k_1,k_2,k_3,k_4,D,\psi\right)
\,,
\notag \\
\end{eqnarray}
\noindent where $\Delta D = D_\mathrm{max}-D_\mathrm{min}$.
Given a $(k_1,k_2,k_3,k_4)$ set, three different shapes can be obtained by changing the order of the sides, this being the reason for the average in equation \ref{eq:tk_int_real}. 

\begin{figure}[tbp]
\centering 
\includegraphics[width=0.99\textwidth]%,trim=0 380 0 200,clip]
{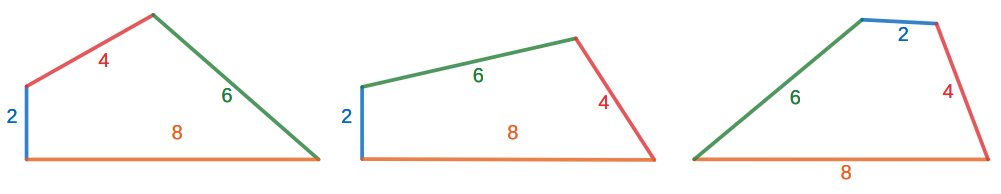}
\caption{\label{fig:quad_diff_shapes} Different quadrilateral shapes allowed for a given choice of $(k_1,k_2,k_3,k_4)$. The estimator from equation \ref{eq:tk_est_text} automatically averages over all these configurations and that is the reason of the sum in equation \ref{eq:tk_int_real} and \ref{eq:tk_int_rsd}.} 
\end{figure}

This can be understood by looking at the sketch in figure \ref{fig:quad_diff_shapes}.  The three different shapes  are obtained by  changing the ordering of the sides $(k_1,k_2,k_3,k_4)$;  for simplicity quadrilaterals shown lie on a plane i.e., no folding around the diagonal is shown in the sketch, where the diagonal $D$ is always defined from the first two entries of $T$'s arguments.

From equation \ref{eq:tk_int_real} is easy to appreciate the improvement over   \cite{Sefusatti:2004xz} comparing with their equation 34. While they  approximate the theoretical model for the integrated trispectrum estimator using an amplitude factor, here we  perform  explicitly the full multi-dimensional integration in Fourier space.

The corresponding estimator applied on data will be later defined in equation \ref{eq:tk_est_text}. 
In appendix \ref{app:theo_models_rsd} the theoretical models for redshift space  power spectrum, bispectrum and trispectrum are briefly reported.

%%%%%%%%%%%%%%%%%%%%%%%%%%%%%%%%%%%%%%%%%%%%
%%%%%%%%%%%%%%%%%%%%%%%%%%%%%%%%%%%%%%%%%%%%
%%%%%%%%%%%%%%%%%%%%%%%%%%%%%%%%%%%%%%%%%%%%

%%%%%%%%%%%%%%%%%%%%%%%%%%%%%%%%%%%%%%%%%%%%%

\subsubsection{Redshift space i-trispectrum: coordinates}
\label{sec:Tcoordredshiftspace}
When redshift space distortions are present, and in the partial sky/plane parallel approximation,  the signal is not isotropic anymore  \cite{Kaiser:1987qv,Hamilton:1997zq}. The preferred direction is given by the line of sight, which, by convention, we align with the $\hat{z}$ direction.  Hence, compared to the real space case discussed above, 
in order to define  the orientation with respect to the line of sight of all the possible quadrilaterals (defined by the set of coordinates $(k_1,k_2,k_3,k_4,D,\psi)$),  only two additional parameters are necessary. 

First we define an angle $\phi_{12}$ to describe the rotation of the triangle $\mathbf{D}+\mathbf{k}_1+\mathbf{k}_2=0$ away from the $xz$-plane (illustrated in the right part of figure \ref{fig:coordinates_tk}). 

As remaining parameter we choose the cosine of the angle between the diagonal, $\mathbf{D}$, and the line of sight, $\mu_D$.
Then by rotating each vector by an angle $\omega_D=\arccos{\mu_D}$, together with the choice of angles $(\psi,\phi_{12})$, all the possible orientations/shapes can be defined (and spanned when integrating over).

It is indeed important to notice that the orientation with respect to the quadrilateral line of sight, together with its shape, can also be modified by varying both angles $\phi_{12}$ and $\psi$. These three variables $(\psi,\phi_{12},\mu_D)$ may be used as a natural basis for an expansion in terms of Legendre polynomials of the full anisotropic i-trispectrum signal.  The redshift space i-trispectrum monopole $\mathcal{{T}}^{(0)}$ can be obtained by performing the four-dimensional integration of the redshift space trispectrum, $T^\mathrm{s}$:
\begin{eqnarray}
\label{eq:tk_int_rsd}
\mathcal{T}^{\rm (0)}\left(k_1,k_2,k_3,k_4\right)
&&=\,
\dfrac{1}{3}\sum_{\substack{k_1,k_2,k_3,k_4 \\ k_1,k_3,k_2,k_4 \\ k_1,k_2,k_4,k_3}}
\dfrac{1}{8\pi^2 \Delta D}
\int^{D_\mathrm{max}}_{D_\mathrm{min}}dD\int^{+1}_{-1}\,d\mu_D\int^{2\pi}_0\,d\phi_{12}\int^{2\pi}_0\,d\psi
\notag \\
&&\times\,
T^\mathrm{s}\left(k_1,k_2,k_3,k_4,D,\mu_D,\phi_{12},\psi\right)
\,.
\end{eqnarray}

%%%%%%%%%%%%%%%%%%%%%%%%%%%%%%%%%%%%%%%%%%%%%
%%%%%%%%%%%%%%%%%%%%%%%%%%%%%%%%%%%%%%%%%%%%%
%%%%%%%%%%%%%%%%%%%%%%%%%%%%%%%%%%%%%%%%%%%%%
\subsection{Estimators: measuring P, B, \texorpdfstring{$\mathbf{\mathcal{T}}$}{} of an overdensity field}
\label{sec:estim}

Given an overdensity field, we need a procedure to estimate the statistics of interest: power spectrum, bispectrum and i-trispectrum.  The overdensity field, as sampled for example by particles or haloes in an N-body simulation or by galaxies in a survey, is a discrete distribution, which contains a shot-noise contribution to the measured  statistics. The first step is then to distribute through a mass assignment scheme (see for example \cite{2010PhDT.........4J,Sefusatti:2015aex}) each particle's contribution to a cell in a three-dimensional discrete grid, in order to derived a pixelised overdensity distribution $\delta_\mathbf{x}$. Then, estimators should be adapted to take advantage of discrete Fast Fourier Transforms algorithms (FFT) \cite{10.1145/1464291.1464352}.   
This procedure is described in detail in appendices \ref{App:estimators} (and \ref{app:shot_noise} for what concerns the shot-noise subtraction). We review here the key steps and present  the main results derived following the approach introduced by Refs. \cite{Scoccimarro:2000sn,Scoccimarro:2015bla} and used also in Ref. \cite{Tomlinson:2019bjx}. 
We will start here writing integrals (as in an idealized case where the fields were continuous and the volume infinite)  and not discrete sums (realistic case for pixelized fields and finite volumes) to define the estimators in Fourier space to better show the fundamental connection between theoretical modelling and numerical implementation of the estimators. Given a cubic realisation of size $L$ and volume $V=L^3$, the overdensity $\delta_{\bf x}$ is usually pixelised within $N^3$ cells whose volume is $V/N^3$.

Starting from  the expressions for the direct and inverse Fourier transforms of the matter density perturbation field,
\begin{eqnarray}
\delta_\mathbf{k} = \int d\mathbf{x}^3\, \delta_\mathbf{x}\,e^{i\mathbf{kx}}
\quad
\mathrm{and}
\quad
\delta_\mathbf{x} =\dfrac{1}{(2\pi)^3} \int d\mathbf{k}^3\, \delta_\mathbf{k}\,e^{-i\mathbf{kx}} \,,
\end{eqnarray}
\noindent where the integration is over the whole volume in configuration and Fourier space, the following quantities can be defined:
\begin{eqnarray}
\label{eq:IJ_def_maintext}
I_{k}(\mathbf{x}) = \int_{k}\dfrac{d\mathbf{q}^3}{(2\pi)^3}\,\delta_{\mathbf{q}}
e^{-i\mathbf{x}\mathbf{q}}
\quad
\mathrm{and}
\quad
J_{k}(\mathbf{x}) = \int_{k}\dfrac{d\mathbf{q}^3}{(2\pi)^3}\,e^{-i\mathbf{x}\mathbf{q}}\,,
\end{eqnarray}
\noindent where the integration is over the spherical shell with radius  defined by the $k$-vector module and with thickness $\Delta k$. Given these ingredients, we can define the integral version of the power spectrum estimator (bandpower evaluated in a thin shell around $k_1$):
\begin{eqnarray}
\hat{P}\left(k_1\right)
&&=\,\dfrac{(2\pi)^{-3}}{N_\mathrm{p}}
\int_{k_1}d\mathbf{q}_1^3
\int_{k_2}d\mathbf{q}_2^3\,
\delta_{\mathbf{q}_1}\,\delta_{\mathbf{q}_2}\,
\delta^D\left(\mathbf{q}_1 + \mathbf{q}_2\right)
\notag \\
&&=\,
\dfrac{k_\mathrm{f}^3}{(2\pi)^{3}}
\dfrac
{\int_{k_1}d\mathbf{q}_1^3
\int_{k_1}d\mathbf{q}_2^3\,
\delta_{\mathbf{q}_1}\,\delta_{\mathbf{q}_2}\,
\delta^D\left(\mathbf{q}_1 + \mathbf{q}_2\right)}
{\int_{k_1}d\mathbf{q}_1^3
\int_{k_1}d\mathbf{q}_2^3\,
\delta^D\left(\mathbf{q}_1 + \mathbf{q}_2\right)}
\,,
\end{eqnarray}
\noindent where $\delta^D$ is the Dirac's delta and where each integral is performed over a spherical shell of radius $k_i$ for $i=1,2$ and thickness $\Delta k$, the width of the bandpower. $N_\mathrm{p}=V_\mathrm{p}/k_\mathrm{f}^3$ is the number of pairs found inside the integration volume in Fourier space, where $V_\mathrm{p}=4\pi k^2\Delta k$ is the integration volume in Fourier space while $k_\mathrm{f}=2\pi/L$ is the box's fundamental frequency.  
 Converting the integrals into discrete sums as described in appendix \ref{App:estimators} the estimator becomes
\begin{align}
\label{eq:pk_est_text}
\hat{P}\left(k_1\right)
 &= \dfrac{L^3}{N^6}
  \sum_{\imath=1}^{N^3}\, I^D_{k_1}(x_\imath)\, I^D_{k_1}(x_\imath)
 \times
\left[\sum_{\jmath=1}^{N^3}
 J^D_{k_1}(y_\jmath)\, J^D_{k_1}(y_\jmath) \right]^{-1} \,,
\end{align}
\noindent 

where $I^D$ and $J^D$ (full expression reported in equation \ref{eq:IJ_discrete}) denote the pixelized ($D$ stands for discrete as a pixelized  version of a continuous field is often referred to as ``discretized") of the quantities defined in equation \ref{eq:IJ_def_maintext} for a grid with $N^3$ cells.

Similarly, for the bispectrum we start from the unbiased estimator 
% naively 
\begin{eqnarray}
\hat{B}\left(k_1,k_2,k_3\right)
&&=\,
\dfrac{V(2\pi)^{-6}}{N_\mathrm{t}}
\int_{k_1}d\mathbf{q}_1^3
\int_{k_2}d\mathbf{q}_2^3
\int_{k_3}d\mathbf{q}_3^3\,
\delta_{\mathbf{q}_1}\,\delta_{\mathbf{q}_2}\,\delta_{\mathbf{q}_3}\,
\delta^D\left(\mathbf{q}_1 + \mathbf{q}_2 + \mathbf{q}_3\right)
\notag \\
&&=\,
\dfrac{Vk_\mathrm{f}^6}{(2\pi)^{6}}
\dfrac
{\int_{k_1}d\mathbf{q}_1^3
\int_{k_2}d\mathbf{q}_2^3
\int_{k_3}d\mathbf{q}_3^3\,
\delta_{\mathbf{q}_1}\,\delta_{\mathbf{q}_2}\,\delta_{\mathbf{q}_3}\,
\delta^D\left(\mathbf{q}_1 + \mathbf{q}_2 + \mathbf{q}_3\right)}
{\int_{k_1}d\mathbf{q}_1^3
\int_{k_2}d\mathbf{q}_2^3
\int_{k_3}d\mathbf{q}_3^3\,
\delta^D\left(\mathbf{q}_1 + \mathbf{q}_2 + \mathbf{q}_3\right) }
\,,
\end{eqnarray}
\noindent where $N_\mathrm{t}=V_\mathrm{t}/k_\mathrm{f}^6$ is the number of triangles included in the integration volume in Fourier space ($V_\mathrm{t}=8\pi^2 k_1k_2k_3\Delta k^3$ is the integration volume in Fourier space for the bispectrum). As for the power spectrum, the above estimator can be expressed in terms of discrete sums
(see equation \ref{eq:bk_expansion0}):
\begin{align}
\label{eq:bk_est_text}
\hat{B}\left(k_1,k_2,k_3\right)&=
  \left(\dfrac{L^6}{N^9}\right)
 \sum_{\imath=1}^{N^3}
 \, I^D_{k_1}(x_\imath)\, I^D_{k_2}(x_\imath) \, I^D_{k_3}(x_\imath) 
\times
\left[\sum_{\jmath=1}^{N^3}
 J^D_{k_1}(y_\jmath)\, J^D_{k_2}(y_\jmath)\, J^D_{k_3}(y_\jmath) \right]^{-1}
\,.
\end{align}

%

%%%%%%%%%%%%%%%%%%%%%%%%%%%%%%%%%%%%%%%%%%%%%
%%%%%%%%%%%%%%%%%%%%%%%%%%%%%%%%%%%%%%%%%%%%%
\subsubsection{i-trispectrum}
\label{sec:tk_estimator}

The idea to define an estimator for the trispectrum which depends only on the magnitude of the four $k$-modes defining the quadrilateral was pioneered by  \cite{Sefusatti:2004xz}.  Here, using the approach of \cite{Scoccimarro:2015bla} we make the estimator fast enough to be applicable to a large set of independent synthetic realisations and eventually on data and, by performing explicitly the angular integrations, make it  transparently connected to the theoretical model for the trispectrum.

Analogously to the power spectrum and bispectrum, the  estimator for the four-point correlator could  be defined as a quantity proportional to a suitable  average of  four $\delta_{\bf k}$ modes. However, differently from the power spectrum and bispectrum cases, this will not be a particularly useful i-trispectrum estimator; when taking the ensemble average of such quantity, two terms appear --according to Wick's theorem--: a connected (c) and an unconnected (u) part.

So the  estimator  simply identified with the four-point correlator estimator will include both these contributions:
\begin{eqnarray}
\label{eq:tk_est_text_tot}
&&\hat{\mathcal{T}}_\mathrm{c+u}\left(k_1,k_2,k_3,k_4\right)
=\,
\dfrac{V^2(2\pi)^{-9}}{N_\mathrm{q}}
\int_{k_1}d\mathbf{q}_1^3
\int_{k_2}d\mathbf{q}_2^3
\int_{k_3}d\mathbf{q}_3^3
\int_{k_4}d\mathbf{q}_4^3\,
\notag \\
&&\times
\delta_{\mathbf{q}_1}\,\delta_{\mathbf{q}_2}\,
\delta_{\mathbf{q}_3}\,\delta_{\mathbf{q}_4}\,
\delta^D\left(\mathbf{q}_1 + \mathbf{q}_2 + \mathbf{q}_3 + \mathbf{q}_4\right)
\notag \\
&&=\,
\dfrac{V^2k_\mathrm{f}^9}{(2\pi)^{9}}
\dfrac{
\int_{k_1}d\mathbf{q}_1^3
\int_{k_2}d\mathbf{q}_2^3
\int_{k_3}d\mathbf{q}_3^3
\int_{k_4}d\mathbf{q}_4^3\,
\delta_{\mathbf{q}_1}\,\delta_{\mathbf{q}_2}\,
\delta_{\mathbf{q}_3}\,\delta_{\mathbf{q}_4}\,
\delta^D\left(\mathbf{q}_1 + \mathbf{q}_2 + \mathbf{q}_3 + \mathbf{q}_4\right)
}
{
\int_{k_1}d\mathbf{q}_1^3
\int_{k_2}d\mathbf{q}_2^3
\int_{k_3}d\mathbf{q}_3^3
\int_{k_3}d\mathbf{q}_4^3\,
\delta^D\left(\mathbf{q}_1 + \mathbf{q}_2 + \mathbf{q}_3+ \mathbf{q}_4\right) 
}
\,,
\notag \\
\end{eqnarray}
\noindent where $N_\mathrm{q}=V_\mathrm{q}/k_\mathrm{f}^9$ is the number of skew-quadrilaterals in the integration volume in Fourier space (see \cite{Gualdi:2020ymf}'s appendix for a derivation of the trispectrum integration volume in Fourier space $V_\mathrm{q}=16\pi^3k_1k_2k_3k_4\Delta k^5$). Notice that this estimator automatically averages over all the possible shapes and permutations included in the model of equation \ref{eq:tk_int_real}.

In fact, 
\begin{align}
\label{eq:tk_unbiased}
\langle\hat{\mathcal{T}}_\mathrm{c+u}\left(k_1,k_2,k_3,k_4\right)\rangle
&=
\dfrac{V^2(2\pi)^{-9}}{N_\mathrm{q}}
\int_{k_1}d\mathbf{q}_1^3
\int_{k_2}d\mathbf{q}_2^3
\int_{k_3}d\mathbf{q}_3^3
\int_{k_4}d\mathbf{q}_4^3\,
\notag \\
&\times
\left[
\langle
\delta_{\mathbf{q}_1}\,\delta_{\mathbf{q}_2}\,
\delta_{\mathbf{q}_3}\,\delta_{\mathbf{q}_4}
\rangle_\mathrm{c}
\,+\, 
\langle\delta_{\mathbf{q}_1}\,\delta_{\mathbf{q}_2}\rangle
\langle\delta_{\mathbf{q}_3}\,\delta_{\mathbf{q}_4}\rangle \, + \,2\,\mathrm{p.}
\right]\,
\delta^D\left(\mathbf{q}_1 + \mathbf{q}_2 + \mathbf{q}_3 + \mathbf{q}_4\right)\,,
\end{align}
\noindent  where $\langle
\delta_{\mathbf{q}_1}\,\delta_{\mathbf{q}_2}\,
\delta_{\mathbf{q}_3}\,\delta_{\mathbf{q}_4}
\rangle_\mathrm{c}$ denotes the connected i-trispectrum part, and the three permutations of the product of two terms proportional to the power 
spectrum ($\langle\delta_{\mathbf{q}_i}\,\delta_{\mathbf{q}_j}\rangle\langle\delta_{\mathbf{q}_{i'}}\,\delta_{\mathbf{q}_{j'}}\rangle$) are the unconnected part. The 
unconnected part is not of particular interest  here, since it is completely determined by the field's power spectrum. 

The signal of interest and the one we want to isolate is the connected part, the i-trispectrum. 
In other words, the i-trispectrum estimator should be obtained by subtracting the unconnected part from the total signal measured by the four-point correlator estimator in equation \ref{eq:tk_est_text_tot}:
\begin{eqnarray}
\label{eq:tk_est_text}
\hat{\mathcal{T}}\left(k_1,k_2,k_3,k_4\right)\equiv \hat{\mathcal{T}}_\mathrm{c}\left(k_1,k_2,k_3,k_4\right)=\hat{\mathcal{T}}_\mathrm{c+u}\left(k_1,k_2,k_3,k_4\right)-\hat{\mathcal{T}}_\mathrm{u}\left(k_1,k_2,k_3,k_4\right)\,.
\end{eqnarray}

\noindent In the cosmological context of interest and for Gaussian initial conditions, 
perturbation theory predicts the unconnected part (which is intrinsically a linear order quantity) to be dominant with respect to the connected one (which is intrinsically third order). This will later be proven using the simulations measurements, with the unconnected part resulting to be two orders of magnitude larger than the connected one.

As derived in appendix \ref{app:tk_connected} the discretised estimator for the total signal of the four-point correlator in Fourier space is\footnote{Recall that we describe the i-trispectrum using only the four sides of the quadrilateral as coordinates and not also the diagonal (as introduced to compute the theoretical expression for the estimator). This is motivated by  numerical computation time considerations. If we had used the diagonal $D$ as additional coordinate, by writing the estimator with the closure conditions $\delta^D\left(\mathbf{q}_1 + \mathbf{q}_2 + \mathbf{D}\right) \times
\delta^D\left(\mathbf{q}_3+ \mathbf{q}_4 - \mathbf{D}\right)$, in the above expression there would have been a double sum over each grid element, implying a computation scaling as $\propto N^6$ instead of $\propto N^3$. Introducing the diagonal here is not necessary because the estimator defined as a function of $(k_1,k_2,k_3,k_4)$ automatically integrates over all possible shapes given by the allowed values of the diagonal $D$ (and the folding angle $\psi$).}
 \begin{align}
 \label{eq:tk_total_est}
\hat{\mathcal{T}}_\mathrm{c+u}\left(k_1,k_2,k_3,k_4\right)&=
 \dfrac{L^9}{N^{12}}
 \sum_{\imath=1}^{N^3}
 \, I^D_{k_1}(x_\imath)\, I^D_{k_2}(x_\imath) \, I^D_{k_3}(x_\imath) \, I^D_{k_4}(x_\imath) 
 \notag \\
&\times
\left[\sum_{\jmath=1}^{N^3}
 J^D_{k_1}(y_\jmath)\, J^D_{k_2}(y_\jmath)\, J^D_{k_3}(y_\jmath)\, J^D_{k_4}(y_\jmath) \right]^{-1}
\,.
\end{align}
In order to isolate the unconnected term and to subtract it from the total signal measured from the simulations  we define the estimator
\begin{align}
\label{eq:tku_est_text}
\hat{\mathcal{T}}_\mathrm{u}\left(k_1,k_2,k_3,k_4\right)
&=\dfrac{V(2\pi)^{-9}}{N_\mathrm{q}}
\int_{k_1}d\mathbf{q}_1^3
\int_{k_2}d\mathbf{q}_2^3
\,\delta_{\mathbf{q}_1}\,\delta_{\mathbf{q}_2}\,
\delta^D\left(\mathbf{q}_1 + \mathbf{q}_2\right)
\notag \\
&\times
\int_{k_3}d\mathbf{q}_3^3
\int_{k_4}d\mathbf{q}_4^3\,
\delta_{\mathbf{q}_3}\,\delta_{\mathbf{q}_4}\,
\delta^D\left(\mathbf{q}_3 + \mathbf{q}_4\right)
+ \quad 2\quad\mathrm{p.}
\,,
\end{align}
\noindent 

\noindent where the above expression captures the fact that the unconnected part by definition is non-zero when at least one of the terms $\propto P^2$ -- each involving product of the two-point correlators  $\langle\delta\delta\rangle\langle\delta \delta\rangle$-- is non-zero.
This contribution then appears only for those quadrilaterals sets with two pairs of equal sides (or all four equal sides). The corresponding discretised estimator is (appendix \ref{sec:tk_unconnected}) 
\begin{align}
\label{eq:tku_expansion1_text}
\hat{\mathcal{T}}_\mathrm{u}\left(k_1,k_2,k_3,k_4\right)
 &=
  \dfrac{L^9}{N^{15}}\,
 \sum_{\imath=1}^{N^3}
 \, I^D_{k_1}(x_\imath)\, I^D_{k_2}(x_\imath)\,
  \sum_{\ell=1}^{N^3}
 \, I^D_{k_3}(z_\ell)\, I^D_{k_4}(z_\ell) 
 \notag \\
 &\times
\left[\sum_{\jmath=1}^{N^3}
 J^D_{k_1}(y_\jmath)\, J^D_{k_2}(y_\jmath)\, J^D_{k_3}(y_\jmath)\, J^D_{k_4}(y_\jmath) \right]^{-1}
\,.
\end{align}
\noindent From equation \ref{eq:tku_est_text} we can also derive an analytical expression to model the unconnected signal of the four-point correlator in Fourier space $\mathcal{T}^\mathrm{th}_\mathrm{u}$ (see equation \ref{eq:tk_unbiased_unconnected_ana} for the derivation):
\begin{align}
\label{eq:tk_unbiased_unconnected_ana_text}
\langle\hat{\mathcal{T}}\left(k_1,k_2,k_3,k_4\right)_\mathrm{u}\rangle
&\approx
\,.
\end{align}
\noindent where $\delta^K_{ij}$ denotes the Kronecker delta for $\mathbf{k}_i-\mathbf{k}_j$ and $S_{1234}$ is a symmetry factor whose value is one in case the quadrilateral has two pairs of equal sides, three if the four sides are the same and zero otherwise. It is important to notice that since the quadrilateral diagonal $D$ is not a fixed parameter given the four sides, $\Delta D=D_\mathrm{max}-D_\mathrm{min}$. $\Delta k_i$ are the bin sizes chosen for the $k$-modes and the diagonal $D$. These will be later defined in section \ref{sec:analysis_setup} in terms of the fundamental frequency $k_\mathrm{f}$ given by the analysis set-up.

We are aware that another option would be to simply discard all the symmetric configurations and by doing so avoiding the quadrilaterals sets whose i-trispectrum has an unconnected part.
While an interesting avenue to explore, we find that it would imply a loss of constraining power, therefore here we proceed with the estimator of equation  \ref{eq:tk_est_text}, since it allows harvesting the information contained in the i-trispectrum of symmetric configurations.

%%%%%%%%%%%%%%%%%%%%%%%%%%%%%%%%%%%%%%%%%%%%
%%%%%%%%%%%%%%%%%%%%%%%%%%%%%%%%%%%%%%%%%%%%
%%%%%%%%%%%%%%%%%%%%%%%%%%%%%%%%%%%%%%%%%%%%
\begin{figure}[tbp]
\centering 
\includegraphics[width=1.\textwidth]%,trim=0 380 0 200,clip]
{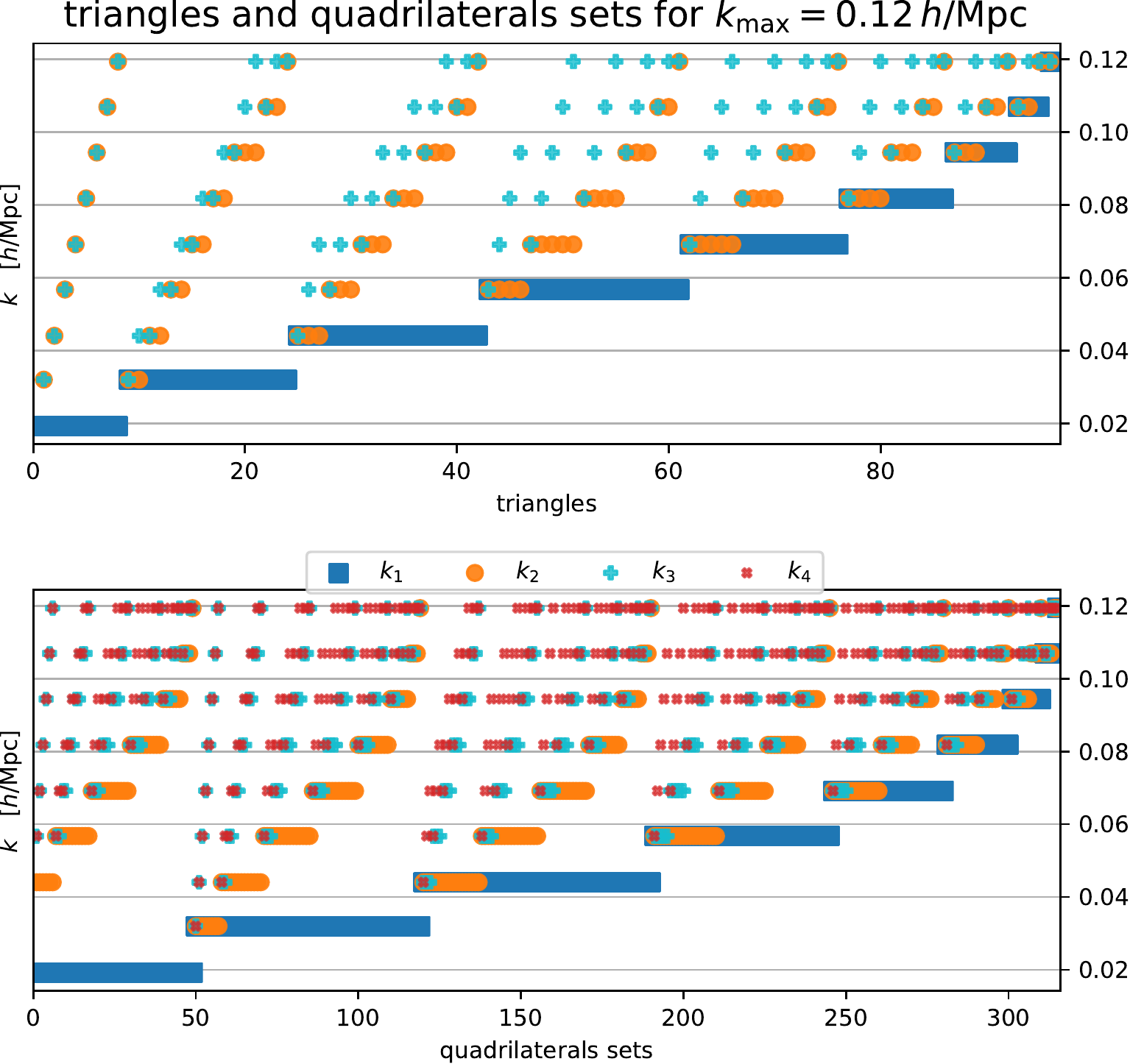}
\caption{\label{fig:bk_tk_configurations}Illustration of the convention adopted  for the ordering of triangles and quadrilaterals sets in the bispectrum and i-trispectrum data-vectors, respectively. $k_1$ increases from left to right while the other sides follow the ordering $k_1\leq k_2\leq k_3$ and $k_1\leq k_2\leq k_3\leq k_4$ for both triangles and quadrilaterals sets.}
\end{figure}

%%%%%%%%%%%%%%%%%%%%%%%%%%%%%%%%%%%%%%%%%%%%%
%%%%%%%%%%%%%%%%%%%%%%%%%%%%%%%%%%%%%%%%%%%%%
\subsection{Covariance matrix estimation and modelling}
\label{sec:cov_methodology}
 
The power spectrum, bispectrum and i-trispectrum for different $k$-bins, triangles and quadrilaterals sets can be organised into a data-vector. 
In our case, the measured power spectrum, bispectrum and i-trispectrum joint data-vector is given by the elements
\begin{eqnarray}
\mathbf{d}=\left(\hat{P}_{k_1},...\hat{P}_{k_\mathrm{max}^P};\hat{B}_{k_1k_1k_1},...\hat{B}_{k_\mathrm{max}^Bk_\mathrm{max}^Bk_\mathrm{max}^B};\hat{\mathcal{T}}_{k_1k_1k_1k_1},...\hat{\mathcal{T}}_{k_\mathrm{max}^\mathcal{T}k_\mathrm{max}^\mathcal{T}k_\mathrm{max}^\mathcal{T}k_\mathrm{max}^\mathcal{T}}\right)\,,
\end{eqnarray}
\noindent where each statistic is characterised by its own maximum $k$-value, $k_\mathrm{max}$.
The theoretical counterpart of this data-vector is composed by the adopted model for each of the statistics, which we label ${\bf d}_{\rm th}$. 
Sometimes for brevity we will refer to $\mathbf{d}(k_{\rm max})$ as the data-vector defined above measured for the same $k_{\rm max}$ for all statistics. 
The configurations ordering for both bispectrum and i-trispectrum in terms of the $k$-values of the triangle and quadrilaterals sets sides is shown in figure \ref{fig:bk_tk_configurations}.
 In order to avoid double-counting shapes, once $k_1$ is chosen the other sides respect the ordering  $k_1\leq k_2\leq k_3$ and $k_1\leq k_2\leq k_3\leq k_4$ for triangles and quadrilaterals sets respectively.
In the figure, $k_1$ (in blue) increases from left to right and the $x$-axis reports the label (index) of the configuration. 
For each configuration the figure shows the length of $k_2$ (orange), $k_3$ (cyan) and $k_4$ (red).

Besides the estimator for the statistic of choice (which comprises the elements of the data-vector $\mathbf{d}$) and the theoretical expression of the estimator
(elements of ${\bf  d}_{\rm th}$),
the remaining key ingredient of all the parameter constraints analyses and forecasts is the covariance matrix, which encodes the uncertainty and correlation present in the data-vector. The uncertainty is related to the error associated with the individual element measurement while correlation quantifies  how dependent to each other the different elements of the data-vector are. 

The covariance matrix of the i-trispectrum and the cross-covariance with power spectrum and bispectrum are fundamental ingredients for assessing the impact of adding this four-point statistic to a parameter constraints analysis. In what follows, for brevity we refer to the full joint power spectrum, bispectrum and i-trispectrum covariance, or data-vector, as PB$\mathcal{T}$. Usually in this kind of forecasts studies, an analytical template is used and often the covariance is assumed to be diagonal (i.e., no correlation between different $k$-bins), with only a Gaussian component (i.e., ignoring contributions from non-linear clustering), together with assuming no cross-correlation between different $n$-point statistics.

Alternatively, given a set of measurements from independent realisations of the data-vector $\mathbf{d}$ of interest, the covariance matrix $\mathbf{Cov}$ can be numerically estimated by 
\begin{eqnarray}
\mathbf{Cov}_{\bf d} =  \dfrac{1}{N_\mathrm{sim}-1}\sum^{N_\mathrm{sim}}_{i=1}
\left(\mathbf{d}_i-\langle\mathbf{d}\rangle\right)
\left(\mathbf{d}_i-\langle\mathbf{d}\rangle\right)^\intercal\,,
\end{eqnarray}
\noindent where $N_\mathrm{sim}$ is the number of independent realisations
(here  we use 5000 realisations see section~\ref{sec:meas_vs_model})
used to measure the data-vector and $\langle\mathbf{d}\rangle$ denotes the average across the whole set of realisations and $\intercal$ denotes the transpose.

The inverse of a covariance matrix $\mathbf{Cov}_\mathbf{\mathbf{d}}^{-1}$ estimated from a finite number of realisation is known to be biased \cite{Hartlap:2006kj,Sellentin:2015waz}.
We correct this by using the Hartlap factor $h_\mathrm{f} = (N_\mathrm{sim} - N_\mathbf{d} -2)/(N_\mathrm{sim}-1)$ \cite{Hartlap:2006kj}, where $N_\mathbf{d}$ is the data-vector's size while $N_\mathrm{sim}$ is the number of simulations used to estimate the covariance (5000 in our case\footnote{While the \textsc{Quijote} \cite{Villaescusa-Navarro:2019bje}  suite offers more than 5000 realisations, we only use 5000 for each redshift  for computational reasons. Given the maximum length of our data-vector, this number is more than enough to estimate the inverse of the covariance matrix. In fact the Hartlap factor is $h_\mathrm{f}=0.51$ or higher.}). Notice that $h_\mathrm{f}\rightarrow1$ when $N_\mathrm{sim}\gg N_\mathbf{d}$. 
A more accurate correction to the inverse of the covariance matrix  estimated from a finite number of independent realisations and its effect on the resulting likelihood  was introduced by \cite{Sellentin:2015waz}. We tested that for the purpose of evaluating the added benefit of including the i-trispectrum in the analysis, the difference between the two corrections is negligible. 

Often the lack of a large number of independent realisations is the reason why resorting to an analytical model for the covariance is appealing. 
In particular, if the overdensity field can be  approximated by a Gaussian random field, the covariance matrix  can be modelled analytically (especially for simulated cubic boxes).  We refer to this approximation as Gaussian and its non-zero terms in the covariance as Gaussian terms. 
 Recent studies on the power spectrum and bispectrum covariance matrices and their analytical models can be found in Refs. \cite{Sugiyama:2019ike,Wadekar:2019rdu,Gualdi:2020ymf}. 

For what concerns the i-trispectrum, we compute, for the first time, the Gaussian terms at leading order following the approach developed, for the power spectrum and bispectrum,  in \cite{Gualdi:2017iey,Gualdi:2018pyw}. 

The  real space matter field i-trispectrum covariance matrix is a specific case of the more general covariance in redshift space for biased tracers, whose derivation is reported in the appendix \ref{app:tk_cov_rsd} at equation \ref{eq:t0_gg_cov}. 
Since the kernels of the real space matter field do not depend on the relative orientation of the $k$-vectors, and there is no dependence of each $k$-vector orientation with respect to the line of sight, no integration is required. Consequently, the analytical expression for the diagonal elements of ${\bf Cov}$ in the Gaussian approximation (see equation \ref{eq:t0_gg_cov} for full expression) reduces to
\begin{eqnarray}
\label{eq:t0_gg_cov_matter}
&&\mathrm{C}^{\mathcal{T}_u\mathcal{T}_u}_{\mathrm{G}}\left(k_1,k_2,k_3,k_4\right) =
% \notag\\
% &=&
\dfrac{(2\pi)^{9}\mathrm{R_{1234}}}{V_\mathrm{s}V^{\mathrm{q}}_{1234}}
\times
P\left(k_1\right)
P\left(k_2\right)
P\left(k_3\right)
P\left(k_4\right)\,,
\end{eqnarray}{}
\noindent where 
the expression of the Fourier integration volume,  $V^\mathrm{q}_{1234}$, is  given in equation \ref{eq:pk_tk_int_volumes_ana};
$\mathrm{R}_{1234}$ is a symmetry factor (given by  $n!$, where $n$ is the number of repeated $k$ modules)  that counts the number of possible permutations of equal sides between the two identical quadrilaterals sets, and its values for a given symmetry are given in equation \ref{eq:R_sym_fac};
 $V_\mathrm{s}$ is the effective survey volume.

We will compare the diagonal elements of the covariance numerically estimated from the simulations with the above theoretical model later in section \ref{sec:cov_results} to show that for the i-trispectrum the Gaussian approximation is only reasonable at high redshifts.

%%%%%%%%%%%%%%%%%%%%%%%%%%%%%%%%%%%%%%%%%%%%%
%%%%%%%%%%%%%%%%%%%%%%%%%%%%%%%%%%%%%%%%%%%%%
\subsection{Primordial non-Gaussianity imprint on the i-trispectrum}
\label{sec:png_methodology}

Deviations from Gaussianity of the primordial gravitational potential \cite{Bardeen:1980kt} are a powerful tool to constrain models of inflation \cite{Bartolo:2004if} describing how the large-scale structure we currently observe was generated. 
In this work we focus on primordial non-Gaussianities (PNG) of the local type which are parametrised by an amplitude parameter $f^\mathrm{local}_\mathrm{nl}$ (we will refer to it as $f_\mathrm{nl}$ from now on for simplicity).
The Bardeen's gravitational potential $\Phi$ \cite{Bardeen:1980kt} can be written as a polynomial expansion in terms of a Gaussian field $\phi$ with coefficients $f_\mathrm{nl}$ and $g_\mathrm{nl}$ up to third order:

\begin{eqnarray}
\Phi(\mathbf{x})=\phi(\mathbf{x})+\dfrac{f_{\mathrm{nl}}}{c^2}\left[\phi^2(\mathbf{x})-\langle\phi^2(\mathbf{x})\rangle\right]+\dfrac{g_{\mathrm{nl}}}{c^4}\left[\phi^3(\mathbf{x})-3\phi(\mathbf{x})\langle\phi^2(\mathbf{x})\rangle\right] \;+\;.\,.\,. 
\end{eqnarray}{}

\noindent where $c$ is the speed of light which is needed in the expansion since in this formalism $\phi$ has units of $c^2$. For more details see appendix \ref{sec:app_png}.

A detection of $|f_\mathrm{nl}|\gtrsim 1$ would rule out single field inflationary models since they predict a much smaller value of the same parameter \cite{Maldacena:2002vr,Creminelli:2004yq}, while at the same time it would favour multi-field inflationary models or alternatives to inflation.
At the moment of writing, the tightest constraints, from the \textit{Planck} analysis of CMB data, are $f_\mathrm{nl} =  0.9 \pm  5.1$ (at $68\%$ confidence
level) \cite{Akrami:2019izv}.
For large-scale structure, using the DR14 eBOSS quasars sample and exploiting the scale dependent bias effect induced by primordial non-Gaussianity \cite{Dalal:2007cu,Matarrese:2008nc} (see \cite{Barreira:2020ekm} for a recent study on the impact of galaxy bias uncertainties on PNG constraints), the state of the art constraints are $-26<f_\mathrm{nl}<14$ (at $68\%$ confidence level) \cite{Castorina:2019wmr}.

To date, only the bispectrum has been thoroughly studied in the literature  for its sensitivity to primordial non-Gaussianity signatures at the lowest-order in perturbation theory \cite{Verde:1999ij,Komatsu:2001rj,Sefusatti:2009qh,Jeong:2009vd,Scoccimarro:2003wn}. 
Recently Ref.~\cite{Gualdi:2020ymf} suggested that the bispectrum anisotropic signal could boost even further the late-time constraints on $f_\mathrm{nl}$. The expressions for the primordial non-Gaussianity  contributions  to the matter power spectrum and bispectrum are reported in appendix \ref{sec:app_png}.

The large-scale structure trispectrum as a test for Gaussianity of the initial conditions was proposed in \cite{Verde:2001pf}. Since then the trispectrum has been mainly considered for CMB analyses, targeted to produce constraints on primordial non-Gaussianity \cite{Munshi:2009wy,Mizuno:2010by,Kamionkowski:2010me,Regan_2015,Izumi:2011di,Smith:2015uia,Fergusson:2010gn}, in particular with \textit{Planck} \cite{Feng:2015pva} deriving constraints on both $\tau_\mathrm{nl}=0.4\pm0.9\times10^4$ and $g_\mathrm{nl}=-1.2\pm2.8\times10^5$. For what concerns late-time analyses, the 21-cm background anisotropies trispectrum was studied in Ref. \cite{Cooray:2008eb}, while Ref. \cite{Bellomo:2018lew} included the trispectrum of LSS to measure the energy-scale of inflation through its relation with primordial non-Gaussianities.

As described in detail in appendix \ref{sec:subsec_tkpng}, and following \cite{Jeong:2009vd,Sefusatti:2009qh}, one can see that in the presence of primordial non-Gaussianity, when expanding the four-point correlator (with shorthand notation $\langle \delta \delta\delta\delta\rangle$)  whose connected component is the trispectrum, four terms appear:
\begin{eqnarray}
\langle\delta\delta\delta\delta\rangle =
T_{(1111)}+T_{(1112)}+T_{(1122)}+T_{(1113)}\,. 
\end{eqnarray}{}
\noindent The third and fourth terms, $T_{(1122)}$ and $T_{(1113)}$, represent the standard trispectrum due to gravitational collapse (when limiting the expansion to order $\propto\delta^6$, or in terms of the Gaussian primordial potential $\propto\phi^6$). The first term $T_{(1111)}$ produces two terms proportional to $f_\mathrm{nl}^2$ and $g_\mathrm{nl}$, respectively. The second term $T_{(1112)}$ is a mixture of primordial non-Gaussianity and gravitational non-linear evolution and it is proportional to $f_\mathrm{nl}$. Therefore the model for the primordial non-Gaussianity imprint on the  matter trispectrum  in real space, $T^\mathrm{PNG}$, is given by\footnote{The corresponding redshift space quantity for the matter field is simply obtained by replacing the matter case kernels $F^{(i)}$ with the redshift space ones $Z^{(i)}$ as done in appendix \ref{app:theo_models_rsd} for the purely gravitational terms $T_{(1122)}$ and $T_{(1113)}$.},
\begin{eqnarray}
\label{eq:png_tk_text}
&&T^{\mathrm{PNG}}(\mathbf{k}_1,\mathbf{k}_2,\mathbf{k}_3,\mathbf{k}_4) =T_{(1111)}+ T_{(1112)}
\notag\\[2mm]
&&=\,
\times\,
\Bigg\{\dfrac{f^2_{\mathrm{nl}}}{c^4}4\,
\dfrac{\mathcal{M}_{k_3}\mathcal{M}_{k_4}}{\mathcal{M}_{k_1}\mathcal{M}_{k_2}}
P({k_1})P({k_2})
\left[ 
\dfrac{P({|\mathbf{k}_1+\mathbf{k}_3|})}
{{\mathcal{M}_{|\mathbf{k}_1+\mathbf{k}_3|}^2}}
+
\dfrac{P({|\mathbf{k}_1+\mathbf{k}_4|})}
{{\mathcal{M}_{|\mathbf{k}_1+\mathbf{k}_4|}^2}}
\right]\;+\;5\,\mathrm{p.}\notag \\
[2mm]
&&+\,
\dfrac{g_{\mathrm{nl}}}{c^4}
\left[6
\mathcal{M}_{k_4}
\dfrac{P({k_1})P({k_2})P({k_3})}
{\mathcal{M}_{k_1}\mathcal{M}_{k_2}\mathcal{M}_{k_3}}
\;+\;3\,\mathrm{p.}\right]\Bigg\}\,
\notag \\[2mm]
&&+\,
\dfrac{f_{\mathrm{nl}}}{c^2}
\times\,
\Bigg\{
\left[
4 
\dfrac{\mathcal{M}_{k_1}}{\mathcal{M}_{k_2}}
P({k_2})P({k_3})
\dfrac{P({|\mathbf{k}_3+\mathbf{k}_4|})}
{\mathcal{M}_{|\mathbf{k}_3+\mathbf{k}_4|}}
F^{(2)}\left[-\mathbf{k}_3,\mathbf{k}_3+\mathbf{k}_4\right]\; + \;5\,\mathrm{p.}\right]
\notag\\[2mm]
&&+\,
\left[
2
\dfrac{\mathcal{M}_{|\mathbf{k}_3+\mathbf{k}_4|}}{\mathcal{M}_{k_1}\mathcal{M}_{k_2}}
P({k_1})P({k_2})P({k_3}) F^{(2)}\left[\mathbf{k}_3+\mathbf{k}_4,-\mathbf{k}_3\right]\;+\;2\,\mathrm{p.}
\right]
\Bigg\} \;+\;3\,\mathrm{p.}
\,.
\notag \\
\end{eqnarray}{}
\noindent where  $\mathcal{M}_k=\left(3D_+/5\Omega_\mathrm{m}H_0^2\right)\,k^2\mathbb{T}_k$, with $D_+$ being the linear growth factor, $\Omega_\mathrm{m}$ the matter density parameter, $H_0$ the Hubble constant and $\mathbb{T}(k)$ the transfer function. 
From the above equation we can see that the term that has the potential to significantly improve the bispectrum constraining power for  $f_\mathrm{nl}$ is $T_{(1112)}$. This is because $T_{(1112)}$ has a similar functional form and $k$-dependence as the matter bispectrum PNG correction (equation \ref{eq:png_bk} in the appendix).

%%%%%%%%%%%%%%%%%%%%%%%%%%%%%%%%%%%%%%%%%%%%
%%%%%%%%%%%%%%%%%%%%%%%%%%%%%%%%%%%%%%%%%%%%
%%%%%%%%%%%%%%%%%%%%%%%%%%%%%%%%%%%%%%%%%%%%
\begin{figure}[tbp]
\centering 
\includegraphics[width=0.99\textwidth]%,trim=0 380 0 200,clip]
{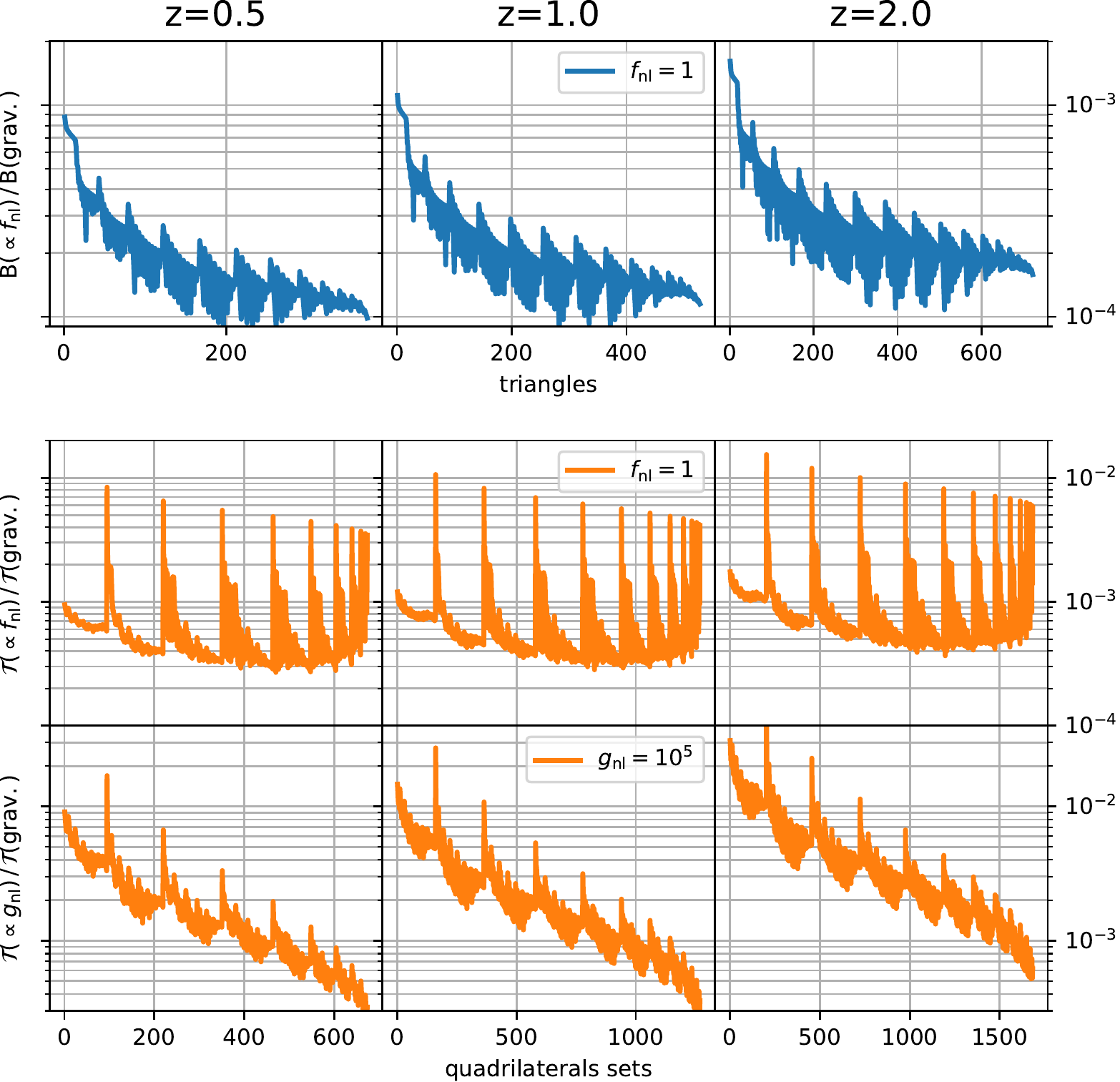}
\caption{\label{fig:bktk_png} Primordial imprint on the matter bispectrum and i-trispectrum in real space.
The first two rows from the top show the ratio between the primordial signal proportional to $f_\mathrm{nl}$ and the gravitational component for the bispectrum and i-trispectrum, for $f_\mathrm{nl}=1$. In the bottom row the i-trispectrum signal proportional to $g_\mathrm{nl}$, also divided by the gravitational part, is displayed for  $g_\mathrm{nl}=10^5$. The analytical expression for these terms is given in equation \ref{eq:png_tk_text}. In particular for the quadrilateral symmetric configurations, the integrals in the model of equation \ref{eq:tk_int_real} have been checked to converge for $k_\mathrm{min}\lesssim10^{-4}$ $h/\mathrm{Mpc}$.
}
\end{figure}

Figure \ref{fig:bktk_png} shows, for matter bispectrum and i-trispectrum in real space,  the ratio between primordial and gravitational components for $f_\mathrm{nl}=1$ and $g_\mathrm{nl}=10^5$. For the bispectrum, as expected, the ratio is ${\cal O}\sim 10^{-3}-10^{-4}$ depending on scale, for the i-trispectrum there are configurations where this ratio is larger and reaches ${\cal O} \sim 10^{-2}$.  These are symmetric  configurations of the i-trispectrum and the boost is driven by the  sum of $k$-vectors appearing at the denominator of one of the two components of $T_{(1112)}$ reported in equation \ref{eq:png_tk_text}.

%%%%%%%%%%%%%%%%%%%%%%%%%%%%%%%%%%%%%%%%%%%%
%%%%%%%%%%%%%%%%%%%%%%%%%%%%%%%%%%%%%%%%%%%%
%%%%%%%%%%%%%%%%%%%%%%%%%%%%%%%%%%%%%%%%%%%%
\subsection{Fisher-based forecasts}
\label{sec:fisher_method}
Before developing all the required theoretical and technical tools necessary to employ a new statistics in the analysis of  real survey data,  there are mainly two ways to assess the additional information harvested by measuring this new statistics on a cosmological field. One method is to look at the general improvement of the full data-vector signal-to-noise (including both old and new statistics). The other, more specific, alternative is to forecast the improvements on the parameters of interest constraints obtained by employing the new statistics. 

With the covariance matrix ${\bf Cov}$ for a given data-vector in hand, we can define two  quantities for these purposes.
These are the cumulative signal-to-noise (S/N) and the Fisher information matrix (F); from the latter, both conditional and marginalised errors on the model parameters can be estimated.

The (expected) cumulative signal-to-noise as a function of the maximum $k$-value, given a (theoretical description of a) data-vector  and the relative covariance matrix $\mathbf{Cov}_\mathbf{d}$, is computed as
\begin{eqnarray}
\label{eq:signal_noise}
\left(\mathrm{S/N}\right)\quad = \quad\sqrt{ \mathbf{d}^\intercal_\mathrm{th}\;\mathbf{Cov}_\mathbf{d}^{-1} \; \mathbf{d}_\mathrm{th}}\,.
\end{eqnarray}
\noindent where the data-vectors ${\bf d}$ should be interpreted as functions of $k_{\rm max}$.

Given a theoretical model description of the  data-vector, ${\bf d}_{\rm th}(\theta_i)$, with dependence on model parameters $\theta_i$, the Fisher information matrix is the Hessian matrix of the associated log likelihood, $\mathcal{L}$, with the derivatives taken with respect to the model parameters,
\begin{eqnarray}
F_{ij}\quad=\quad -\Bigg\langle\dfrac{\partial^2\mathcal{L}}{\partial\theta_i\partial\theta_j} \Bigg\rangle\Bigg|_{\theta_\mathrm{ml}}\,
\end{eqnarray}
\noindent where  $\theta_\mathrm{ml}$ indicates that the derivatives are  evaluated at  the  likelihood maximum.

In the case of a single-parameter estimate, the minimum achievable error is given by $\Delta\theta^i_\mathrm{min}=1/\sqrt{F_{ii}}$. In the general multi-parameter case, this is the conditional error as it assumes that all other parameters are perfectly known.  To obtain  the marginal error, in the multi-parameter case the inverse of the Fisher information matrix needs to be used 
$\Delta\theta^i_\mathrm{min}=\sqrt{F_{ii}^{-1}}$ \cite{Tegmark:1996bz}.
Using the data-vector's covariance matrix and its derivatives with respect to the model parameters, each element of the Fisher information matrix can be computed as
\begin{eqnarray}
\label{eq:fish_info}
F_{ij} = \dfrac{\partial \mathbf{d}_{\rm th}^\intercal}{\partial\theta_i}\,\mathbf{Cov}_\mathbf{d}^{-1}\,\dfrac{\partial \mathbf{d}_{\rm th}}{\partial\theta_j}
\end{eqnarray}
 since here (as in most Fisher forecasts and in most large-scale structure studies) we adopt  a Gaussian likelihood  with fixed covariance matrix (see e.g., \cite{Carron:2012pw,Kalus:2015lna}).

In what follows we will make use of another related quantity: the cumulative $\chi^2$ as a function of  $k_\mathrm{max}$. Given a measured data-vector ${\bf d}$ and its theoretical model ${\bf d}^{\rm th}$, the cumulative $\chi^2$ can be computed using the full covariance matrix, $\chi^2_{\rm Cov}$, or  can be approximated (as often done in practice) by considering only the diagonal elements, $\chi^2_{\sigma^2}$. These expressions read,
\begin{eqnarray}
\label{eq:chi_squared}
\chi^2_\mathrm{Cov}(k_\mathrm{max}) &=&
\left(\mathbf{d}_\mathrm{th}(k_\mathrm{max}) - \langle \mathbf{d}(k_\mathrm{max})\rangle\right)^\intercal 
\,\mathbf{Cov}_\mathbf{d}^{-1}(k_\mathrm{max})\,
\left(\mathbf{d}_\mathrm{th}(k_\mathrm{max}) - \langle \mathbf{d}(k_\mathrm{max})\rangle\right)
\notag \\
\notag \\
\chi^2_{\sigma^2}(k_\mathrm{max})
&=& \sum_{i|\forall d^i\in\mathbf{d}_\mathrm{th}(k_\mathrm{max})} 
\dfrac{\left({d}^i_\mathrm{th}(k_\mathrm{max}) - \langle {d}^i(k_\mathrm{max})\rangle\right)^2}
{\sigma_i^2(k_\mathrm{max})}
\quad\mathrm{where,}
\notag \\
\sigma^2_i(k_\mathrm{max})
&=&\dfrac{1}{N_\mathrm{sim}-1}
\sum_{\ell=1}^{N_\mathrm{sim}}
\left({d}^i_\ell(k_\mathrm{max}) - \langle {d}^i(k_\mathrm{max})\rangle\right)^2\,.
\notag \\
\end{eqnarray}
\noindent
$\mathbf{d}_\mathrm{th} (k_{\rm max})\,,\,\,\mathbf{d}(k_\mathrm{max})$ is a shorthand for the data-vector with configurations limited by $k_\mathrm{max}$ (same for the covariance $\mathbf{Cov}_\mathbf{d}^{-1}(k_\mathrm{max})$),  while $\langle d^i(k_\mathrm{max})\rangle$ and $\sigma^2_i(k_\mathrm{max})$ are the mean and the variance for the $i$-th mode (with $i$ running over all the elements of the data-vector $\mathbf{d}(k_\mathrm{max})$) estimated from the set of simulations.
Here we assume (as often done) that the number of degrees of freedom, $N_{\rm dof}$,  corresponds to the length of the data-vector up to a certain $k_\mathrm{max}$ minus one.
$\chi^2_\mathrm{Cov}$ includes the cross correlations among different elements of the data vector,
$\chi^2_{\sigma^2}$, on the other hand, assumes the different data-vector elements to be uncorrelated and therefore uses only the variance.  The $\chi^2$ is  popular  because it offers a relatively quick, although often inaccurate, goodness of fit test. 

Notice that while in our analysis we will adopt the mean of the measured data-vector as reference for the $\chi^2$-test, we will not use the error on the mean (which would be computed by dividing the variance in equation \ref{eq:chi_squared} by the number of realisations used), but the one relative to a single realisation. This is because deriving a dark matter statistics model which would be accurate enough to return a good $\chi^2$ for an average over thousands of $\mathrm{Gpc}^3$ is beyond the scope of this work.

%%%%%%%%%%%%%%%%%%%%%%%%%%%%%%%%%%%%%%%%%%%%%
%%%%%%%%%%%%%%%%%%%%%%%%%%%%%%%%%%%%%%%%%%%%%
%%%%%%%%%%%%%%%%%%%%%%%%%%%%%%%%%%%%%%%%%%%%%
\section{Results}
\label{sec:results}
%%%%%%%%%%%%%%%%%%%%%%%%%%%%%%%%%%%%%%%%%%%%%
%%%%%%%%%%%%%%%%%%%%%%%%%%%%%%%%%%%%%%%%%%%%%
\subsection{Real space: measure vs. model}
\label{sec:meas_vs_model}

In this section  we  compare  the measurements of the power spectrum, bispectrum and i-trispectrum  estimators from the set of \textsc{Quijote}  simulations \cite{Villaescusa-Navarro:2019bje} to the respective theoretical models (section \ref{sec:theo_models}) for three different redshift snapshots: $z=\{0.5$, $1$, $2\}$. 
We also  measure the i-trispectrum auto and cross (with power spectrum and bispectrum) covariance matrix, and compare its diagonal elements with an analytical model.

%%%%%%%%%%%%%%%%%%%%%%%%%%%%%%%%%%%%%%%%%%%%%
%%%%%%%%%%%%%%%%%%%%%%%%%%%%%%%%%%%%%%%%%%%%%
\subsubsection{Analysis set-up}
\label{sec:analysis_setup}
We use 5000 realisations of the N-body simulations from the \textsc{Quijote} suite \cite{Villaescusa-Navarro:2019bje}. Given the lenght of our data vector we estimate that $N_\mathrm{sim}=5000$ offer an optimum balance between speed and computational resources and accuracy of the  estimate of the covariance matrix and its inverse. The simulations follow the gravitational evolution of $512^3$ dark matter particles, in a periodic cubic box with size $L=1 h^{-1}\textrm{Gpc}$. The initial conditions for the simulations were generated at $z=127$ using 2LPT \cite{Springel:2002uv,Crocce:2006ve,Scoccimarro:2011pz}, and we concentrate on the snapshots at redshifts $z=0.5$, $1$ and $2$. 

The underlying cosmology of the \textsc{Quijote} simulations is a flat $\Lambda$CDM model (consistent with the latest CMB constraints  \cite{collaboration2018planck}). Specifically, the matter and baryon density parameters are $\Omega_\mathrm{m}=0.3175$, $\Omega_\mathrm{b}=0.049$, and the dark energy equation of state parameter is $w=-1$; the reduced Hubble parameter is $h=0.6711$, the late-time dark matter fluctuations amplitude parameter is $\sigma_8=0.834$, the scalar spectral index is $n_\mathrm{s}=0.9624$, and neutrinos are massless, i.e. $M_\nu=0.0$ eV.

We discretise the box in $256^3$ grid cells, and  consider a bin size of $\Delta k=2k_\mathrm{f}$, where $k_\mathrm{f}=0.00625$ $h/\mathrm{Mpc}$ is the fundamental frequency. Since non-linearities increase as the redshift decreases, perturbation theory, and hence our model, breaks down at different scales for different redshifts. In the following analysis the quadrilaterals sets sides for the i-trispectrum are limited to $k_\mathrm{max}(z=0.5)=0.15$ $h/\mathrm{Mpc}$, $k_\mathrm{max}(z=1)=0.17$ $h/\mathrm{Mpc}$ and $k_\mathrm{max}(z=2)=0.19$ $h/\mathrm{Mpc}$. 
For both power spectrum and bispectrum we display results for a $k_\mathrm{max}$ which is one third larger than the one for the i-trispectrum at each redshift (i.e. $k_\mathrm{max}(z=0.5)=0.2$ $h/\mathrm{Mpc}$, $k_\mathrm{max}(z=1)=0.22$ $h/\mathrm{Mpc}$ and $k_\mathrm{max}(z=2)=0.25$ $h/\mathrm{Mpc}$). 
In what follows, the triangles (for the bispectrum)  and quadrilaterals sets (for the i-trispectrum) are ordered and  labelled  according to the convention illustrated in  figure \ref{fig:bk_tk_configurations}.

We assume a Poissonian shot-noise and we subtract it from the measured signal using the procedure described in appendix \ref{app:shot_noise}. This consists in using measured quantities (such as power spectra and bispectra) to build the shot-noise corrections given in the literature e.g., \cite{Verde:2001pf}.

%%%%%%%%%%%%%%%%%%%%%%%%%%%%%%%%%%%%%%%%%%%%
%%%%%%%%%%%%%%%%%%%%%%%%%%%%%%%%%%%%%%%%%%%%
%%%%%%%%%%%%%%%%%%%%%%%%%%%%%%%%%%%%%%%%%%%%
\begin{figure}[tbp]
\centering 
\includegraphics[width=0.9\textwidth]%,trim=0 380 0 200,clip]
{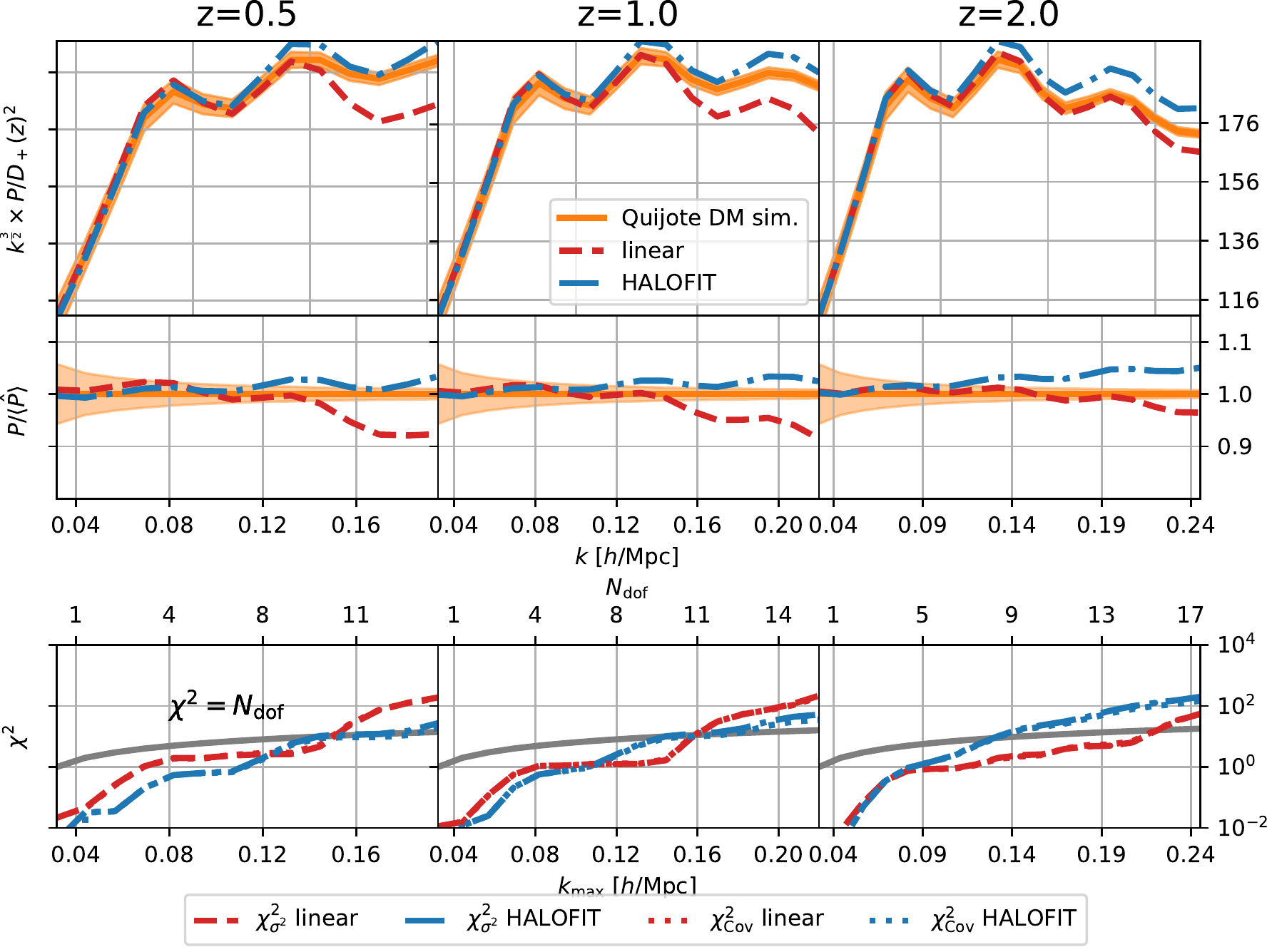}
\caption{\label{fig:pk_fit_15kmax}Matter power spectra (real space) measured from the \textsc{Quijote} simulations at redshifts $z=0.5,\, 1,\, 2$ compared with the theoretical models. The orange line is the power spectrum measured from the simulations, with the shaded orange area being the standard deviation estimated from a set of 5000 different realisations, while the red and blue lines are respectively the linear and \textsc{halofit} power spectrum given by \textsc{class} \cite{Lesgourgues:2011re} using the same underlying cosmology of the simulations \cite{Villaescusa-Navarro:2019bje}. The top row shows the quantity $k^{3/2}P(k)D_+^{-2}(z)$ where $D_+$ denotes the linear growth rate. The second row shows the ratio of the power spectrum to the one measured from the simulation output. The orange shaded area indicate the simulations' standard deviation.
On the lower part, the cumulative $\chi^2$ using both linear and \textsc{halofit} for ${\bf d}_{\rm th}$,  is plotted, along with the number of degrees of freedom (see text for more details). Dashed and dotted lines overlap because the covariance for the power spectrum  at these scales is to a good approximation diagonal. Notice that the variances and covariances used for the $\chi^2$-tests are relative to a single realisation of  $1\,\mathrm{Gpc}^3$ volume. 
}
\end{figure}

%%%%%%%%%%%%%%%%%%%%%%%%%%%%%%%%%%%%%%%%%%%%
%%%%%%%%%%%%%%%%%%%%%%%%%%%%%%%%%%%%%%%%%%%%
%%%%%%%%%%%%%%%%%%%%%%%%%%%%%%%%%%%%%%%%%%%%
\begin{figure}[tbp]
\centering 
\includegraphics[width=0.99\textwidth]%,trim=0 380 0 200,clip]
{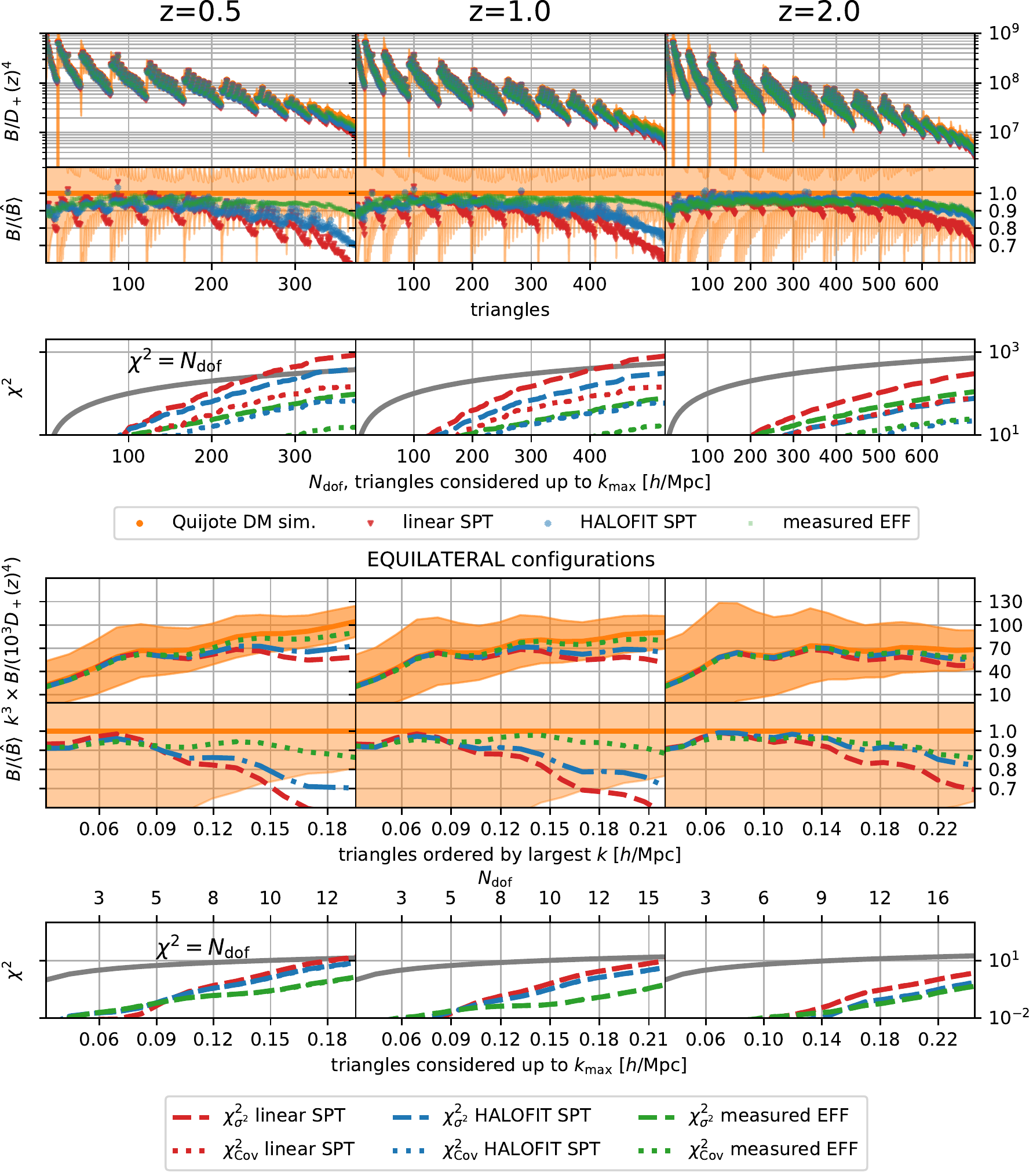}
\caption{\label{fig:bk_fit_15kmax} 
Same as figure \ref{fig:pk_fit_15kmax} but for the bispectrum. The top half of the plot shows all configurations and the bottom half shows only equilateral triangle configurations to make the dependence on scale  easier to interpret.
In addition to standard perturbation theory (SPT) prediction using linear (red) and \textsc{halofit} (blue) power spectra as input, the green lines correspond to the $B^{\rm EFF}$ bispectrum prediction using the effective kernels \cite{GilMarin:2011ik,Gil-Marin:2014pva} and for non-linear power spectrum the  average of the simulations measurements. 
In both cases the cumulative $\chi^2$ and the degrees of freedom curves are reported.
The difference between $\chi^2_{\sigma^2}$ and $\chi^2_\mathrm{Cov}$ from equation \ref{eq:chi_squared} is visible only in the ``all shapes'' case (top half of the plot); equilateral configurations are  only very weakly correlated.} 
\end{figure}

%%%%%%%%%%%%%%%%%%%%%%%%%%%%%%%%%%%%%%%%%%%%%
%%%%%%%%%%%%%%%%%%%%%%%%%%%%%%%%%%%%%%%%%%%%%
\subsubsection{Power spectrum and bispectrum}
In figure \ref{fig:pk_fit_15kmax} the mean and the standard deviation  of the power spectrum  measurements from the \textsc{Quijote} simulations (orange solid lines) are compared to the theoretical model predictions. The adopted $k$-bin size is $\Delta k=0.0126$ $h/\mathrm{Mpc}$ and the {\it rms} is computed by the scatter among the 5000 realisations.  Both linear theory (red dashed) and \textsc{halofit} (blue dot-dashed) $P(k)$ predictions  are shown. 
In the first row the power spectra are multiplied by a factor of $k^\frac{3}{2}$ and normalised by the linear growth factor squared at each snapshot's redshift, $D_+(z)^2$, also computed using \textsc{class}.
The next row shows the ratio between the theoretical models and the simulation measurement
(simulation output in orange, \textsc{halofit} in blue dot-dashed and linear theory prediction in red dashed). The third row shows the cumulative  $\chi^2$ computed as described in equation \ref{eq:chi_squared}.
It is possible to appreciate  the redshift dependence of the $k_\mathrm{max}$ up to which the models fit the data well. While for $z=0.5$ and $1$  \textsc{halofit} performs better than the linear power spectrum prediction, at $z=2$ the opposite happens and the linear model has a good $\chi^2$ up to $k\sim0.20\,h/\mathrm{Mpc}$.
This  effect is understood as follows. In this range of scales at this redshift  many $k$-modes are undergoing gravitational collapse simultaneously (for those $k$ values such that $k^3P(k)/(2\pi)^3\simeq1$, which is the regime where structure formation is more complex to describe with halo-model/semi-analytic approaches. On the other hand, at later times, when the structure formation may be more nonlinear (and naively more difficult to model),  the collapse happens in a more ordered way, and it is easier for \textsc{Halofit} to capture it. One should also recall that  \textsc{Halofit} is designed to match the power spectrum shape up to $k\simeq10\,h/{\rm Mpc}$, so it is not surprising that its description of the power spectrum is not too accurate at relatively weakly non-linear scales.

The cumulative chi-square computed with  $\chi^2_{\sigma^2}$ and $\chi^2_\mathrm{Cov}$ from equation \ref{eq:chi_squared} almost perfectly overlap: at these scales the power spectrum  modes are not very correlated and the $P(k)$ covariance matrix is quasi-diagonal.

Figure \ref{fig:bk_fit_15kmax} shows the same information for the bispectrum. In the top half of the plot all triangle configurations are shown, while the lower half displays only  equilateral  configurations in order to make  the scale dependence easier to interpret.

In addition to the SPT models using the linear and \textsc{halofit} power spectra as input ($B_{\rm SPT}$ and  $B_{\rm SPT-NL}$), we  also present the comparison of the  measurements with  the effective model $B_{\rm EFF}$ (see equation \ref{eq:BSPT} and Refs. \cite{GilMarin:2011ik,Gil-Marin:2014pva}).
In this case  we use the mean of the power spectrum measurements from the simulations as non-linear power spectrum input for the effective model ("measured EFF" in the figure). The SPT model with \textsc{halofit} non-linear power spectrum  always outperforms the standard SPT one (with linear power spectrum) and becomes closer to the effective bispectrum model as the redshift increases. At all redshifts, the effective bispectrum model significantly increases  the maximum $k$ at which the model fits the measurements.
It is interesting to note how the $B_{\rm EFF}$ model, with the effective kernel  calibrated for only specific shapes  from N-body simulations,
offers a better description than SPT for all triangle shapes. 

The difference between the cumulative $\chi^2$ assuming uncorrelated bispectrum modes and accounting for the full covariance, $\chi^2_{\sigma^2}$ and $\chi^2_\mathrm{Cov}$ (equation \ref{eq:chi_squared}), is visible in the “all shapes" case (top half of the plot): different bispectrum modes (triangles) are in general correlated;  equilateral configurations  on the other hand are only very weakly correlated.

%%%%%%%%%%%%%%%%%%%%%%%%%%%%%%%%%%%%%%%%%%%%%
%%%%%%%%%%%%%%%%%%%%%%%%%%%%%%%%%%%%%%%%%%%%%
%%%%%%%%%%%%%%%%%%%%%%%%%%%%%%%%%%%%%%%%%%%%%
%%%%%%%%%%%%%%%%%%%%%%%%%%%%%%%%%%%%%%%%%%%%%
\begin{figure}[tbp]
\centering 
\includegraphics[width=\textwidth]%,trim=0 380 0 200,clip]
{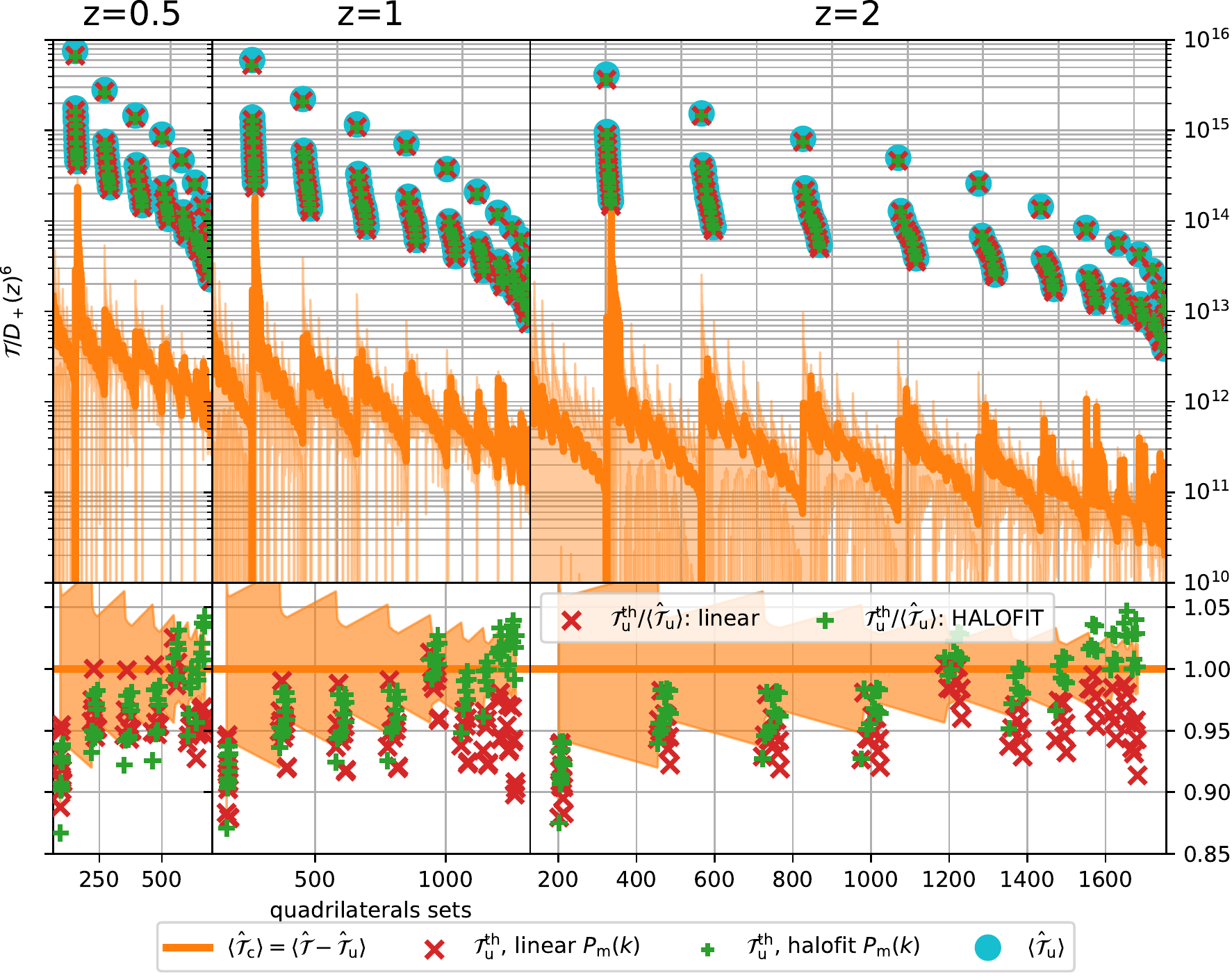}
\caption{\label{fig:unconnected_vs_connected} 
Connected ($\hat{\mathcal{T}}_\mathrm{c}$) and unconnected ($\hat{\mathcal{T}}_\mathrm{u}$) components of the total average four-point correlator  signal $\hat{\mathcal{T}}_\mathrm{c+u}$ measured from  the 5000 dark matter simulations,
using the estimator from equation \ref{eq:tk_est_text} 
at redshifts $z=0.5,1,2$. Since the $k$-range of interest varies with redshift ($k_\mathrm{max}$ increases with $z$ as the field is more linear) the number of configurations shown also depends on redshift (see text for details).
The light-blue dots (with error bars too small to be visible in this plot) correspond to the  unconnected term $\hat{\mathcal{T}}_\mathrm{u}$ measured  from the simulations using the estimator described in equation \ref{eq:tku_expansion1_text}.
The green ($+$)  and red ($\times$) signs represent the theory prediction for the unconnected signal (equation \ref{eq:tk_unbiased_unconnected_ana_text}) computed  respectively using  linear and \textsc{halofit} matter power spectra. 
The orange line corresponds to the mean of 5000 measurements of the total signal minus the unconnected part estimated also from 5000 measurements (equation \ref{eq:tku_est_text}), the orange shaded area indicates the standard deviation (scatter among the 5000 simulations) for each mode. ${\cal T}_\mathrm{u}$ is about  two orders of magnitude larger than ${\cal T}_\mathrm{c}$: in fact 
the  unconnected part is of order $\propto\delta^4$ while the connected one is  $\propto\delta^6$.
The fact that for both linear and \textsc{halofit} versions the ratio between model and estimated unconnected part oscillates by approximately $\sim10\%$ justifies the choice of subtracting the measured unconnected term from the total signal estimated by the four-point correlator (equation \ref{eq:tk_est_text}).
}
\end{figure}

%%%%%%%%%%%%%%%%%%%%%%%%%%%%%%%%%%%%%%%%%%%%
%%%%%%%%%%%%%%%%%%%%%%%%%%%%%%%%%%%%%%%%%%%%
%%%%%%%%%%%%%%%%%%%%%%%%%%%%%%%%%%%%%%%%%%%%
\begin{figure}[tbp]
\centering 
\includegraphics[width=0.99\textwidth]%,trim=0 380 0 200,clip]
{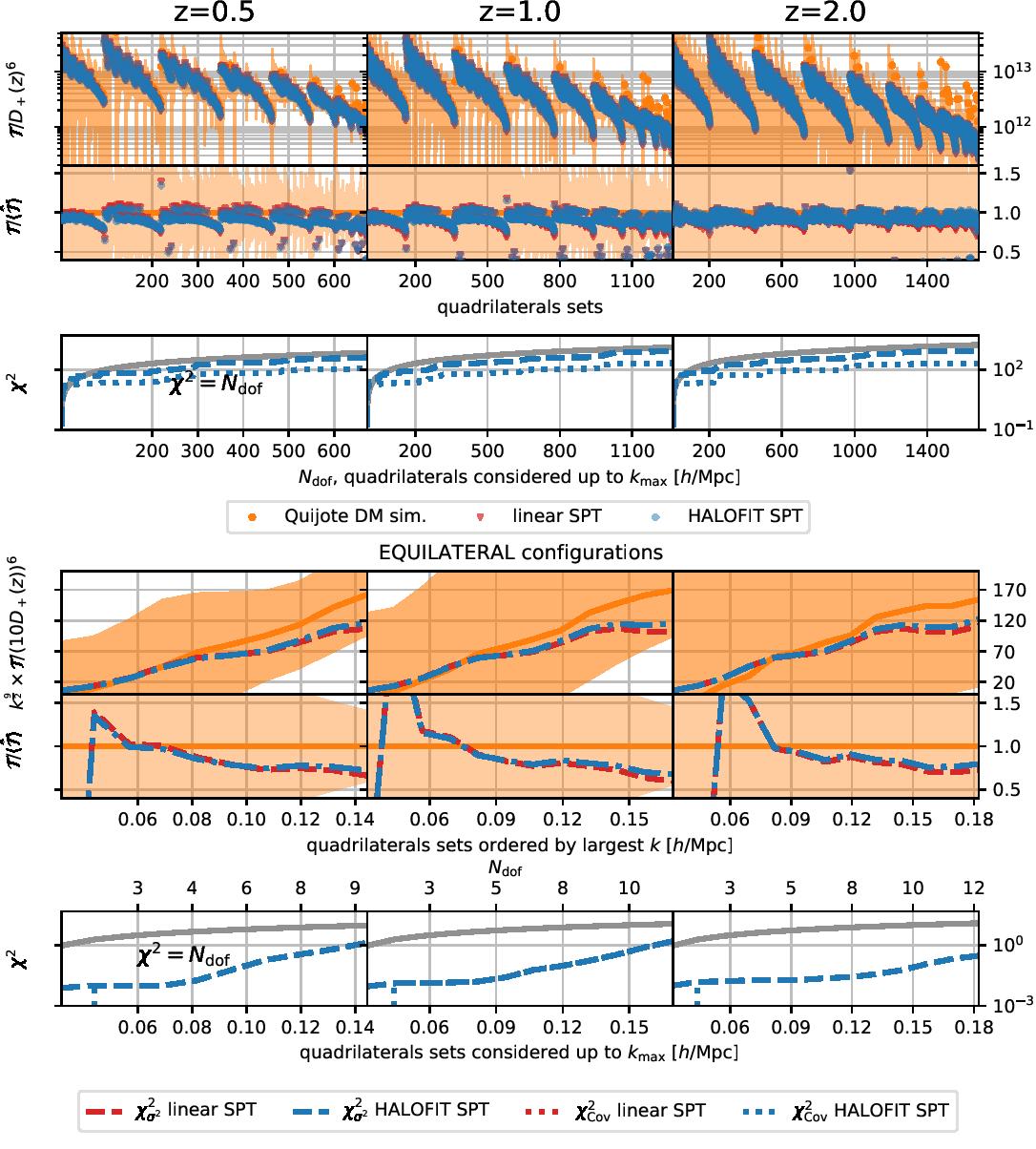}
\caption{\label{fig:tk_fit_15kmax}
Same as figures \ref{fig:pk_fit_15kmax} and \ref{fig:bk_fit_15kmax} but for the i-trispectrum.
The theoretical model $\mathcal{T}^{\rm th}_{\rm c}$ recovers the mean of the measurements $\langle \hat{\mathcal{T}}_{\rm c}\rangle$ well within the estimated variance.  The cumulative $\chi^2$, by closely following the number of degrees of freedom's curve, confirms the good match between theory and measurements at all the considered redshifts ($z=0.5,1,2$) up to each respective smallest scale considered $k_\mathrm{max}=0.15,0.17,0.19$ $h/
\mathrm{Mpc}$.
Similarly to the bispectrum in figure \ref{fig:bk_fit_15kmax}, the difference between $\chi^2_{\sigma^2}$ and $\chi^2_\mathrm{Cov}$ from equation \ref{eq:chi_squared} is significant only when all the possible quadrilaterals sets are considered (equilateral configurations appear to be significantly  less correlated than generic ones and that is the reason why the different $\chi^2$-lines almost perfectly overlap).}
\end{figure}

%%%%%%%%%%%%%%%%%%%%%%%%%%%%%%%%%%%%%%%%%%%%%
%%%%%%%%%%%%%%%%%%%%%%%%%%%%%%%%%%%%%%%%%%%%%
\subsubsection{\lowercase{i}-trispectrum}

As described in section \ref{sec:tk_estimator}, in order to obtain the i-trispectrum signal, the unconnected part needs to be subtracted from the total signal measured using the estimator  presented in equation \ref{eq:tk_est_text_tot}.
In figure \ref{fig:unconnected_vs_connected} we compare the unconnected signal theory template $\mathcal{T}_{\rm u}^{\rm th}$, (equation \ref{eq:tk_unbiased_unconnected_ana_text}) with the measurement performed using the estimator $\hat{\mathcal{T}}_u$ in equation \ref{eq:tku_expansion1_text}. The unconnected part is approximately two orders of magnitude larger than the connected part (which is expected from standard perturbation theory, being the unconnected part a lower order term with respect than the connected one).

To avoid systematic errors due to limitations in the perturbation theory description of the unconnected part of the signal, we prefer to  estimate the unconnected contribution  using $\hat{\mathcal{T}}_u$ equation \ref{eq:tku_expansion1_text} instead of the analytical model. This is to be subtracted from the total signal $\hat{\cal T}_{\rm c+u}$ to obtain the i-trispectrum $\hat{\cal T}_{\rm c}$ according to equation \ref{eq:tk_est_text}.
In the lower panel of figure \ref{fig:unconnected_vs_connected} the ratio between the models for the i-trispectrum unconnected part using linear and \textsc{halofit} matter power spectra can be seen to diverge from one as the size of the quadrilateral sides $(k_1,k_2,k_3,k_4)$ increases, as expected from the breaking down of perturbation theory at non-linear scales (see figure \ref{fig:bk_tk_configurations} to visualise how the sides vary among configurations).

In figure \ref{fig:tk_fit_15kmax}  we show the comparison between the resulting measured i-trispectrum $\hat{\cal T}_{\rm c}$, and the theoretical model, ${\cal T}^{\rm th}_{\rm c}$ defined in equations \ref{eq:tk_3d_matter_model} and \ref{eq:tk_int_real} (same conventions as for figure \ref{fig:bk_fit_15kmax}).
For easier interpretation, in the lower part only equilateral configurations are shown.

\begin{figure}[tbp!]
\centering 
\includegraphics[width=0.95\textwidth]%,trim=0 380 0 200,clip]
{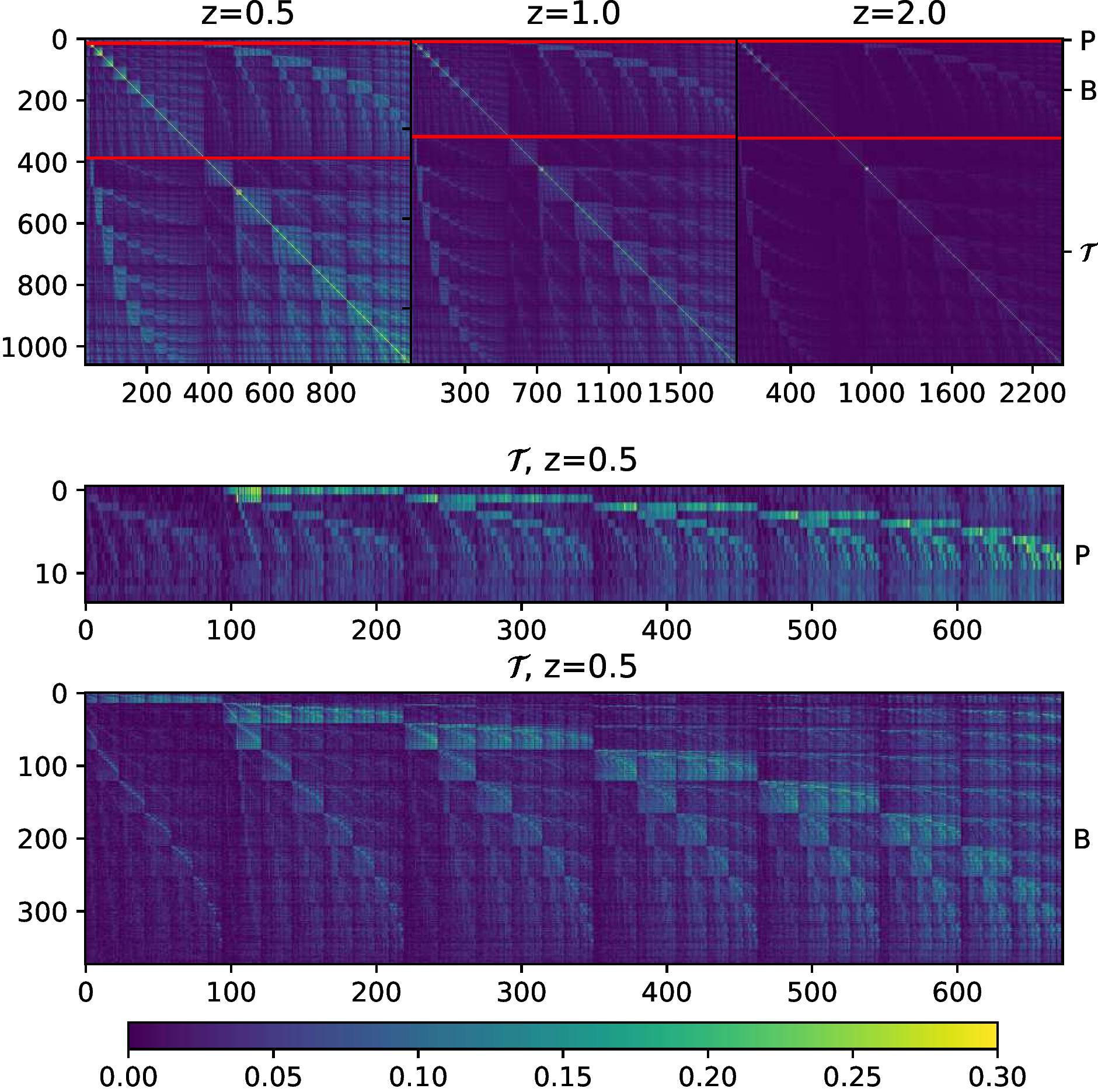}
\caption{\label{fig:measured_cov}
Reduced covariance matrix, $r_{ij}={\mathrm{Cov}^{ij}_\mathbf{d}}/\sqrt{{{\rm Cov}^{ii}_\mathbf{d}}{{\rm Cov}^{jj}_\mathbf{d}}}$, estimated from the set of 5000 simulations at each redshift. 
In the top row the joint PB$\mathcal{T}$ covariance matrices are displayed while  the central and bottom rows show a zoom-in of the cross correlation between i-trispectrum and  power spectrum / bispectrum at $z=0.5$. The ordering of configurations is the same as seen in figures \ref{fig:bk_fit_15kmax} and \ref{fig:tk_fit_15kmax}.
From left to right the different columns show the results for $z=0.5,1,2$ with $k_\mathrm{max}=0.15,0.17,0.19$ $h/\mathrm{Mpc}$, respectively.
The horizontal red lines guide the eye to recognise the covariance parts corresponding to power spectrum, bispectrum and i-trispectrum.
The covariance matrices acquire more structure as $z$ decreases, which is a sign of non-linear growth and stronger mode coupling among different configurations/data-vector's elements. 
Especially for low redshifts ($z=0.5,1$), the above plots show that both the correlation between different modes of the same statistics and cross-correlation between modes of different statistics are not negligible and reach up to $\sim30\%$ the value of the diagonal elements. A Fisher forecast using a theoretically computed diagonal covariance matrix under the Gaussian approximation (as it is often done in the literature) would in this case produce too optimistic constraints by underestimating the redundancy in the data-vector due to correlation among different elements. 
}
\end{figure}

The integrated theoretical model (equations \ref{eq:tk_3d_matter_model} and \ref{eq:tk_int_real}), both for linear and \textsc{halofit} power spectrum, performs reasonably well.
When all the quadrilaterals sets are considered, the cumulative $\chi^2$ closely follows the line for our adopted $N_{\rm dof}$, number of degrees of freedom. 
Notice that the different quadrilaterals sets making up the i-trispectrum data-vector are correlated, as it can be inferred from the covariance matrix shown in figure \ref{fig:measured_cov}, therefore the effective number of degrees of freedom is effectively lower than the number of configurations used.
 
An improvement in the fit could be obtained by extending the effective model of the kernels to the third-order ones needed to compute the i-trispectrum. At the same time the second order kernels could be fitted using both bispectrum and i-trispectrum. We leave this to future work.

%%%%%%%%%%%%%%%%%%%%%%%%%%%%%%%%%%%%%%%%%%%%%
%%%%%%%%%%%%%%%%%%%%%%%%%%%%%%%%%%%%%%%%%%%%%
%%%%%%%%%%%%%%%%%%%%%%%%%%%%%%%%%%%%%%%%%%%%%
%%%%%%%%%%%%%%%%%%%%%%%%%%%%%%%%%%%%%%%%%%%%%

%%%%%%%%%%%%%%%%%%%%%%%%%%%%%%%%%%%%%%%%%%%%%
%%%%%%%%%%%%%%%%%%%%%%%%%%%%%%%%%%%%%%%%%%%%%
%%%%%%%%%%%%%%%%%%%%%%%%%%%%%%%%%%%%%%%%%%%%%
%%%%%%%%%%%%%%%%%%%%%%%%%%%%%%%%%%%%%%%%%%%%%
\begin{figure}[tbp!]
\centering 
\includegraphics[width=\textwidth]%,trim=0 380 0 200,clip]
{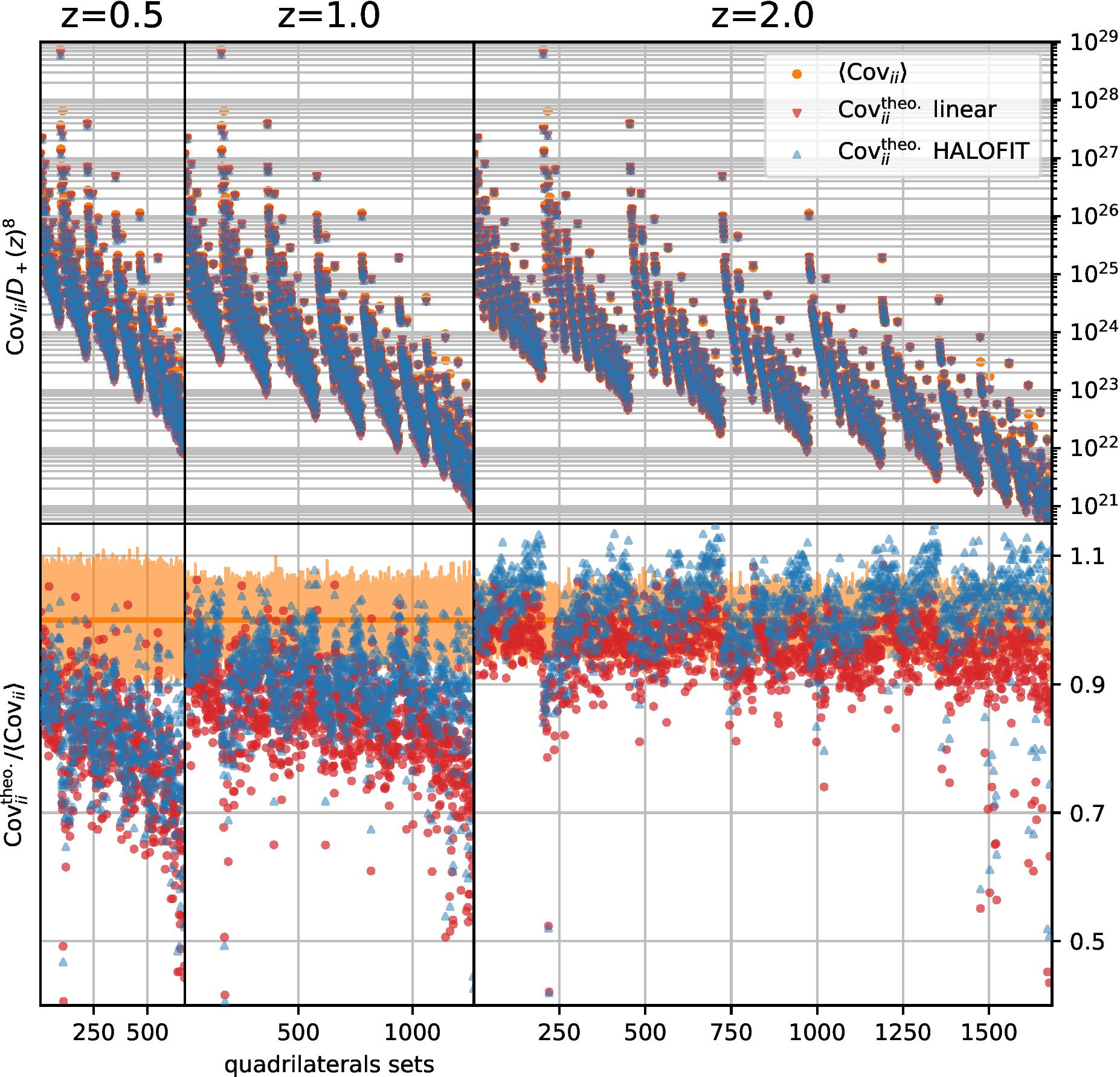}
\caption{\label{fig:ana_cov}
Comparison between the diagonal elements of the numerically-estimated i-trispectrum  covariance matrix and  the theoretical model (only  Gaussian term) of equation \ref{eq:t0_gg_cov_matter}. 
The  model is shown for  both the linear (red) and \textsc{halofit} (blue) power spectrum.
The orange shaded area is the  {\it rms} estimated by jackknife i.e. standard deviation 
for the diagonal elements obtained by splitting the 5000 simulations into $(16_{z=0.5},\,9_{z=1},\,7_{z=2})$ equally populated subsets and then by estimating the covariance matrix diagonal for each subset. From the different estimates it was then possible to derive a standard deviation for each covariance matrix diagonal element.
In agreement with what is observed in figure \ref{fig:measured_cov}, the Gaussian approximation for the covariance diagonal
becomes increasingly better as the redshift increases.
Together with figure \ref{fig:measured_cov}, this highlights the importance of using the numerically-estimated  covariance matrix and not a Gaussian approximation for redshifts lower than $z=2$. Using only the Gaussian term would indeed significantly underestimate both the correlation between different data-vector's elements (figure \ref{fig:measured_cov}) and the non-linearly induced auto-correlation of each element with itself.
}
\end{figure}
%%%%%%%%%%%%%%%%%%%%%%%%%%%%%%%%%%%%%%%%%%%%%
%%%%%%%%%%%%%%%%%%%%%%%%%%%%%%%%%%%%%%%%%%%%%
\subsubsection{Covariance matrix and Gaussian term analytical model}
\label{sec:cov_results}
The covariance matrix of the i-trispectrum and that of the full   power spectrum, bispectrum  and i-trispectrum data vector (PB$\mathcal{T}$ for short), are fundamental ingredients for assessing the impact of adding this four-point statistic to a parameter constraints analysis. Usually in this kind of forecasts studies, an analytical template is used and often the covariance is assumed to be diagonal with only a Gaussian component, together with assuming zero cross-correlation between different $n$-point statistics.

In this section we show that following the above assumption in the case of the i-trispectrum would induce a significant bias leading to an overestimate of the i-trispectrum constraining power when added to the power spectrum and bispectrum data-vector.
To this purpose, we numerically estimate the  PB$\mathcal{T}$ covariance matrix  from the simulations  as described in section \ref{sec:cov_methodology}.
The  $k_{\rm max}$ at each redshift for each statistics are the same as described in section \ref{sec:analysis_setup}.

Figure \ref{fig:measured_cov} shows the reduced covariance matrix, $r_{ij}={\mathrm{Cov}^{ij}_\mathbf{d}}/\sqrt{{{\rm Cov}^{ii}_\mathbf{d}}{{\rm Cov}^{jj}_\mathbf{d}}}$, numerically evaluated from the 5000 simulations. The top row shows the full (PB$\mathcal{T}$)  covariance (to guide the eye, the red lines indicate the P, B and $\mathcal{T}$ sections).
In the central and bottom rows of figure \ref{fig:measured_cov} the cross-correlations of the i-trispectrum with both power spectrum and bispectrum are displayed at $z=0.5$. 
As the redshift increases it is evident how the different elements of the joint data-vector become less and less correlated since the  field is more linear and the mode-mixing induced by gravitational collapse is less important. At redshift $z=0.5$  for the i-trispectrum, the off-diagonal elements of the reduced covariance become up to $\sim 0.3$. This means that two different configurations can be cross-correlated as much as one third of their auto-correlation value. In other words, the level of redundancy in the data-vector increases as the redshift decreases.

Note also that  the cross-correlations between power spectra of $k$-modes and bispectra of triangle configurations with the i-trispectra of quadrilaterals sets (up to the i-trispectrum $k_\mathrm{max}$) are non-negligible at low redshifts.

A quantitative idea of the importance of the non-Gaussian contributions appearing in the i-trispectrum covariance matrix can be obtained by comparing the analytical model of the diagonal Gaussian term of equation \ref{eq:t0_gg_cov_matter} with the one estimated numerically. This is shown in figure \ref{fig:ana_cov}: the difference increases as the redshift decreases. The shaded area is obtained by estimating the scatter of the covariance diagonal elements as follows. We take measurements from $N_\mathrm{G}(z)=$ $(16_{z=0.5},\,9_{z=1},\,7_{z=2})$  groups of simulations boxes randomly selected from the whole set of 5000 realisations. The $N_\mathrm{G}(z)$ is set by requiring that the number of simulation boxes per group (i.e. $5000/N_\mathrm{G}(z)$) is no smaller than half of the data-vector dimension. 
Note that  the number of realisations used to estimate the covariance can  safely be lower than the respective data-vector's dimension since we are just interested in deriving the diagonal elements, without inverting the covariance \cite{Hartlap:2006kj}.
This confirms what is seen in figure \ref{fig:measured_cov}, that for the i-trispectrum data-vector the non-Gaussian terms in the covariance are not negligible, in particular at lower redshifts.

The visible difference between $\chi^2_{\sigma^2}$ and $\chi^2_\mathrm{Cov}$ (equation \ref{eq:chi_squared}) in figure \ref{fig:tk_fit_15kmax} when all the configurations are considered further supports the importance of going beyond the Gaussian diagonal covariance approximation for the i-trispectrum (and also previously in figure \ref{fig:bk_fit_15kmax} for the bispectrum) also for  evaluating the goodness of fit of a theoretical model.

%%%%%%%%%%%%%%%%%%%%%%%%%%%%%%%%%%%%%%%%%%%%
%%%%%%%%%%%%%%%%%%%%%%%%%%%%%%%%%%%%%%%%%%%%
%%%%%%%%%%%%%%%%%%%%%%%%%%%%%%%%%%%%%%%%%%%%
\begin{figure}[tbp]
\centering 
\includegraphics[width=0.99\textwidth]%,trim=0 380 0 200,clip]
{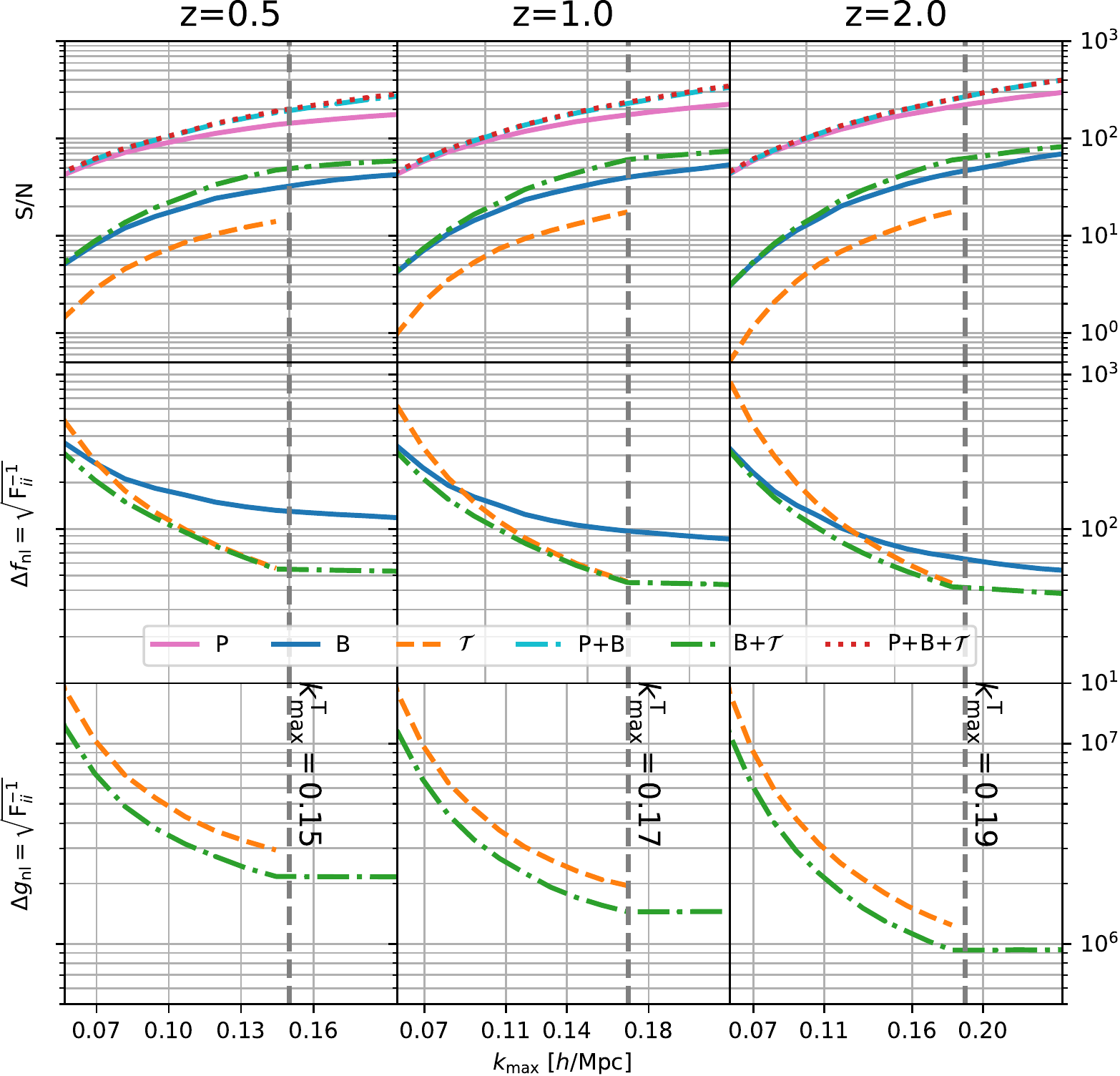}
\caption{\label{fig:sn_png_constraints} 
Signal-to-noise ratio and primordial non-Gaussianity constraints forecasts.
The cumulative signal-to-noise ratio computed using equation \ref{eq:signal_noise} is displayed on the top row for the following statistics combinations: power spectrum (pink), bispectrum (blue), i-trispectrum (orange), power spectrum plus bispectrum (cyan), bispectrum plus i-trispectrum (green), power spectrum plus bispectrum plus i-trispectrum (red).
In the second and third rows the forecasts for the constraints (equation \ref{eq:fish_info}) on the primordial non-Gaussianity parameters $f_\mathrm{nl}$ and $g_\mathrm{nl}$ are shown as a function of the maximum $k$-values considered. The dashed vertical lines indicate the i-trispectrum $k_\mathrm{max}$, beyond which no additional quadrilaterals sets configurations are added to the joint data-vector as  $k_\mathrm{max}$ increases.
Even if adding the i-trispectrum to the joint power spectrum and bispectrum data-vector does not produce a significative change in terms of cumulative signal-to-noise, the benefits are evident when looking at the forecast for the constraints on the primordial non-Gaussianity parameters $f_\mathrm{nl}$ and $g_\mathrm{nl}$.
}
\end{figure}

%%%%%%%%%%%%%%%%%%%%%%%%%%%%%%%%%%%%%%%%%%%%
%%%%%%%%%%%%%%%%%%%%%%%%%%%%%%%%%%%%%%%%%%%%
%%%%%%%%%%%%%%%%%%%%%%%%%%%%%%%%%%%%%%%%%%%%
\subsection{Primordial non-Gaussianity constraints: forecasts}
\label{sec:png_forecasts}
\noindent
For local primordial non-Gaussianity, a first assessment of the additional constraining power given by the  i-trispectrum can be made via a Fisher-based forecast, considering the real space matter field. Promising results in this idealised case can motivate a more realistic analysis (e.g., for biased discrete tracers, in redshift space). 

Moreover, in this conservative scenario, we only consider conditional errors on the non-Gaussianity parameters, i.e., assuming all other parameters are fixed. If these were let free to vary, we expect the improvements on the constraints given by the i-trispectrum to be larger due to the reduction of the degeneracies present in the parameters space. This expectation is motivated by the analogy with the findings of \cite{Sefusatti:2006pa,Chan:2016ehg,Yankelevich:2018uaz,Hahn:2019zob,Agarwal:2020lov} obtained when adding the bispectrum to the power spectrum constraining power.

In particular for the forecasted constraints on $f_\mathrm{nl}$  and $g_\mathrm{nl}$, when considering only the bispectrum we report $\Delta f_\mathrm{nl}^\mathrm{min}=1/\sqrt{F_{f_\mathrm{nl}f_\mathrm{nl}}}$ while for the i-trispectrum and bispectrum plus i-trispectrum we use $\Delta f_\mathrm{nl}^\mathrm{min}=\sqrt{F_{f_\mathrm{nl}f_\mathrm{nl}}^{-1}}$ and $\Delta g_\mathrm{nl}^\mathrm{min}=\sqrt{F_{g_\mathrm{nl}g_\mathrm{nl}}^{-1}}$ (i.e., we  report the error on a non-Gaussianity parameter marginalised over the other one accounting for the covariance between $g_\mathrm{nl}$ and $f_\mathrm{nl}$).

 In figure \ref{fig:sn_png_constraints} we show both cumulative signal-to-noise ratio (equation \ref{eq:signal_noise}) and Fisher forecasts (equation \ref{eq:fish_info}) for the constraints on $f_\mathrm{nl}$ and $g_\mathrm{nl}$ as a function of $k_\mathrm{max}$ for each of the redshifts considered in the analysis.
The colours/line-styles are as follows: power spectrum only in magenta, bispectrum in blue, i-trispectrum in orange (dashed), joint power spectrum and bispectrum in cyan (dashed), joint bispectrum and i-trispectrum in  green (dot-dashed)  and full PB$\mathcal{T}$ in red (dotted).
In the cumulative signal-to-noise (top row) plots one can appreciate the increase in signal generated by adding the i-trispectrum to the bispectrum, especially in the mildly non-linear regime (as $k$ increases). Because of the logarithmic scale, it is difficult to notice the improvement when also the power spectrum is considered. The relative magnitude of the i-trispectrum effect when added to the bispectrum is similar to that observed when adding the bispectrum to the power spectrum.

Similarly, the middle row of figure \ref{fig:sn_png_constraints}  shows the substantial improvement obtained by using the i-trispectrum together with the bispectrum in order to constrain $f_\mathrm{nl}$. We assumed in this work that the power spectrum sensitivity to and constraining power for primordial non-Gaussianity is negligible compared to bispectrum and i-trispectrum. This is because we are considering the dark matter particles as tracers. When moving to haloes this is no longer a reasonable approximation since PNG leaves a distinctive signature in the halo power spectrum through the scale dependent bias \cite{Dalal:2007cu,Matarrese:2008nc}. Moreover the inclusion of the power spectrum would be fundamental for constraining additional cosmological or nuisance parameters, which in this analysis have been kept fixed.

Note, perhaps not unexpectedly \cite{Verde:2001pf},  that in the very mildly non-linear regime ($k>0.1 \,h/$Mpc at $z=0.5$) virtually all the constraining power for $f_{\rm nl}$ comes from the i-trispectrum.
Even if the bispectrum is not sensitive at the considered order in perturbation theory ($\propto\delta^4$ for the bispectrum) to $g_\mathrm{nl}$,   when used together with the i-trispectrum it helps in reducing the degeneracy present in $T_{(1111)}$ between the two terms proportional to $f_\mathrm{nl}^2$ and $g_\mathrm{nl}$ (see bottom row of figure \ref{fig:sn_png_constraints}).  

The vertical dashed lines in figure \ref{fig:sn_png_constraints} mark the maximum $k$-value used to build the quadrilaterals sets for the i-trispectrum. We considered larger $k$-values for power spectrum and bispectrum in order to show that even if the i-trispectrum would be employed up to a lower $k_\mathrm{max}$ (similarly to what happens between power spectrum and bispectrum), its effect is still significant.

Finally in table \ref{tab:png_forecasts} we summarise these findings.
The table reports values for the forecasted 1D $68\%$ confidence interval regions for both $f_\mathrm{nl}$ and $g_\mathrm{nl}$. Especially at lower redshifts, adding the i-trispectrum produces constraints on $f_\mathrm{nl}$ that are two times tighter than the ones produced by the bispectrum alone.

The results displayed in figure \ref{fig:sn_png_constraints} and table \ref{tab:png_forecasts} are encouraging for the prospect of large scale structures surveys, such as DESI \cite{Levi:2013gra},  which are expected to produce constraints on local primordial non-Gaussianity parameters  which will be competitive  and complementary  to the ones obtained up now by CMB experiments such as \textit{Planck} \cite{Akrami:2019izv}.
While certainly encouraging, it would be naive to conclude that these findings translate not just qualitatively but also quantitatively to realistic surveys. Real world issues such as survey geometry, galaxy bias and redshift space distortions may affect the above conclusions. In what follows, we present a first step towards more realistic estimates.

%%%%%%%%%%%%%%%%%%%%%%%%%%%%%%%%%%%%%%%%%%%%
%%%%%%%%%%%%%%%%%%%%%%%%%%%%%%%%%%%%%%%%%%%%
\renewcommand{\arraystretch}{2.2}
\begin{table}[tbp]
\centering
\begin{tabular}{|c|c|ccc|ccc|ccc|ccc|}
\hline
\multicolumn{2}{|c|}{$k_\mathrm{max}$ $[h/\mathrm{Mpc}]\rightarrow$}&  \multicolumn{3}{c|}{0.12} & \multicolumn{3}{c|}{0.15}
& \multicolumn{3}{c|}{0.17} & \multicolumn{3}{c|}{0.19} \\
\hline
  & $\mathbf{d}$ &  \multicolumn{3}{c|}{$z=0.5,1,2$} & \multicolumn{3}{c|}{$z=0.5,1,2$} & \multicolumn{3}{c|}{$z=0.5,1,2$} & \multicolumn{3}{c|}{$z=0.5,1,2$}\\
\hline 
\multirow{3}{*}{$\Delta f_\mathrm{nl}$}
&$B$   & 149 & 124 & 102 & 132 & 106 & 81 & 124 & 98 & 69 & 121 & 94 & 66 \\ 
&$\mathcal{T}$   & 80 & 87 & 109 & 56 & 60 & 71 & -  & 44 & 51 & -  & -  & 44 \\
&$B+\mathcal{T}$ & 78 & 81 & 83  & 55 & 58 & 60 & 54 & 45 & 47 & 53 & 45 & 42 \\
\hline 
1 - $\dfrac{\Delta f_\mathrm{nl}^{B+\mathcal{T}}}{\Delta f_\mathrm{nl}^B}$
& $\left[\%\right]$ 
& \textbf{48} & \textbf{35} & \textbf{19} 
& \textbf{58} & \textbf{45} & \textbf{26} 
& \textbf{56} & \textbf{53} & \textbf{32} 
& \textbf{56} & \textbf{52} & \textbf{36} \\
\hline
$\Delta g_\mathrm{nl}$ 
& $\mathcal{T}$ & 
368 & 304 & 252 & 
296 & 231 & 179 & 
-   & 195 & 138 & 
-   & -   & 124 \\
$(\times10^{-4})$
&$B+\mathcal{T}$ 
& 273 & 225 & 183 
& 219 & 174 & 130 
& 218 & 145 & 103 
& 218 & 144 & 93 \\
\hline
1 - $\dfrac{\Delta g_\mathrm{nl}^{B+\mathcal{T}}}{\Delta g_\mathrm{nl}^{\mathcal{T}}}$
& $\left[\%\right]$ 
& \textbf{14} & \textbf{12} & \textbf{11}  
& \textbf{19} & \textbf{17} & \textbf{16} 
& -           & \textbf{20} & \textbf{16} 
& -           & -           & \textbf{17} \\
\hline
\end{tabular}
\caption{\label{tab:png_forecasts} 
1D $68\%$ forecasted credible regions for both $f_\mathrm{nl}$ and $g_\mathrm{nl}$ as a function of $k_\mathrm{max}$ for the bispectrum, i-trispectrum and bispectrum plus i-trispectrum in real space. The highlighted numbers correspond to the improvement on the parameters constraints given by employing both bispectrum and i-trispectrum, with respect to using only the bispectrum. All the values have been obtained through a Fisher forecast where the covariance matrix has been estimated from 5000 measurements on simulations and its inverse corrected by the corresponding Hartlap factor \cite{Hartlap:2006kj}. Each simulation's volume is $1\,[{\rm Gpc}/h]^3$.
}
\end{table}

%%%%%%%%%%%%%%%%%%%%%%%%%%%%%%%%%%%%%%%%%%%%%
%%%%%%%%%%%%%%%%%%%%%%%%%%%%%%%%%%%%%%%%%%%%%
\subsection{Redshift space}
\label{sec:rsd_results}
While the real space analysis presented so far indicates that in principle there is additional, useful information in the i-trispectrum, realistic observations are affected by redshift space distortions. To assess whether the real space results also hold in redshift space, in this section we present the same analysis performed on the redshift space matter density field. 
For this purpose, for each statistics we limit ourselves to the monopole signal only. Of course, there is potentially a lot of additional information enclosed in redshift space multipoles, but this will be presented elsewhere.

The theoretical modelling for the quantities in redshift space is presented in appendix \ref{app:theo_models_rsd}.
Notice that, as normally done for power spectrum and bispectrum, in redshift space $T^\mathrm{s}$ is corrected by a term $D_\mathrm{FoG}^T$ modelling the Fingers-of-God effect (hereafter FoG) \cite{Jackson:2008yv} (see appendix \ref{app:theo_models_rsd}).

\begin{figure}[tbp]
\centering 
\includegraphics[width=1.0\textwidth]%,trim=0 380 0 200,clip]
{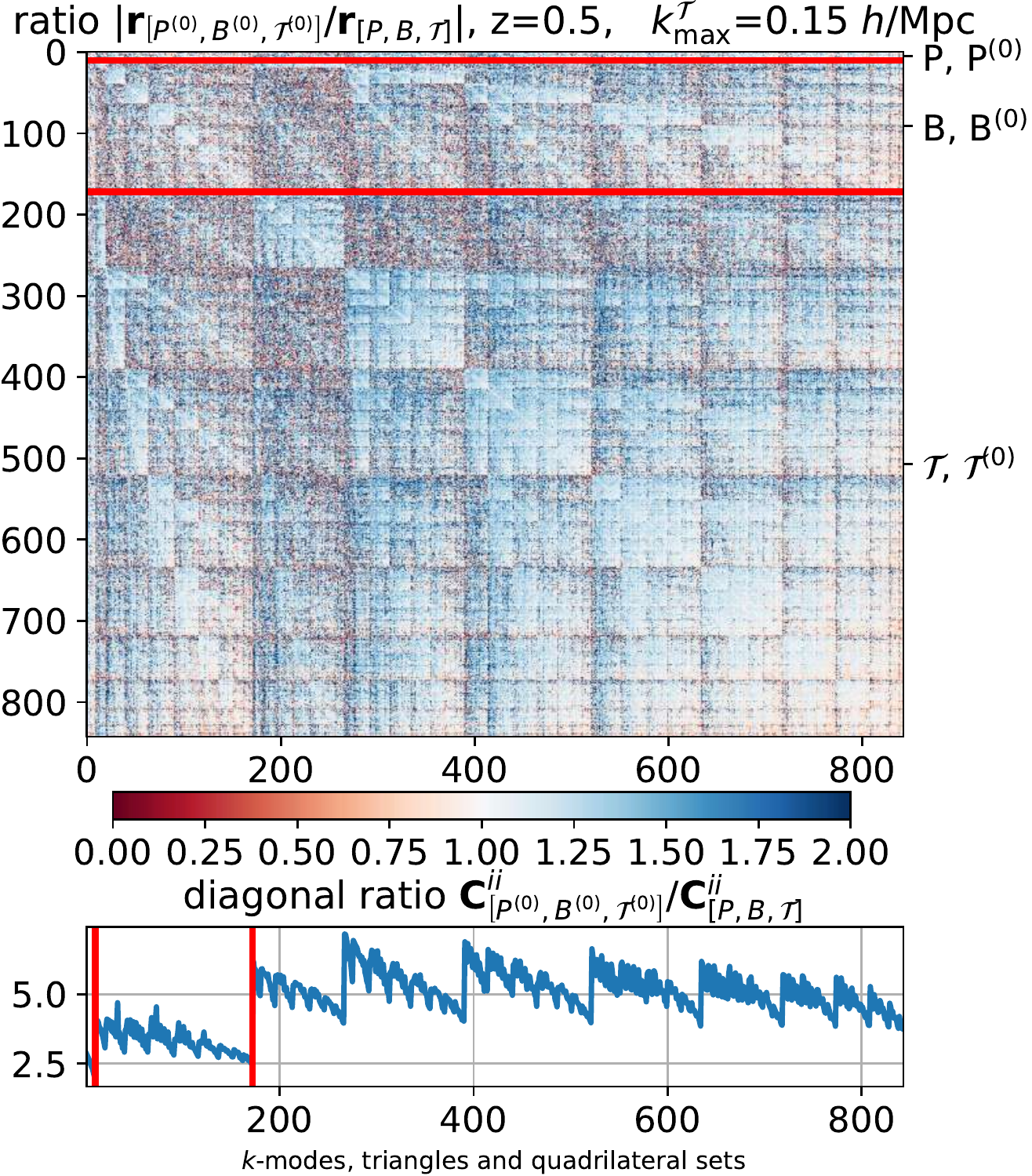}
\caption{\label{fig:rsd_cov_ratio} 
Absolute value of the ratio between the reduced covariance matrices for the data-vector measured in both real and redshift space for the $z=0.5$ case. The absolute value helps with noticing how the off-diagonal cross-correlation between different data-vectors elements increases when the measurement is performed in redshift space. This seems to be relevant especially for the configurations relative to the largest scales considered (top left corner of each auto-covariance matrix, see figure \ref{fig:bk_tk_configurations} for better visualising the ordering of the configurations for bispectrum and i-trispectrum as a function of the scale). Indeed for the bispectrum, for which $k_\mathrm{max}^B=1.3\times k_\mathrm{max}^\mathcal{T}$, the bottom corner of the auto-covariance shows the opposite effect, a reduction of the cross-correlation between different triangles having sides with the largest $k$-values (smallest scales).
}
\end{figure}

We choose to focus on the redshift $z=0.5$ case, which as it can be seen from figures \ref{fig:pk_fit_15kmax}, \ref{fig:bk_fit_15kmax} and \ref{fig:tk_fit_15kmax}, is the case where the modelling is most severely tested: the field is more non-linear and  smaller error bars are derived from the measurements on simulations (compared to the $z=1,2$ cases).

We begin by comparing the covariance matrix for the full data-vector in redshift space to the real space one. The overall structure of the redshift space matrix is very similar to that shown in the top panel of figure \ref{fig:measured_cov}, however in details, the redshift space covariance shows more coupling between different data-vector elements. To better appreciate this,  figure \ref{fig:rsd_cov_ratio} shows the absolute value of the ratio (redshift to real space) of the elements of the reduced covariance matrices at $z=0.5$. The off-diagonal cross-correlation between different data-vectors elements increases when the measurement is performed in redshift space, especially at the largest scales (top left region of each auto-covariance block). For the small-scales bispectrum on the other hand, the smallest scales (bottom right region of the relevant blocks) show a reduction of the cross-correlation. 

We attribute this effect to two factors. First the $k_\mathrm{max}$ for the bispectrum is higher than for the i-trispectrum, therefore proportionally more modes are affected by FoG effect and in the stable-clustering regime. Secondly the i-trispectrum, as it was described in section \ref{sec:Tcoordredshiftspace}, is by definition an integrated quantity over many different quadrilateral shapes and hence it shows increased correlations.

Before we can proceed to show the comparison between simulation output and theoretical modelling for the data-vector, we recall that the parameters describing the small-scales FoG damping are ultimately phenomenological parameters that must be directly fit or calibrated on N-body simulations (and possibly marginalised over).

We use a $\chi^2_\mathrm{Cov}$ minimisation (equation \ref{eq:chi_squared})
to find the values for the small-scales fingers-of-God  parameters $\sigma_P$, $\sigma_B$ and $\sigma_T$ (equations \ref{eq:fog_parameters_pkbk} and \ref{eq:fog_parameters_tk}). This is illustrated in figure \ref{fig:rsd_chi2} in appendix \ref{app:theo_models_rsd}. In what follows, we adopt the values of $\sigma_P$, $\sigma_B$ and $\sigma_T$ that minimise the respective $\chi^2_\mathrm{Cov}$. These FoG parameters are kept fixed in the Fisher forecast analysis.

Figures \ref{fig:pk_fit_15kmax} and \ref{fig:bk_fit_15kmax} haven shown that in real space at $z=0.5$ the \textsc{halofit} matter power spectrum model perform better than the linear one, also as input for the bispectrum model. Therefore for all the results presented in this section we use the \textsc{halofit} matter power spectrum as input for computing the theoretical models. In order to obtain the best possible fit, we employ the redshift space version of the effective kernel $Z^{(2)}$ \cite{Gil-Marin:2014pva} for both bispectrum and i-trispectrum models (equation \ref{eq:pkbktk_rsd}).

\begin{figure}[tbp]
\centering 
\includegraphics[width=\textwidth]%,trim=0 380 0 200,clip]
{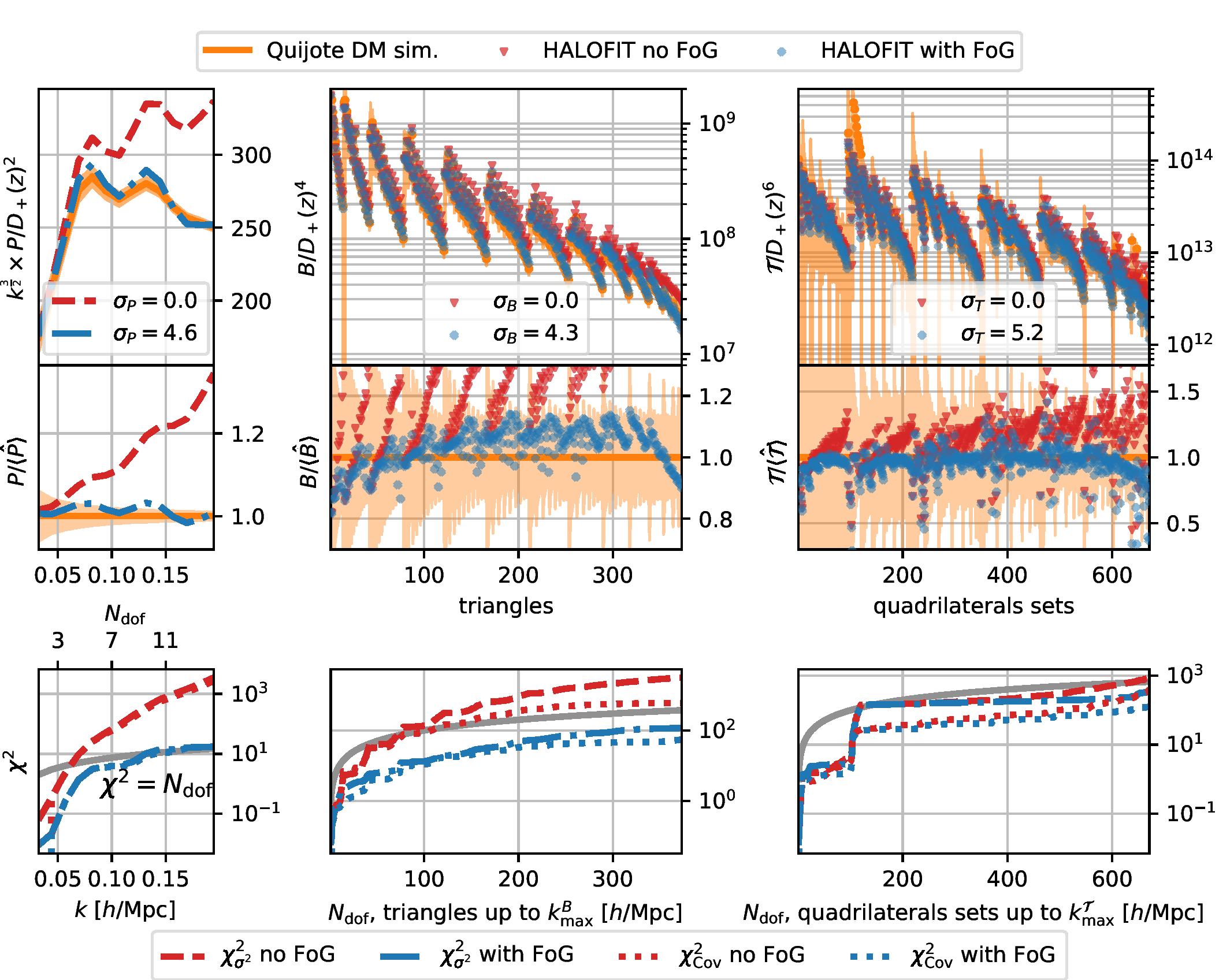}
\caption{\label{fig:rsd_fit_p0b0t0} 
Comparison between measurements of the power spectrum, bispectrum and i-trispectrum monopoles from the \textsc{Quijote} simulations in redshift space at $z=0.5$ and their relative theoretical models. Differently from what shown in the real space case in figure \ref{fig:bk_fit_15kmax}, here we use the effective kernels for the bispectrum with both the linear and \textsc{halofit} fiducial matter power spectrum as input. Also in this case we have that $k_\mathrm{max}^P=k_\mathrm{max}^B=1.3\times k_\mathrm{max}^\mathcal{T}$, where $k_\mathrm{max}^\mathcal{T}(z=0.5)=0.15\,h/$Mpc.
As expected for the chosen redshift,the power spectrum monopole tree level model (equation \ref{eq:pkbktk_rsd}) computed using the \textsc{halofit} prescription for the matter power spectrum achieves a much better fit than the when using the linear one.
The effective model of the redshift second order perturbation theory kernels \cite{Gil-Marin:2014pva} allows the bispectrum monopole model to have a very good fit up to the maximum $k$-value.
Using the effective second order kernel $Z^{(2)}$ also for the i-trispectrum model returns a very good fit. Especially when the FoG damping is used the ratio between theory and average measurement stabilises around unity for the entire  $k$-range considered.
}
\end{figure}

Figure \ref{fig:rsd_fit_p0b0t0} is the redshift space monopole equivalent of the $z=0.5$ panels of figures \ref{fig:pk_fit_15kmax}, \ref{fig:bk_fit_15kmax} and \ref{fig:tk_fit_15kmax}. Together with the lines showing the models for the power spectrum, bispectrum and i-trispectrum models computed using the FoG parameters best-fit values, the models without FoG correction are also shown. As expected, the FoG term becomes more important as $k$ increases and hence in particular for power spectrum and bispectrum whose $k_\mathrm{max}$ is larger than the i-trispectrum one.
In the i-trispectrum case, the FoG correction helps in stabilising the ratio between model and mean of the measurements around unity for all the considered quadrilaterals sets. 

\begin{figure}[tbp]
\centering 
\includegraphics[width=\textwidth]%,trim=0 380 0 200,clip]
{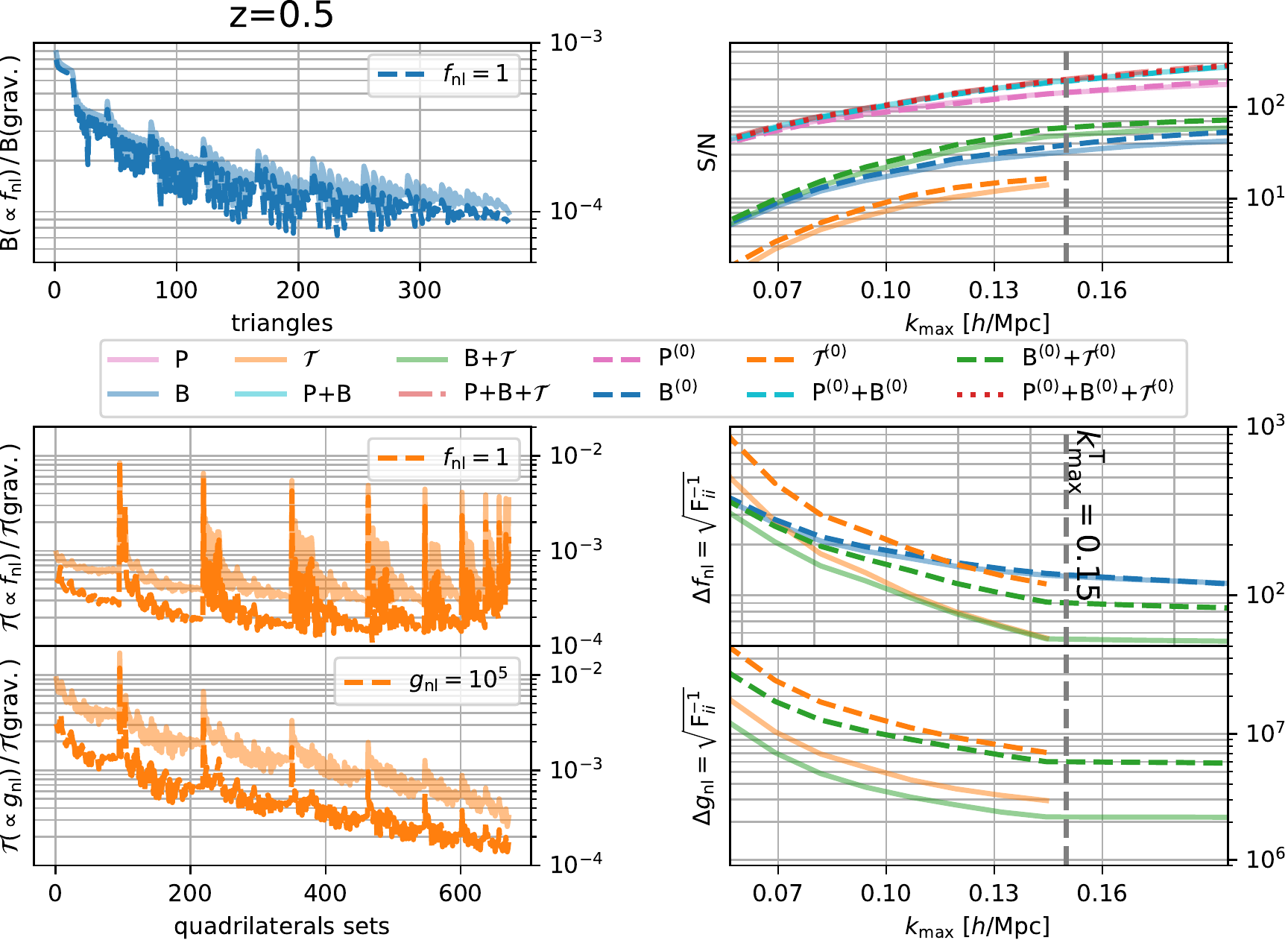}
\caption{\label{fig:rsd_sn_png_forecasts} 
Equivalent of figures \ref{fig:bktk_png} (left side) and \ref{fig:sn_png_constraints} (right side) for the redshift space case ($z=0.5$). The real space quantities are reported in lighter tones. Comparing with figure \ref{fig:bktk_png} relative to the real space case, we see that while the ratio between primordial and gravitational part for the bispectrum is basically left unchanged, the same ratio for the i-trispectrum monopole is on average slightly lower for both terms proportional to $f_\mathrm{nl}$ and $g_\mathrm{nl}$ when passing from real (lighter tone) to redshift (darker tone) space. This reflects into a lower impact in improving the constraints on $f_\mathrm{nl}$ when adding the i-trispectrum signal to the bispectrum with respect to the bispectrum monopole alone. Even if the parameter constraints improvements for both $f_\mathrm{nl}$ and $g_\mathrm{nl}$ are smaller than the ones reported in table \ref{tab:png_forecasts} for the real case, the i-trispectrum monopole added value is still significant (see table \ref{tab:png_forecasts_rsd} for the exact values) and increases for increasing $k_\mathrm{max}^\mathcal{T}$.
}
\end{figure}

The forecasted constraints for primordial non-Gaussianity parameters and their improvement when adding the i-trispectrum to the bispectrum are shown in figure \ref{fig:rsd_sn_png_forecasts} and reported in table \ref{tab:png_forecasts_rsd}. In the left side of figure \ref{fig:rsd_sn_png_forecasts} the ratio between primordial non-Gaussianity and gravitational components for both bispectrum and i-trispectrum is displayed in the redshift space case (dashed lines) and for comparison also in real space (solid line, half transparent identical colours). 

Clearly the measurement in redshift space suppresses the strength of the primordial non-Gaussian component of the signal with respect to the gravitational one. This effect appears to be stronger in the i-trispectrum than for the bispectrum. 

This  can be understood as follows. The redshift-space distortions  on large scales  are gravity-driven and give a larger boost to the gravitational signal than to the PNG signal; an effect we refer to as  Kaiser-boost effect. This can be easily appreciated  in the left column of figure \ref{fig:sn_png_constraints} where the change from real to redshift space is shown by using for the same colour both a lighter and a darker tone, respectively. Being a higher order statistic, the Kaiser-boost is naturally larger for the trispectrum than for the bispectrum as it is highlighted by the larger shift in the ratio between primordial and gravitational components. To further visualize this, in  figure \ref{fig:rsd_vs_real_kaiserboost} of  appendix \ref{sec:app_png} we focus on the Kaiser boost for  equilateral configurations for both $B$ and ${\cal T}$. Because of cosmic variance, this boost also increases the errors (via the covariance matrix).

In the top-right corner of the cumulative signal-to-noise plot, the reduction in the ratio (S/N) is evident only for the i-trispectrum alone before the $k_\mathrm{max}^\mathcal{T}(z=0.5)=0.15\,h/$Mpc threshold. For the bispectrum something similar happens at smaller scales, around $k_\mathrm{max}\sim 0.18\, h/$Mpc.

Finally the bottom right corner showing the forecasted constraints on both $f_\mathrm{nl}$ and $g_\mathrm{nl}$, for the different statistics combinations, connects all the elements appearing in the previous results of this section regarding the measurement, modelling and forecasts in redshift space. The increased cross-correlation between different quadrilaterals sets, highlighted by the ratio between redshift and real space covariance matrices in figure \ref{fig:rsd_cov_ratio}, together with the decrease for the i-trispectrum of both the relevance of the primordial term with respect to the gravitational one and of the cumulative signal-to-noise ratio, result in a smaller impact in redshift space of the i-trispectrum in improving the constraints on both $f_\mathrm{nl}$ and $g_\mathrm{nl}$ with respect to the bispectrum alone. This is quantitatively described in table \ref{tab:png_forecasts_rsd}.

Nevertheless the improvements are still significant, reaching for $f_\mathrm{nl}$ a $\sim 32\%$ reduction of the 68$\%$ 1D confidence interval when both bispectrum and i-trispectrum are employed in the analysis. The improvement for $f_\mathrm{nl}$ become larger as $k_\mathrm{max}^\mathcal{T}$ increases. This implies that improving the modelling of the signal to extend the $k$-range to include smaller scales could return even tighter constraints on $f_\mathrm{nl}$.

%%%%%%%%%%%%%%%%%%%%%%%%%%%%%%%%%%%%%%%%%%%%%
%%%%%%%%%%%%%%%%%%%%%%%%%%%%%%%%%%%%%%%%%%%%%
%%%%%%%%%%%%%%%%%%%%%%%%%%%%%%%%%%%%%%%%%%%%%
%%%%%%%%%%%%%%%%%%%%%%%%%%%%%%%%%%%%%%%%%%%%%

%%%%%%%%%%%%%%%%%%%%%%%%%%%%%%%%%%%%%%%%%%%%%
%%%%%%%%%%%%%%%%%%%%%%%%%%%%%%%%%%%%%%%%%%%%%
%%%%%%%%%%%%%%%%%%%%%%%%%%%%%%%%%%%%%%%%%%%%%
%%%%%%%%%%%%%%%%%%%%%%%%%%%%%%%%%%%%%%%%%%%%%
%%%%%%%
%%%%%%%%%%%%%%%%%%%%%%%%%%%%%%%%%%%%%%%%%%%%%
%%%%%%%%%%%%%%%%%%%%%%%%%%%%%%%%%%%%%%%%%%%%%
%%%%%%%%%%%%%%%%%%%%%%%%%%%%%%%%%%%%%%%%%%%%%

%%%%%%%%%%%%%%%%%%%%%%%%%%%%%%%%%%%%%%%%%%%%%
%%%%%%%%%%%%%%%%%%%%%%%%%%%%%%%%%%%%%%%%%%%%%
%%%%%%%%%%%%%%%%%%%%%%%%%%%%%%%%%%%%%%%%%%%%%
%%%%%%%%%%%%%%%%%%%%%%%%%%%%%%%%%%%%%%%%%%%%%

%%%%%%%%%%%%%%%%%%%%%%%%%%%%%%%%%%%%%%%%%%%%%
%%%%%%%%%%%%%%%%%%%%%%%%%%%%%%%%%%%%%%%%%%%%%
\renewcommand{\arraystretch}{2.3}
\begin{table}[tbp]
\centering
\begin{tabular}{|c|c|cc|cc|cc|cc|}
\hline
\multicolumn{2}{|c|}{$z=0.5$, $k_\mathrm{max}$ $[h/\mathrm{Mpc}]\rightarrow$}&  \multicolumn{2}{c|}{0.10} & \multicolumn{2}{c|}{0.12}
& \multicolumn{2}{c|}{0.15} & \multicolumn{2}{c|}{0.17} \\
\hline
 & $\mathbf{d}$ &  \multicolumn{2}{c|}{RSD (real)} & \multicolumn{2}{c|}{RSD (real)} & \multicolumn{2}{c|}{RSD (real)} & \multicolumn{2}{c|}{RSD (real)}\\
\hline 
\multirow{3}{*}{$\Delta f_\mathrm{nl}$}
&$B^{(0)}$             & 194 & (183) & 157 & (149) & 135 & (132) & 125 & (124) \\ 
& $\mathcal{T}^{(0)}$  & 240 & (136) & 153 & (80)  & 117 & (56)  &  -  &   -   \\
&$B+\mathcal{T}^{(0)}$ & 164 & (122) & 118 & (78)  & 91  & (55)  & 88  & (54)  \\
\hline 
1 - $\dfrac{\Delta f_\mathrm{nl}^{B^{(0)}+\mathcal{T}^{(0)}}}{\Delta f_\mathrm{nl}^{B^{(0)}}}$
& $\left[\%\right]$ 
& \textbf{15} & (34) & \textbf{25} & (48) &  \textbf{32}  & (58) &  \textbf{30} & (56)\\
\hline
$\Delta g_\mathrm{nl}$
& $\mathcal{T}^{(0)}$        & 1404 & (545) & 938 & (368) & 710 & (296) & - &   -   \\
$(\times10^{-4})$
&$B^{(0)}+\mathcal{T}^{(0)}$ & 1049 & (379) & 767 & (273) & 589 & (219) & 583 & (218) \\
\hline
1 - $\dfrac{\Delta g_\mathrm{nl}^{B^{(0)}+\mathcal{T}^{(0)}}}{\Delta g_\mathrm{nl}^{\mathcal{T}^{(0)}}}$
& $\left[\%\right]$ & \textbf{8} & (11) &  \textbf{4} & (14) &  \textbf{5} & (19)  & \textbf{-} & -\\
\hline
\end{tabular}
\caption{\label{tab:png_forecasts_rsd} 
1D $68\%$ forecasted credible regions for both $f_\mathrm{nl}$ and $g_\mathrm{nl}$ as a function of $k_\mathrm{max}$ for the bispectrum, i-trispectrum and bispectrum plus i-trispectrum in redshift space. Each simulation's volume is $1\,[{\rm Gpc}/h]^3$.
}
\end{table}

%%%%%%%%%%%%%%%%%%%%%%%%%%%%%%%%%%%%%%%%%%%%%
%%%%%%%%%%%%%%%%%%%%%%%%%%%%%%%%%%%%%%%%%%%%%
%%%%%%%%%%%%%%%%%%%%%%%%%%%%%%%%%%%%%%%%%%%%%
\section{Conclusions}
\label{sec:conclusions}

In this work we have undertaken the first step towards
employing the four-point correlation function's Fourier transform, the trispectrum, in cosmological analyses  of current and future galaxy clustering data-sets.

The major challenge associated to the trispectrum is its high-dimensionality: six degrees of freedom are necessary to describe a skew-quadrilateral (eight in redshift space); this makes the trispectrum algorithmically and numerically prohibitive.

We propose here to  overcome this difficulty by using a compressed version of the trispectrum signal, which we refer to as  the i-trispectrum. The i-trispectrum  integrates  the signal over  all the skew-quadrilaterals defined by a set of four $k$-modes moduli $(k_1,k_2,k_3,k_4)$.
As such, the i-trispectrum  provides a  solution to the trispectrum challenge by reducing the number of degrees of freedom down to four.

For the first time we model and measure  the i-trispectrum  both in real and redshift space.
We present the i-trispectrum estimator (equation \ref{eq:tk_est_text}) which we then use to measure  the signal from the \textsc{Quijote} simulations  suite at different redshifts, and compare it  with a theoretical model of the i-trispectrum (equation \ref{eq:tk_int_real}) based on perturbation theory.
 We find very good agreement between i-trispectrum model and measurements (figure \ref{fig:tk_fit_15kmax}) up to a maximum $k$ that, as expected,  depends on redshift ($k_\mathrm{max}^{z=0.5}=0.15$ $h/\mathrm{Mpc}$, $k_\mathrm{max}^{z=1}=0.17$ $h/\mathrm{Mpc}$ and $k_\mathrm{max}^{z=2}=0.19$ $h/\mathrm{Mpc}$).

It is important to point out that the unconnected component of the four-point correlator ($\mathcal{T}_\mathrm{u}$) must be estimated and subtracted from the total measured  four-point signal to isolate the i-trispectrum.
The unconnected part is far from being negligible for symmetric configurations (figure \ref{fig:unconnected_vs_connected}), and its removal is fundamental in order to isolate the
the signal (i-trispectrum) containing  cosmological information of the field not already present in the power spectrum. 

From 5000 \textsc{Quijote} simulations we  also estimate and present the  i-trispectrum  covariance matrix and its cross-correlation with power spectrum and bispectrum (figure \ref{fig:measured_cov}). Comparing it with the simplest covariance analytical model (Gaussian field), we show that non-Gaussian and off-diagonal terms are only  negligible above $z\sim2$ (figure \ref{fig:ana_cov}). At lower redshifts, where most of the volume of  present and forthcoming surveys is located, the Gaussian covariance approximation should not be used. 

In analogy to the findings for the Cosmic Microwave Background anisotropies, we envisage the  i-trispectrum to be particularly useful to improve the constraints  on primordial non-Gaussianity (PNG) arising from  lower-order statistics. We thus derive an  analytical model for the  local PNG signature in the i-trispectrum (equation \ref{eq:png_tk_text} and figure \ref{fig:bktk_png}), and produce realistic (using a numerically estimated covariance matrix to account for all the cross-correlations), idealized (matter field, hence low shot noise) but incomplete (non including the scale-dependent bias effect) Fisher forecasts on the PNG amplitude parameters $f_\mathrm{nl}$ and $g_\mathrm{nl}$ constraints.

In fact, additional  information  is indeed expected to be enclosed in the statistics of biased tracers showing the scale-dependent bias effect \cite{Dalal:2007cu}, which for the bispectrum and hence by analogy for the trispectrum, is not just present at very large scales but it is spread over many configurations \cite{Giannantonio:2009ak,Tellarini:2016sgp,Sefusatti:2011gt}.

Including the  i-trispectrum in the  power spectrum and bispectrum analysis has a significant impact in the resulting constraints (figure \ref{fig:sn_png_constraints} and table \ref{tab:png_forecasts}).
In particular, in real space the $68\%$ marginalised credible intervals for $f_\mathrm{nl}$ are approximately halved when the i-trispectrum constraining power is added to the bispectrum.

%----
 The redshift space results --monopole only-- (section \ref{sec:rsd_results}, figure \ref{fig:rsd_fit_p0b0t0}) are qualitatively similar to the real space ones. However the redshift space covariance matrix shows an increased correlation between the i-trispectra of different quadrilaterals sets at the largest scales (figure \ref{fig:rsd_cov_ratio}). Only at high $k$-values this trend inverts, possibly because of the impact of the Finger-of-God effects.
 This is why the i-trispectrum added value in terms of $f_\mathrm{nl}$ and $g_\mathrm{nl}$ constraints (figure \ref{fig:rsd_sn_png_forecasts} and table \ref{tab:png_forecasts_rsd}) is  reduced compared to the  real space  case. Nevertheless  the inclusion of the i-trispectrum provides a significant $\sim 30\%$ improvement on  $f_{\rm nl}$'s $68\%$ 1D marginalised credible intervals.

There are some conservative aspects to
our analysis, since by considering the matter field there is no scale dependent bias effect \cite{Dalal:2007cu,Matarrese:2008nc} boosting the PNG signal.
Moreover we consider a parameter space limited to the two primordial non-Gaussianity amplitudes $f_\mathrm{nl}$ and $g_\mathrm{nl}$. It is reasonable to expect that  when considering
haloes/galaxies in redshift space, thus constraining a larger parameter set, the inclusion of the  i-trispectrum to the  data-vector can provide more significant improvements.  For example the  constraints on the growth rate $f$, the bias coefficients, the amplitude of dark matter clustering  $\sigma_8$,  could be also  significantly tightened by including the i-trispectrum.
Therefore the i-trispectrum has the potential of reducing degeneracies between nuisance (e.g., galaxy bias) and cosmological parameters usually constrained by clustering analysis.

It is important to highlight that both in real and redshift space, the i-trispectrum added value becomes larger as $k_\mathrm{max}$ increases. 
This motivates an update of the standard perturbation theory effective model \cite{GilMarin:2011ik,Gil-Marin:2014pva} using the i-trispectrum together with the bispectrum to fit the required parameters (Novell et al. in preparation).

The next step in order to bring the i-trispectrum into contact with real LSS data, is to model the signal of haloes or galaxies as biased tracers of the underlying dark matter field.
For this we need to  extend to third order the multivariate bias expansion necessary to account for the scale-dependent bias effect. This has already been done in redshift space at second order for the bispectrum \cite{Giannantonio:2009ak,Baldauf:2010vn,Tellarini:2015faa,Tellarini:2016sgp}.
We plan to do this in future work.

An advantage of using higher-order statistics such as the bispectrum and i-trispectrum is to derive constraints on $f_\mathrm{nl}$ and $g_\mathrm{nl}$ highly complementary to CMB ones, without needing the signal from very large scales ($k\lesssim0.001$ $h/\mathrm{Mpc}$) often affected by observational systematic errors, while at the same time including small scales modes. This may not be competitive for the quasars (where volume is very large but the signal to noise ratio is low), but it is interesting for emission line galaxies (ELGs) and luminous red galaxies (LRGs) which cover less volume but have higher signal to noise.

Finally, even more than for the bispectrum \cite{Gualdi:2019sfc}, an optimal compression algorithm will be needed in order to make it feasible to exploit the i-trispectrum full potential by using the maximum number of quadrilaterals sets allowed by the survey specifications.

We envision that, even if focused mainly on spectroscopic surveys and dark matter tracers, the estimator and the modelling presented here will be of relevance to broader sections of cosmology.

%%%%%%%%%%%%%%%%%%%%%%%%%%%%%%%%%%%%%%%%%%%%%%%%%%%%%%%%%%%%%%%%%%%%%
%%%%%%%%%%%%%%%%%%%%%%%%%%%%%%%%%%%%%%%%%%%%%%%%%%%%%%%%%%%%%%%%%%%%%
\appendix
%%%%%%%%%%%%%%%%%%%%%%%%%%%%%%%%%%%%%%%%%%%%%
%%%%%%%%%%%%%%%%%%%%%%%%%%%%%%%%%%%%%%%%%%%%%
%%%%%%%%%%%%%%%%%%%%%%%%%%%%%%%%%%%%%%%%%%%%%
\section{Theoretical models for power spectrum, bispectrum and i-trispectrum}
\label{app:theo_models_rsd}
Below we report the standard perturbation theory (SPT) expressions used in the modelling of the data-vector for the analysis presented in the main text. For completeness we write the models for power spectrum, bispectrum and i-trispectrum  for biased tracers in redshift space, with dependence on the orientation with respect to  the line of sight ($\mu=\cos(\theta)$ with $\theta$ the angle of the $k$-vector with respect to the line of sight). The expressions for the  real space  matter field are simply obtained by setting  the linear bias parameter to unity  in the SPT kernels while all the higher-order bias parameters $b_{i>1}$ together with the logarithmic growth rate parameter $f$ are set to zero.
In particular, the halo power spectrum in real space  is related at first order to the matter one by $P_\mathrm{h}(k)=b_1^2P(k)$;  the real space expressions for the  matter bispectrum  are reported in the main text in equations \ref{eq:BSPT}, the real space, matter SPT expression for the i-trispectrum is given in the main text  in equation \ref{eq:TSPT}.
In redshift space and for biased fields we have
\begin{eqnarray}
\label{eq:pkbktk_rsd}
P^\mathrm{s}(\mathbf{k}) 
&=&
Z^{(1)}\left[\mu,b_1,f\right]^2P(k)
\notag\\
B^\mathrm{s}(\mathbf{k}_1,\mathbf{k}_2,\mathbf{k}_3)
&=&
Z^{(1)}\left[\mu_1,b_1,f\right]Z^{(1)}\left[\mu_2,b_1,f\right]
Z^{(2)}\left[\mathbf{k}_1,\mathbf{k}_2,b_1,b_2,b_{s_2},f\right]
P(k_1)P(k_2)
\,+\,2\,\mathrm{p}.
\notag \\
T^\mathrm{s}(\mathbf{k}_1,\mathbf{k}_2,\mathbf{k}_3,\mathbf{k}_4)
&=&
4\,Z^{(1)}\left[\mu_1,b_1,f\right]Z^{(1)}\left[\mu_2,b_1,f\right]
P(k_1)P(k_2)
\notag \\
&\times&
\Big\{
Z^{(2)}\left[\mathbf{k}_1,-\mathbf{k}_{13},b_1,b_2,b_{s_2},f\right]
Z^{(2)}\left[\mathbf{k}_2, \mathbf{k}_{13},b_1,b_2,b_{s_2},f\right]
P(k_{13})
\notag \\
&+&
\,\,\,\,Z^{(2)}\left[\mathbf{k}_1,-\mathbf{k}_{14},b_1,b_2,b_{s_2},f\right]
Z^{(2)}\left[\mathbf{k}_2, \mathbf{k}_{14},b_1,b_2,b_{s_2},f\right]
P(k_{14})
\Big\}
\,+\,5\,\mathrm{p}.
\notag \\
&+&
6\,Z^{(1)}\left[\mu_1,b_1,f\right]Z^{(1)}\left[\mu_2,b_1,f\right]Z^{(1)}\left[\mu_3,b_1,f\right]
\notag \\
&\times&
\,\,\,\,Z^{(3)}\left[\mathbf{k}_{1},\mathbf{k}_2, \mathbf{k}_{3},b_1,b_2,b_{s_2},b_3,f\right]
P(k_1)P(k_2)P(k_3)
\,+\,3\,\mathrm{p}.
\end{eqnarray}

\noindent where the redshift-space distortions kernels can be found  for example in Ref. \cite{Karagiannis:2018jdt}'s appendix and a specific study on a more accurate bias expansion at cubic order was done by the authors of Ref.~ \cite{Abidi:2018eyd}. For what concerns the primordial non-Gaussianity terms reported in equation \ref{eq:png_tk_text} and appendix \ref{sec:app_png} for matter field in real space, the equivalent redshift space expressions are obtained by simply replacing the matter field perturbation theory kernels $F^{(i)}$ with the redshift space ones $Z^{(i)}$. Notice that this can be done only for the matter field since when considering haloes or galaxies, the scale-dependent bias effect \cite{Dalal:2007cu,Matarrese:2008nc} introduces additional non-negligible terms. 
As it can be seen from the above expression (see also appendix \ref{sec:subsec_tkpng}), the i-trispectrum is composed of two different terms d expansions in perturbation theory up to order $\propto\delta^6$ \cite{Fry:1983cj}:
\begin{eqnarray}
\label{eq:tk_spt_decomposition}
T^\mathrm{s}(\mathbf{k}_1,\mathbf{k}_2,\mathbf{k}_3,\mathbf{k}_4)
= T^\mathrm{s}_{(1122)} + T^\mathrm{s}_{(1113)}\,.
\end{eqnarray}

\noindent The isotropic signal component --monopole-- for both power spectrum and bispectrum in redshift space is given by

\begin{eqnarray}
P^{(0)}=\dfrac{1}{2}\int^{1}_{-1}d\mu \,P^\mathrm{s}(k,\mu) 
\quad \mathrm{and} \quad 
B^{(0)} = \dfrac{1}{4}\int^{1}_{-1}d\mu_1d\mu_2\,B^\mathrm{s}(\mathbf{k}_1,\mathbf{k}_2,\mathbf{k}_3)\,,
\end{eqnarray}
\noindent while the i-trispectrum monopole was defined in the main text in equation \ref{eq:tk_int_rsd}.

%%%%%%%%%%%%%%%%%%%%%%%%%%%%%%%%%%%%%%
\begin{figure}[tbp]
\centering 
\includegraphics[width=\textwidth]%,trim=0 380 0 200,clip]
{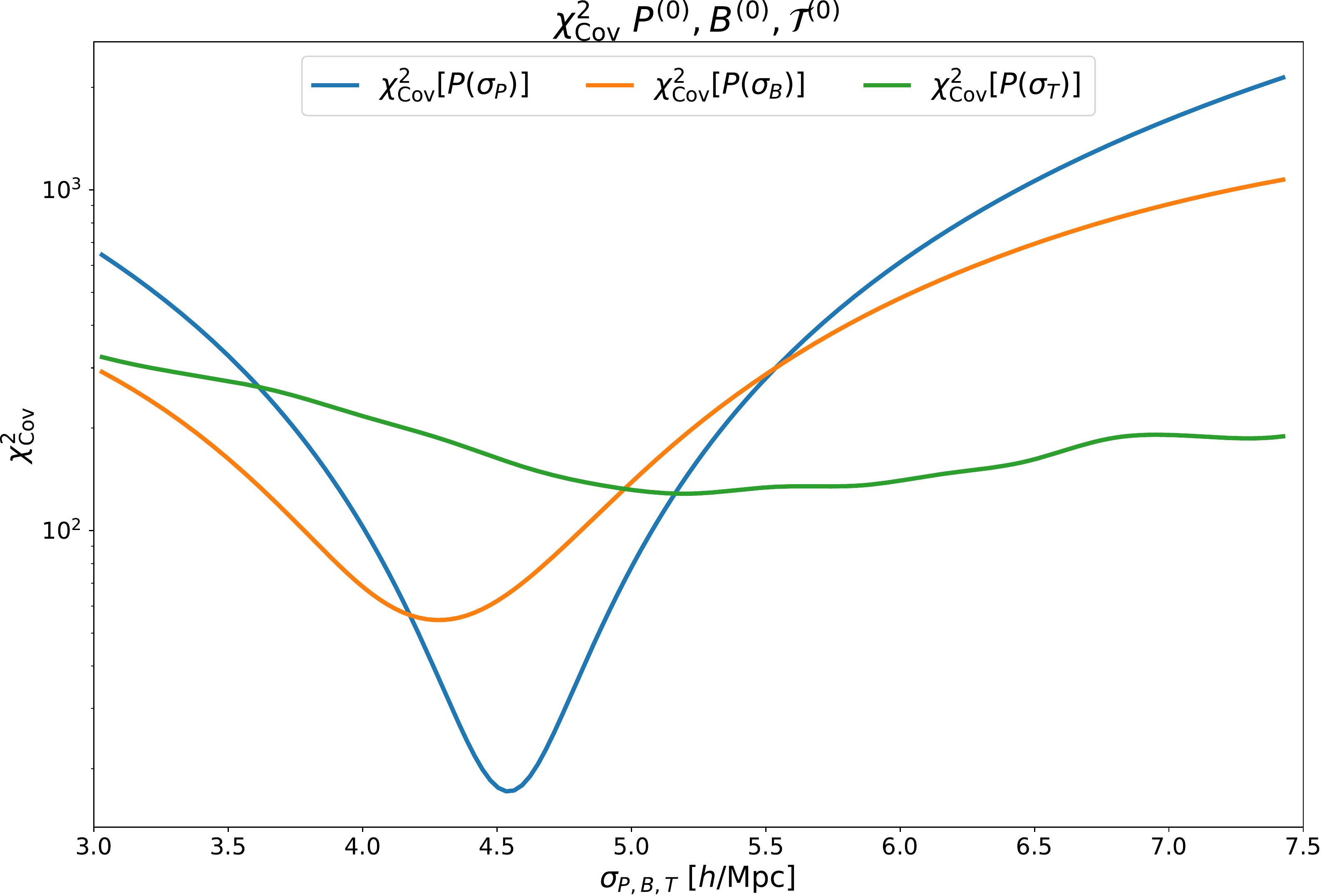}
\caption{\label{fig:rsd_chi2} 
$\chi^2_\mathrm{Cov}$-test (equation \ref{eq:chi_squared}) used to find the best-fitting values for the Fingers-of-God parameters,  $\sigma_P$, $\sigma_B$ and $\sigma_T$ (equations \ref{eq:fog_parameters_pkbk} and \ref{eq:fog_parameters_tk}), for the power spectrum, bispectrum and i-trispectrum of the 5000 realisations of the \textsc{Quijote} N-body simulations.}
\end{figure}
%%%%%%%%%%%%%%%%%%%%%%%%%%%%%%%%%%%%%%

To model the small-scales incoherent velocity dispersion, a damping term is added for each statistics \cite{Jackson:2008yv}.
Similarly to the prescriptions used for power spectrum and bispectrum in \cite{Gil-Marin:2014sta} which require two effective parameters $\sigma_{P}$ and $\sigma_{B}$ entering the Lorentzian damping functions in front of the respective data-vectors
\begin{align}
\label{eq:fog_parameters_pkbk}
   D_\mathrm{FoG}^P&=\dfrac{1}{\left[1 + k^2\mu^2\sigma_{P}^2/2\right]^2}\, \notag \\
    D_\mathrm{FoG}^B&=\,\dfrac{1}{\left[1 + (k_1^2\mu_1^2+k_2^2\mu_2^2+k_3^2\mu_3^2)^2\sigma_{B}^4/2\right]^2 }\,,
\end{align}{}
\noindent in the case of the i-trispectrum we use the {\it ansatz}:
\begin{align}
\label{eq:fog_parameters_tk}
    D_\mathrm{FoG}^T&=\,\dfrac{1}{\left[1 + (k_1^2\mu_1^2+k_2^2\mu_2^2+k_3^2\mu_3^2+k_4^2\mu_4^2)^3\sigma_{T}^6/2\right]^2}\,.
\end{align}{}
The parameters $\sigma_P$, $\sigma_B$, $\sigma_T$, should be seen as effective parameters to be calibrated on simulations (or measured/marginalised over when analysing the data). In our case we find for each statistic the best-fit value of the respective FoG parameter by minimising the $\chi^2_\mathrm{Cov}$ as in  equation \ref{eq:chi_squared}. The results of this procedure are displayed in figure \ref{fig:rsd_chi2} and the best-fit values are $\sigma_P=4.6\, {\rm Mpc}/h$, $\sigma_B=4.3\, {\rm Mpc}/h$ and $\sigma_T=5.2\, {\rm Mpc}/h$.

%%%%%%%%%%%%%%%%%%%%%%%%%%%%%%%%%%%%%%%%%%%%%
%%%%%%%%%%%%%%%%%%%%%%%%%%%%%%%%%%%%%%%%%%%%%
%%%%%%%%%%%%%%%%%%%%%%%%%%%%%%%%%%%%%%%%%%%%%
\section{Estimators definition}
\label{App:estimators}
Assuming a cubic survey volume $V$ of size $L$, the fundamental frequency is $k_\mathrm{f}=\frac{2\pi}{L}$. If we divide the box into $N$ equal volume cubic cells per side, the Nyquist frequency is defined as $k_\mathrm{Ny}=N\times k_\mathrm{f}/2$.

Let's recall that the first step in order to measure statistics in Fourier space from data/simulations is to grid sample the chosen tracers field by using an appropriate mass assignment scheme (see for example \cite{Sefusatti:2015aex}). This returns a value of the density field at each grid-node in configuration space. From this set-up it is afterwards possible to exploit discrete Fast Fourier Transforms algorithms (FFT) \cite{10.1145/1464291.1464352} to obtain the Fourier transform of the density field at each node. The two grids (mass assigment and Fourier transform steps) could in principle be taken to be different but often they coincide (as in our case).

In this section we will refer to the process of converting an estimator in integral form into one written in terms of discrete sums and hence that can be applied onto a pixelated field as "pixelisation". 

Using the density (matter/galaxy) perturbation variable $\delta$ we define the Fourier transform direct and inverse arbitrary convention:

\begin{eqnarray}
\delta_\mathbf{k} = \int d\mathbf{x}^3\, \delta_\mathbf{x}\,e^{i\mathbf{kx}}
\quad
\mathrm{and}
\quad
\delta_\mathbf{x} =\dfrac{1}{(2\pi)^3} \int d\mathbf{k}^3\, \delta_\mathbf{k}\,e^{-i\mathbf{kx}} \,.
\end{eqnarray}

%%%%%%%%%%%%%%%%%%%%%%%%%%%%%%%%%%%%%%%%%%%%%%%%%%%%%%%%%%%%%%%%%%%%%
%%%%%%%%%%%%%%%%%%%%%%%%%%%%%%%%%%%%%%%%%%%%%%%%%%%%%%%%%%%%%%%%%%%%%
\subsection{Power spectrum}
The integral version of the estimator for the power spectrum is defined as \cite{Gualdi:2018pyw}

\begin{eqnarray}
\label{eq:pk_est}
\hat{P}\left(k_1\right)=\dfrac{(2\pi)^{-3}}{N_\mathrm{p}(k_1)}
\int_{k_1}d\mathbf{q}_1^3
\int_{k_1}d\mathbf{q}_2^3\,
\delta_{\mathbf{q}_1}\,\delta_{\mathbf{q}_2}\,
\delta^D\left(\mathbf{q}_{12}\right)\,,
\end{eqnarray}

\noindent where each integral is performed over a spherical shell of radius $k_i$ and thickness $\Delta k$, and ${\bf q}_{12}\equiv {\bf q}_1 + {\bf q}_2$. 
$N_\mathrm{p}$ is the number of modes found inside the integration volume in Fourier space and it is defined as

\begin{eqnarray}
\label{eq:npairs}
N_\mathrm{p}(k_1)=\dfrac{V_\mathrm{p}(k_1)}{k_\mathrm{f}^3}=\dfrac{1}{k_\mathrm{f}^3}
\int_{k_1}d\mathbf{q}_1^3
\int_{k_1}d\mathbf{q}_2^3\,
\delta^D\left(\mathbf{q}_{12} \right) \,.
\end{eqnarray}

\noindent In order to derive the pixelised version of the above estimator we need the Dirac's delta expression as Fourier transform of $1$:

\begin{eqnarray}
\label{eq:Dirac}
\delta^D\left(\mathbf{q}\right) = 
\int \dfrac{d\mathbf{x}^3}{(2\pi)^3}\,e^{i\mathbf{x}\mathbf{q}}\,.
\end{eqnarray}

\noindent Therefore in  equations \ref{eq:pk_est} and \ref{eq:npairs} we can expand the Dirac's delta and rearrange the order of integration. Proceeding with equation \ref{eq:pk_est} by also expanding $N_\mathrm{p}$ we have

\begin{eqnarray}
\label{eq:pk_expansion0}
\hat{P}\left(k_1\right)&=
(2\pi)^{-3} \int d\mathbf{x}^3
\int_{k_1}d\mathbf{q}_1^3\,\delta_{\mathbf{q}_1}\,e^{i\mathbf{x}\mathbf{q}_1}
\int_{k_1}d\mathbf{q}_2^3\,\delta_{\mathbf{q}_2}\,e^{i\mathbf{x}\mathbf{q}_2}
\notag \\
&\times
\left[k_\mathrm{f}^{-3}
\int d\mathbf{y}^3
\int_{k_1}d\mathbf{q}_1^3\,e^{i\mathbf{y}\mathbf{q}_1}
\int_{k_1}d\mathbf{q}_2^3\,e^{i\mathbf{y}\mathbf{q}_2}
\right]^{-1}\,.
\notag \\
\end{eqnarray}

\noindent Now we introduce the two quantities:

\begin{eqnarray}
\label{eq:IJ_def}
I_{k_i}(\mathbf{x}) = \int_{k_i}\dfrac{d\mathbf{q}_i^3}{(2\pi)^3}\,\delta_{\mathbf{q}_i}
e^{i\mathbf{x}\mathbf{q}_i}
\quad
\mathrm{and}
\quad
J_{k_i}(\mathbf{x}) = \int_{k_i}\dfrac{d\mathbf{q}_i^3}{(2\pi)^3}\,e^{i\mathbf{x}\mathbf{q}_i}\,,
\end{eqnarray}

\noindent which applied to equation \ref{eq:pk_expansion0} give,

\begin{align}
\label{eq:pk_expansion1}
\hat{P}\left(k_1\right)=
(2\pi)^3\,k_\mathrm{f}^3\int d\mathbf{x}^3\, I_{k_1}(\mathbf{x})\, I_{k_2}(\mathbf{x}) 
\times
\left[(2\pi)^6\int d\mathbf{y}^3
J_{k_1}(\mathbf{y})\, J_{k_2}(\mathbf{y})\right]^{-1}\,.
\end{align}

\noindent To take advantage of fast Fourier transform techniques,  each continuous integral over the whole spatial volume can be pixelised as $\int d\mathbf{x}^3\longrightarrow \Delta V\sum_{\ell=1}^{N^3}$, where $\Delta V = \dfrac{L^3}{N^3}$ is the volume of each cell. Then equation \ref{eq:pk_expansion1} can be rewritten as,

\begin{align}
\label{eq:pk_expansion2}
\hat{P}\left(k_1\right)&=
(2\pi)^{-3}k_\mathrm{f}^3
\sum_{\imath=1}^{N^3}\, I_{k_1}(x_\imath)\, I_{k_2}(x_\imath) 
\times
\left[\sum_{\jmath=1}^{N^3}
 J_{k_1}(y_\jmath)\, J_{k_2}(y_\jmath) \right]^{-1}
 \notag \\
 &=
 L^{-3}
 \sum_{\imath=1}^{N^3}\, I_{k_1}(x_\imath)\, I_{k_2}(x_\imath) 
\times
\left[\sum_{\jmath=1}^{N^3}
 J_{k_1}(y_\jmath)\, J_{k_2}(y_\jmath) \right]^{-1}
 \,.
\end{align}

\noindent  We start by  pixelising the Fourier transform of the density field $\delta_{\mathbf{q}_i}$:
\begin{eqnarray}
\delta_{\mathbf{q}_i}=
\int d\mathbf{x}^3\, \delta_\mathbf{x}\,e^{i\mathbf{xq}_i} \quad\longrightarrow\quad \left(\dfrac{L}{N}\right)^3\sum^{N^3}_{j=1}e^{iq_ix_j}\delta_{x_j} = \left(\dfrac{L}{N}\right)^3 \,F_{\mathbf{q}_i}\,,
\end{eqnarray}
\noindent where $F_i$ is the quantity computed from the Discrete Fourier Transforms algorithm (in our case FFTW \cite{FFTW05}). Then we can proceed to  pixelise both quantities in equation \ref{eq:IJ_def}:

\begin{align}
\label{eq:IJ_discrete}
I_{k_i}(\mathbf{x}) &= \int_{k_i}\dfrac{d\mathbf{q}_i^3}{(2\pi)^3}\,
e^{i\mathbf{x}\mathbf{q}_i}
\left(\dfrac{L}{N}\right)^3\,F_{\mathbf{q}_i} 
= \int\dfrac{d\mathbf{q}^3}{(2\pi)^3}\,
e^{i\mathbf{x}\mathbf{q}}
\left(\dfrac{L}{N}\right)^3\,F^i_{\mathbf{q}} 
\notag \\
&\rightarrow\,
\left(\dfrac{L}{N}\right)^3 \left(\dfrac{1}{N}\right)^3_\mathrm{IFT}\sum_{\ell=1}^{N^3}e^{i\mathbf{x}q_\ell}\,F^i_{q_\ell}
=
\left(\dfrac{L}{N}\right)^3 \left(\dfrac{1}{N}\right)^3_\mathrm{IFT}I^D_{k_i}(\mathbf{x})
\notag \\\notag\\
J_{k_i}(\mathbf{x}) &= \int_{k_i}\dfrac{d\mathbf{q}_i^3}{(2\pi)^3}\,e^{i\mathbf{x}\mathbf{q}_i}
= 
\int\dfrac{d\mathbf{q}^3}{(2\pi)^3}\,e^{i\mathbf{x}\mathbf{q}}\,S^i_\mathbf{q}
\notag \\
&\rightarrow\,
\left(\dfrac{1}{N}\right)^3_\mathrm{IFT}\sum_{\ell=1}^{N^3}e^{i\mathbf{x}q_\ell}\,S^i_{q_\ell}
=
\left(\dfrac{1}{N}\right)^3_\mathrm{IFT}J^D_{k_i}(\mathbf{x})
\,
\end{align}
\noindent where in the first step the index $i$ is used to indicate that the integrand is non-null only within the shell with radius $k_i$ and thickness $\Delta k$. Hence  $S^i_\mathbf{q}$ is defined so that it is  equal to unity for $\mathbf{q}$ inside the $k_i$-shell and zero outside. The normalisation factor due to the discrete inverse Fourier transform has been specified using the "IFT" subscript (inverse Fourier transform). 

Finally  with the  pixelised result of  equation \ref{eq:IJ_discrete},  \ref{eq:pk_expansion2} becomes:

\begin{align}
\label{eq:pk_expansion3}
\hat{P}\left(k_1\right)
 &=
 L^{-3}
 \sum_{\imath=1}^{N^3}\, I_{k_1}(x_\imath)\, I_{k_2}(x_\imath) 
\times
\left[\sum_{\jmath=1}^{N^3}
 J_{k_1}(y_\jmath)\, J_{k_2}(y_\jmath) \right]^{-1}
 \notag \\
 &\rightarrow
  L^{-3}
 \left(\dfrac{L}{N}\right)^6 \left(\dfrac{1}{N}\right)^6_\mathrm{IFT}
 \sum_{\imath=1}^{N^3}\, I^D_{k_1}(x_\imath)\, I^D_{k_2}(x_\imath)
 \times
\left[\left(\dfrac{1}{N}\right)^6_\mathrm{IFT}\sum_{\jmath=1}^{N^3}
 J^D_{k_1}(y_\jmath)\, J^D_{k_2}(y_\jmath) \right]^{-1}
 \notag\\
 &= \dfrac{L^3}{N^6}
  \sum_{\imath=1}^{N^3}\, I^D_{k_1}(x_\imath)\, I^D_{k_2}(x_\imath)
 \times
\left[\sum_{\jmath=1}^{N^3}
 J^D_{k_1}(y_\jmath)\, J^D_{k_2}(y_\jmath) \right]^{-1} \,.
\end{align}

\noindent By construction, $\hat{P}\left(k_1\right)$ has the appropriate dimension of length to the power of 3.

%%%%%%%%%%%%%%%%%%%%%%%%%%%%%%%%%%%%%%%%%%%%%%%%%%%%%%%%%%%%%%%%%%%%%
%%%%%%%%%%%%%%%%%%%%%%%%%%%%%%%%%%%%%%%%%%%%%%%%%%%%%%%%%%%%%%%%%%%55
\subsection{Bispectrum}
Starting  from the unbiased estimator as defined in \cite{Gualdi:2018pyw}, for the bispectrum we have,
\begin{eqnarray}
\label{eq:bk_est}
\hat{B}\left(k_1,k_2,k_3\right)=\dfrac{V(2\pi)^{-6}}{N_\mathrm{t}(k_1,k_2,k_3)}
\int_{k_1}d\mathbf{q}_1^3
\int_{k_2}d\mathbf{q}_2^3
\int_{k_3}d\mathbf{q}_3^3\,
\delta_{\mathbf{q}_1}\,\delta_{\mathbf{q}_2}\,\delta_{\mathbf{q}_3}\,
\delta^D\left(\mathbf{q}_{123}\right)\,,
\end{eqnarray}
\noindent where ${\bf{q}}_{123}\equiv{\bf{q}}_1+{\bf{q}}_2+{\bf{q}}_3$, $N_\mathrm{t}$ is the number of triangles included in the integration volume in Fourier space,
\begin{eqnarray}
\label{eq:ntrips}
N_\mathrm{t}(k_1,k_2,k_3)=\dfrac{V_\mathrm{t}(k_1,k_2,k_3)}{k_\mathrm{f}^6}=\dfrac{1}{k_\mathrm{f}^6}
\int_{k_1}d\mathbf{q}_1^3
\int_{k_2}d\mathbf{q}_2^3
\int_{k_3}d\mathbf{q}_3^3\,
\delta^D\left({\bf{q}}_{123}\right) \,.
\end{eqnarray}
\noindent Proceeding then as for the power spectrum and decomposing the Dirac's deltas using equation \ref{eq:Dirac} to derive the quantities in equation \ref{eq:IJ_def} and pixelising using equation \ref{eq:IJ_discrete} we have that
\begin{align}
\label{eq:bk_expansion0}
&\hat{B}\left(k_1,k_2,k_3\right)=
V(2\pi)^3\,k_\mathrm{f}^6\int d\mathbf{x}^3
\, I_{k_1}(\mathbf{x})\, I_{k_2}(\mathbf{x}) \,I_{k_3}(\mathbf{x})
\times
\left[(2\pi)^9\int d\mathbf{y}^3
J_{k_1}(\mathbf{y})\, J_{k_2}(\mathbf{y})\, J_{k_3}(\mathbf{y})\right]^{-1}
\notag \\
% &=
%  V\,(2\pi)^{-6}\,k_\mathrm{f}^6\,
%  \sum_{\imath=1}^{N^3}
%  \, I_{k_1}(x_\imath)\, I_{k_2}(x_\imath) \, I_{k_3}(x_\imath) 
% \times
% \left[\sum_{\jmath=1}^{N^3}
%  J_{k_1}(y_\jmath)\, J_{k_2}(y_\jmath)\, J_{k_3}(y_\jmath) \right]^{-1}
%  \notag \\
% &=
%  L^{-3}
%  \sum_{\imath=1}^{N^3}
%  \, I_{k_1}(x_\imath)\, I_{k_2}(x_\imath) \, I_{k_3}(x_\imath) 
% \times
% \left[\sum_{\jmath=1}^{N^3}
%  J_{k_1}(y_\jmath)\, J_{k_2}(y_\jmath)\, J_{k_3}(y_\jmath) \right]^{-1}
% \,.
% \end{align}
% %
% \noindent We can fully pixelise using equation \ref{eq:IJ_discrete}:
% %
% \begin{align}
% \label{eq:bk_expansion1}
% &\hat{B}\left(k_1,k_2,k_3\right)=
%  L^{-3}
%  \sum_{\imath=1}^{N^3}
%  \, I_{k_1}(x_\imath)\, I_{k_2}(x_\imath) \, I_{k_3}(x_\imath) 
% \times
% \left[\sum_{\jmath=1}^{N^3}
%  J_{k_1}(y_\jmath)\, J_{k_2}(y_\jmath)\, J_{k_3}(y_\jmath) \right]^{-1}
% \notag \\
 &\rightarrow
  L^{-3}
  \left(\dfrac{L}{N}\right)^9 \left(\dfrac{1}{N}\right)^9_\mathrm{IFT}
 \sum_{\imath=1}^{N^3}
 \, I^D_{k_1}(x_\imath)\, I^D_{k_2}(x_\imath) \, I^D_{k_3}(x_\imath) 
\notag \\
&\times
\left[\left(\dfrac{1}{N}\right)^9_\mathrm{IFT}\sum_{\jmath=1}^{N^3}
 J^D_{k_1}(y_\jmath)\, J^D_{k_2}(y_\jmath)\, J^D_{k_3}(y_\jmath) \right]^{-1}
 \notag \\
 &=
  \left(\dfrac{L^6}{N^9}\right)
 \sum_{\imath=1}^{N^3}
 \, I^D_{k_1}(x_\imath)\, I^D_{k_2}(x_\imath) \, I^D_{k_3}(x_\imath) 
\times
\left[\sum_{\jmath=1}^{N^3}
 J^D_{k_1}(y_\jmath)\, J^D_{k_2}(y_\jmath)\, J^D_{k_3}(y_\jmath) \right]^{-1}
\,.
\end{align}
\noindent The dimension of $\hat{B}$ is by construction length to the power of 6.

\subsection{I-trispectrum}
In analogy to the bispectrum, for the (integrated) i-trispectrum (see main text for more details) we begin by  defining the unbiased estimator
\begin{eqnarray}
\label{eq:tk_est}
\hat{\mathcal{T}}_{\rm c+u}\left(k_1,k_2,k_3,k_4\right)&=&\dfrac{V^2(2\pi)^{-9}}{N_\mathrm{q}\left(k_1,k_2,k_3,k_4\right)}
\int_{k_1}d\mathbf{q}_1^3
\int_{k_2}d\mathbf{q}_2^3
\int_{k_3}d\mathbf{q}_3^3
\notag \\
&\times&
\int_{k_4}d\mathbf{q}_4^3\,
\delta_{\mathbf{q}_1}\,\delta_{\mathbf{q}_2}\,
\delta_{\mathbf{q}_3}\,\delta_{\mathbf{q}_4}\delta^D\left(\mathbf{q}_{1234} \right)\,,
\notag \\
\end{eqnarray}
\noindent  where ${\bf{q}}_{1234}\equiv{\bf{q}}_1+{\bf{q}}_2+{\bf{q}}_3+{\bf{q}}_4$, $N_\mathrm{q}$ is the number of skew-quadrilaterals in the integration volume in Fourier space, defined as
\begin{eqnarray}
\label{eq:nquad}
N_\mathrm{q}\left(k_1,k_2,k_3,k_4\right)=\dfrac{V_\mathrm{q}\left(k_1,k_2,k_3,k_4\right)}{k_\mathrm{f}^9}=\dfrac{1}{k_\mathrm{f}^9}
\int_{k_1}d\mathbf{q}_1^3
\int_{k_2}d\mathbf{q}_2^3
\int_{k_3}d\mathbf{q}_3^3
\int_{k_3}d\mathbf{q}_4^3\,
\delta^D\left(\mathbf{q}_{1234}\right) \,.
\notag\\
\end{eqnarray}
\noindent We can confirm that it is actually an unbiased estimator, as for the cases of power spectrum and bispectrum, by taking the ensemble average,
\begin{eqnarray}
\label{eq:tk_unbiased_app}
\langle\hat{\mathcal{T}}_{\rm c+u}\left(k_1,k_2,k_3,k_4\right)\rangle
&=&
\dfrac{V^2(2\pi)^{-9}}{N_\mathrm{q}\left(k_1,k_2,k_3,k_4\right)}
\int_{k_1}d\mathbf{q}_1^3
\int_{k_2}d\mathbf{q}_2^3
\int_{k_3}d\mathbf{q}_3^3
\int_{k_4}d\mathbf{q}_4^3\,
\langle\delta_{\mathbf{q}_1}\,\delta_{\mathbf{q}_2}\,
\delta_{\mathbf{q}_3}\,\delta_{\mathbf{q}_4}\rangle\,
\notag \\
&\times&
\delta^D\left(\mathbf{q}_{1234} \right)
\notag \\
&=&
\dfrac{V^2(2\pi)^{-9}}{N_\mathrm{q}\left(k_1,k_2,k_3,k_4\right)}
\int_{k_1}d\mathbf{q}_1^3
\int_{k_2}d\mathbf{q}_2^3
\int_{k_3}d\mathbf{q}_3^3
\notag \\
&\times&
\int_{k_4}d\mathbf{q}_4^3\,
\left[
\langle
\delta_{\mathbf{q}_1}\,\delta_{\mathbf{q}_2}\,
\delta_{\mathbf{q}_3}\,\delta_{\mathbf{q}_4}
\rangle_\mathrm{c}
\,+\, 
\langle\delta_{\mathbf{q}_1}\,\delta_{\mathbf{q}_2}\rangle
\langle\delta_{\mathbf{q}_3}\,\delta_{\mathbf{q}_4}\rangle \, + \,2\,\mathrm{p.}
\right] \delta^D\left(\mathbf{q}_{1234}\right) \,.\notag \\
\end{eqnarray}
\noindent We then separately analyse the connected term and the unconnected one obtained by applying Wick's theorem to complete the proof of the above estimator being unbiased.

\subsubsection{I-trispectrum: connected  part and total signal}
\label{app:tk_connected}
If one considers only the connected part of the total signal estimator introduced in equation \ref{eq:tk_unbiased_app}, it is possible to recover the correspondence with the i-trispectrum signal
\begin{align}
\label{eq:tk_unbiased_connected}
&\langle\hat{\mathcal{T}}_\mathrm{c}\left(k_1,k_2,k_3,k_4\right)\rangle
=
\dfrac{V^2(2\pi)^{-9}k_\mathrm{f}^{9}}{V_\mathrm{q}\left(k_1,k_2,k_3,k_4\right)}
\int_{k_1}d\mathbf{q}_1^3
\int_{k_2}d\mathbf{q}_2^3
\int_{k_3}d\mathbf{q}_3^3
\int_{k_4}d\mathbf{q}_4^3\,
\notag \\
&\times
(2\pi)^3\delta^D\left(\mathbf{q}_{1234} \right)^2
T(\mathbf{q}_1,\mathbf{q}_2,\mathbf{q}_3,\mathbf{q}_4)
\notag \\
&=
\dfrac{(2\pi)^{-3}k_\mathrm{f}^{3}}{V_\mathrm{q}\left(k_1,k_2,k_3,k_4\right)}
\int_{k_1}d\mathbf{q}_1^3
\int_{k_2}d\mathbf{q}_2^3
\int_{k_3}d\mathbf{q}_3^3
\int_{k_4}d\mathbf{q}_4^3\,
(2\pi)^3k_\mathrm{f}^{-3}\delta^D\left(\mathbf{q}_{1234} \right)
T(\mathbf{q}_1,\mathbf{q}_2,\mathbf{q}_3,\mathbf{q}_4)
\notag \\
&=
\dfrac{1}{V_\mathrm{q}\left(k_1,k_2,k_3,k_4\right)}
\int_{k_1}d\mathbf{q}_1^3
\int_{k_2}d\mathbf{q}_2^3
\int_{k_3}d\mathbf{q}_3^3
\int_{k_4}d\mathbf{q}_4^3\,
\delta^D\left(\mathbf{q}_{1234}\right)
T(\mathbf{q}_1,\mathbf{q}_2,\mathbf{q}_3,\mathbf{q}_4)
\notag \\
&\approx
\mathcal{T}_{\rm c}\left(k_1,k_2,k_3,k_4\right)
\dfrac{1}{V_\mathrm{q}\left(k_1,k_2,k_3,k_4\right)}
\int_{k_1}d\mathbf{q}_1^3
\int_{k_2}d\mathbf{q}_2^3
\int_{k_3}d\mathbf{q}_3^3
\int_{k_4}d\mathbf{q}_4^3\,
\delta^D\left(\mathbf{q}_{1234} \right)
\notag \\
&=\mathcal{T}\left(k_1,k_2,k_3,k_4\right)
\,.
\end{align}
\noindent Here, as in \cite{Joachimi:2009zj}, we exploited the fact that  ${\delta^D}^2\approx\delta^D\,k_\mathrm{f}^{-3}$. In the last passage we assumed that if the $k$-shell's thickness $\Delta k$ is small enough, then after the implicit average over the possible diagonals values, $\mathcal{T}\left(k_1,k_2,k_3,k_4\right)$ can be brought outside of the integrals. Under these approximations, this connected part of the estimator is unbiased and equivalent to measuring the i-trispectrum.

\noindent Then the full four-point correlation  estimator can be pixelised (again  using equation \ref{eq:IJ_discrete} for the total estimator trispectrum + unconnected part) as 
\begin{align}
\label{eq:tk_expansion0}
&\hat{\mathcal{T}}_{\rm c+u}\left(k_1,k_2,k_3,k_4\right)=
V^2(2\pi)^3\,k_\mathrm{f}^9\int d\mathbf{x}^3
\, I_{k_1}(\mathbf{x})\, I_{k_2}(\mathbf{x}) 
\,I_{k_3}(\mathbf{x})\,I_{k_4}(\mathbf{x})
\notag \\
&\times
\left[(2\pi)^{12}\int d\mathbf{y}^3
J_{k_1}(\mathbf{y})\, J_{k_2}(\mathbf{y})\, 
J_{k_3}(\mathbf{y})\,J_{k_4}(\mathbf{y})\right]^{-1}
\notag \\
&\rightarrow
 \dfrac{L^9}{N^{12}}
 \sum_{\imath=1}^{N^3}
 \, I^D_{k_1}(x_\imath)\, I^D_{k_2}(x_\imath) \, I^D_{k_3}(x_\imath) \, I^D_{k_4}(x_\imath) 
\times
\left[\sum_{\jmath=1}^{N^3}
 J^D_{k_1}(y_\jmath)\, J^D_{k_2}(y_\jmath)\, J^D_{k_3}(y_\jmath)\, J^D_{k_4}(y_\jmath) \right]^{-1}
\,.
\notag \\
\end{align}

\noindent From the last line of equation \ref{eq:tk_expansion0}  we see that also the pixelised version of the estimator  $\hat{\cal T}_{\rm c+u}$  has by construction dimensions of length to the power of 9.

%%%%%%%%%%%%%%%%%%%%%%%%%%%%%%%%%%%%%%%%%%%%%%%%%%%%%%%%%%%%%%%%%%%55
\subsubsection{Unconnected part}
\label{sec:tk_unconnected}
We now expand the unconnected part of the estimator for one of the three possible permutations: 

\begin{align}
\label{eq:tk_unbiased_unconnected}
&\langle\hat{\mathcal{T}}_{\rm u}\left(k_1,k_2,k_3,k_4\right)\rangle
=
\dfrac{V^2(2\pi)^{-9}k_\mathrm{f}^{9}}{V_\mathrm{q}\left(k_1,k_2,k_3,k_4\right)}
\int_{k_1}d\mathbf{q}_1^3
\int_{k_2}d\mathbf{q}_2^3
\int_{k_3}d\mathbf{q}_3^3
\int_{k_4}d\mathbf{q}_4^3\,
\notag \\
&\times
(2\pi)^3\delta^D\left(\mathbf{q}_{12} \right) P(\mathbf{q}_1)
(2\pi)^3\delta^D\left(\mathbf{q}_{34}\right) P(\mathbf{q}_3)
\delta^D\left(\mathbf{q}_{1234}\right) \,\,(+\,2\,{\rm p.})
\notag \\
&=
\dfrac{(2\pi)^{3}k_\mathrm{f}^{3}}{V_\mathrm{q}\left(k_1,k_2,k_3,k_4\right)}
\int_{k_1}d\mathbf{q}_1^3
\int_{k_2}d\mathbf{q}_2^3
\int_{k_3}d\mathbf{q}_3^3
\int_{k_4}d\mathbf{q}_4^3\,
\notag \\
&\times
\delta^D\left(\mathbf{q}_{12} \right)
\delta^D\left(\mathbf{q}_{34} \right) 
\delta^D\left(\mathbf{q}_{1234}\right)
 P(\mathbf{q}_1)P(\mathbf{q}_3) \,\,({\rm +2 p.})
\notag \\
&=
\dfrac{(2\pi)^{3}k_\mathrm{f}^{3}}{V_\mathrm{q}\left(k_1,k_2,k_3,k_4\right)}
\int_{k_1}d\mathbf{q}_1^3
\int_{k_2}d\mathbf{q}_2^3\,
\delta^D\left(\mathbf{q}_{12}  \right) P(\mathbf{q}_1)
\int_{k_3}d\mathbf{q}_3^3
\int_{k_4}d\mathbf{q}_4^3\,
\delta^D\left(\mathbf{q}_{34} \right)^2 P(\mathbf{q}_3)  \,\,({\rm +\,2 p.})
\notag \\
&=
\dfrac{(2\pi)^{3}}{V_\mathrm{q}\left(k_1,k_2,k_3,k_4\right)}
\int_{k_1}d\mathbf{q}_1^3
\int_{k_2}d\mathbf{q}_2^3\,
\delta^D\left(\mathbf{q}_{12} \right) P(\mathbf{q}_1)
\int_{k_3}d\mathbf{q}_3^3
\int_{k_4}d\mathbf{q}_4^3\,
\delta^D\left(\mathbf{q}_{34} \right)P(\mathbf{q}_3)  \,\,({\rm +\,2 p.})
\,.
\end{align}

\noindent To evaluate \ref{eq:tk_unbiased_unconnected} two alternatives are possible: the analytical approach or the numerical one. In the analytical approach we use the theoretical expressions for the integration volume in Fourier space for both power spectrum and i-trispectrum \cite{Gualdi:2020ymf}. 
The analytical approach derives the theoretical result for the model of the integrated trispectrum unconnected part, since a-priori its precise form is unknown (it is expected of course to be proportional to $\propto P(k_i)P(k_j)$).

The numerical approach on the other end, proves that the estimator defined for the unconnected part of the integrated trispectrum is unbiased.
In this case one can pixelise this unconnected part of the estimator as done for the connected part.
In both cases the unconnected part needs to be subtracted from the measured total signal in order to obtain the i-trispectrum.

%%%%%%%%%%%%%%%%%%%%%%%%%%%%%%%%%%%%%%%%%%%%
\paragraph{Analytical approach}
We start by recalling the two expressions for the integration volumes \cite{Gualdi:2020ymf}:

\begin{eqnarray}
\label{eq:pk_tk_int_volumes_ana}
V_{\mathrm{p}}(k_1) = 4\pi k_1^2 \Delta k_1 \quad \mathrm{and}\quad
V_{\mathrm{q}} (k_1,k_2,k_3,k_4) = 16\pi^3 k_1k_2k_3k_4\Delta k_1\Delta k_2\Delta k_3\Delta k_4 \Delta D \,,
\notag\\
\end{eqnarray}

\noindent where $\Delta k_i$ is the bin-size for the $i$-th $k$-mode. Resuming from the last line of equation \ref{eq:tk_unbiased_unconnected} and assuming that, if the $k$-shells are thin enough, the averaged value of the power spectrum does not differ significantly from the value computed at the centre of the bin, we have that

\begin{align}
\label{eq:tk_unbiased_unconnected_ana}
&\langle\hat{\mathcal{T}}\left(k_1,k_2,k_3,k_4\right)\rangle_\mathrm{u}
\approx
\dfrac{(2\pi)^{3}}{V_\mathrm{q}\left(k_1,k_2,k_3,k_4\right)}
 P(k_1) P(k_3)
\int_{k_1}d\mathbf{q}_1^3
\int_{k_2}d\mathbf{q}_2^3\,
\delta^D\left(\mathbf{q}_{12}  \right)
\notag \\
& \times
\int_{k_3}d\mathbf{q}_3^3
\int_{k_4}d\mathbf{q}_4^3\,
\delta^D\left(\mathbf{q}_{34} \right) (+x\,{\rm p}.)
\notag \\
& =
(2\pi)^{3}
P(k_1) P(k_3)
\delta^K_{12}\delta^K_{34}
\dfrac{4\pi k_1^2 \Delta k_1 \times 4\pi k_3^2 \Delta k_3 }{ 16\pi^3 k_1k_2k_3k_4\Delta k_1\Delta k_2\Delta k_3\Delta k_4 \Delta D} \,+\,(x\,\mathrm{p.})
\notag \\
&=
\dfrac{(2\pi)^{3}\delta^K_{12}\delta^K_{34}P(k_1) P(k_3)}
{ \pi\, \Delta k_2\Delta k_4 \Delta D} \,+\,(x\,\mathrm{p.})
\,.
\end{align}

\noindent where $x=2$ for  the case of a quadrilateral with four equal sides, $k_1=k_2=k_3=k_4$. In the case of quadrilaterals sets with pairs of equal sides there is only one non-null permutation ($x=0$). Given the algorithm we use to generate quadrilaterals sets (see figure \ref{fig:bk_tk_configurations}), in our work the above only happens for example $k_1=k_2$ and $k_3=k_4$.

A quick dimensional analysis shows that this result has the expected dimensions of length to the power of 9, the same as the connected term (trispectrum).

%%%%%%%%%%%%%%%%%%%%%%%%%%%%%%%%%%%%%%%%%%%%
\paragraph{Numerical approach}
By taking the ensemble average it is possible to check that the following estimator is unbiased with respect to the final result of the analytical approach in equation \ref{eq:tk_unbiased_unconnected_ana}
\begin{align}
\label{eq:tku_est}
&\hat{\mathcal{T}}_\mathrm{u}\left(k_1,k_2,k_3,k_4\right)=\dfrac{V(2\pi)^{-9}}{N_\mathrm{q}\left(k_1,k_2,k_3,k_4\right)}
\int_{k_1}d\mathbf{q}_1^3
\int_{k_2}d\mathbf{q}_2^3
\,\delta_{\mathbf{q}_1}\,\delta_{\mathbf{q}_2}\,
\delta^D\left(\mathbf{q}_{12} \right)\,
\notag \\
&\times
\int_{k_3}d\mathbf{q}_3^3
\int_{k_4}d\mathbf{q}_4^3\,
\delta_{\mathbf{q}_3}\,\delta_{\mathbf{q}_4}\,
\delta^D\left(\mathbf{q}_{34} \right)
\quad+ \quad 2\quad\mathrm{p.}
\,,
\end{align}
\noindent which we can pixelised analogously to what done for the connected part for each of the permutations:
\begin{align}
\label{eq:tku_expansion0}
&\hat{\mathcal{T}}_\mathrm{u}\left(k_1,k_2,k_3,k_4\right)=
V(2\pi)^3\,k_\mathrm{f}^9\int d\mathbf{x}^3
\, I_{k_1}(\mathbf{x})\, I_{k_2}(\mathbf{x}) 
\int d\mathbf{z}^3
\,
\,I_{k_3}(\mathbf{z})\,I_{k_4}(\mathbf{z})
\notag \\
&\times
\left[(2\pi)^{12}\int d\mathbf{y}^3
J_{k_1}(\mathbf{y})\, J_{k_2}(\mathbf{y})\, 
J_{k_3}(\mathbf{y})\,J_{k_4}(\mathbf{y})\right]^{-1}
\notag \\
&=
 V\,(2\pi)^{-9}\,k_\mathrm{f}^9\,\Delta V^2\,
 \sum_{\imath=1}^{N^3}
 \, I_{k_1}(x_\imath)\, I_{k_2}(x_\imath)\,
  \sum_{\ell=1}^{N^3}
 \, I_{k_3}(z_\ell)\, I_{k_4}(z_\ell) 
\notag\\
&\times\Delta V\,
\left[\sum_{\jmath=1}^{N^3}
 J_{k_1}(y_\jmath)\, J_{k_2}(y_\jmath)\, J_{k_3}(y_\jmath)\, J_{k_4}(y_\jmath) \right]^{-1}
 \notag \\
&=
 \dfrac{1}{L^3N^3}\,
 \sum_{\imath=1}^{N^3}
 \, I_{k_1}(x_\imath)\, I_{k_2}(x_\imath)\,
  \sum_{\ell=1}^{N^3}
 \, I_{k_3}(z_\ell)\, I_{k_4}(z_\ell) 
\left[\sum_{\jmath=1}^{N^3}
 J_{k_1}(y_\jmath)\, J_{k_2}(y_\jmath)\, J_{k_3}(y_\jmath)\, J_{k_4}(y_\jmath) \right]^{-1}
\,.
\end{align}
\noindent  Once again, pixelising using equation \ref{eq:IJ_discrete}, we have that the i-trispectrum unconnected part estimator becomes
\begin{align}
\label{eq:tku_expansion1}
&\hat{\mathcal{T}}_\mathrm{u}\left(k_1,k_2,k_3,k_4\right)
=
 \dfrac{1}{L^3N^3}\,
 \sum_{\imath=1}^{N^3}
 \, I_{k_1}(x_\imath)\, I_{k_2}(x_\imath)\,
  \sum_{\ell=1}^{N^3}
 \, I_{k_3}(z_\ell)\, I_{k_4}(z_\ell) 
 \notag \\
 &\times
\left[\sum_{\jmath=1}^{N^3}
 J_{k_1}(y_\jmath)\, J_{k_2}(y_\jmath)\, J_{k_3}(y_\jmath)\, J_{k_4}(y_\jmath) \right]^{-1}
 \notag \\
 &\rightarrow
  \dfrac{L^9}{N^{15}}\,
 \sum_{\imath=1}^{N^3}
 \, I^D_{k_1}(x_\imath)\, I^D_{k_2}(x_\imath)\,
  \sum_{\ell=1}^{N^3}
 \, I^D_{k_3}(z_\ell)\, I^D_{k_4}(z_\ell) 
\left[\sum_{\jmath=1}^{N^3}
 J^D_{k_1}(y_\jmath)\, J^D_{k_2}(y_\jmath)\, J^D_{k_3}(y_\jmath)\, J^D_{k_4}(y_\jmath) \right]^{-1}
\,.
\end{align}

\noindent The above result for the pixelised estimator corresponds to the analytical model derived in equation \ref{eq:tk_unbiased_unconnected_ana}.

%%%%%%%%%%%%%%%%%%%%%%%%%%%%%%%%%%%%%%%%%%%%
%%%%%%%%%%%%%%%%%%%%%%%%%%%%%%%%%%%%%%%%%%%%
%%%%%%%%%%%%%%%%%%%%%%%%%%%%%%%%%%%%%%%%%%%%
\begin{figure}[tbp]
\centering 
\includegraphics[width=1.\textwidth]%,trim=0 380 0 200,clip]
{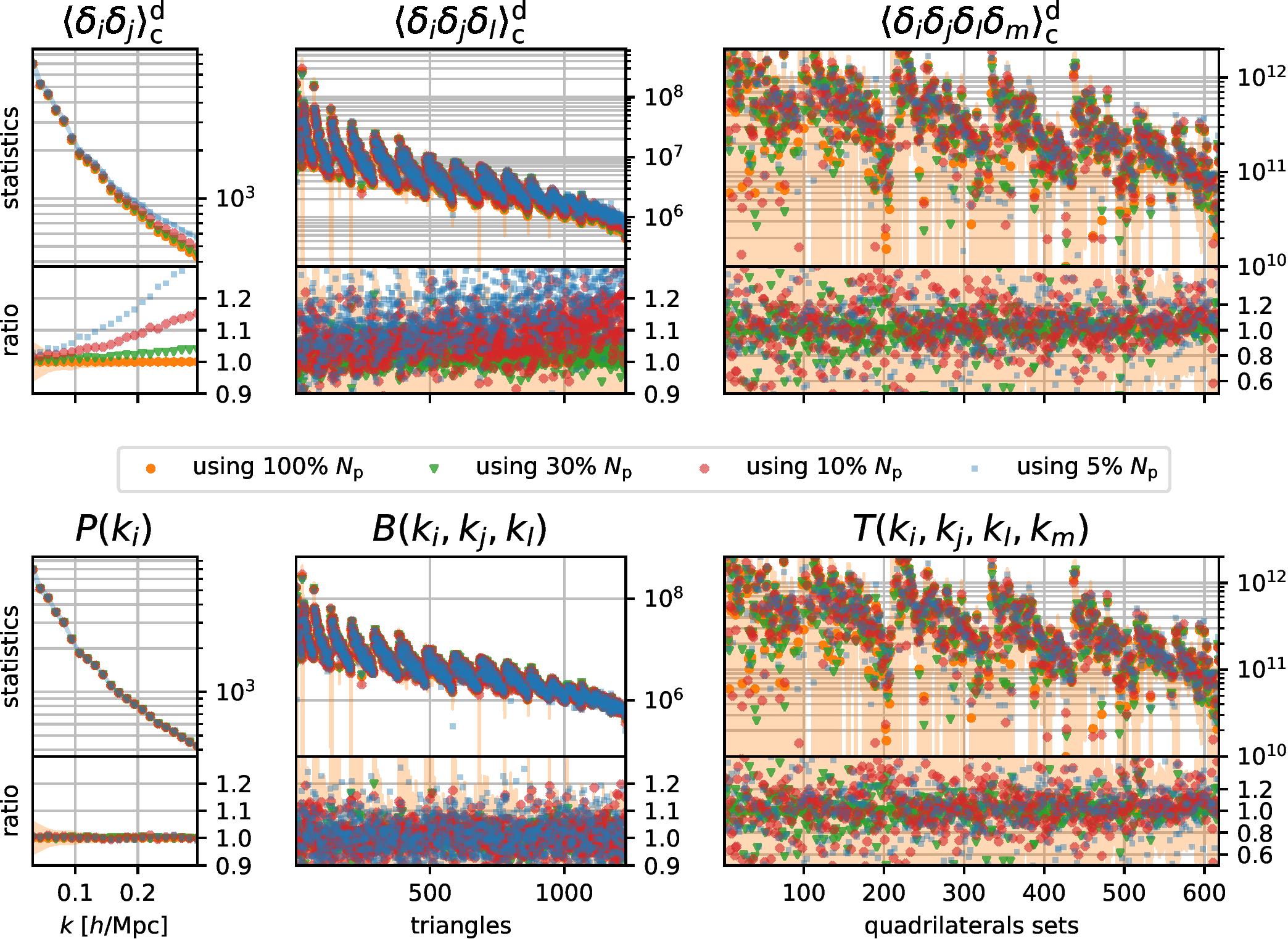}
\caption{\label{fig:shot_noise} 
Comparison between measured connected discrete correlators (upper half, from equations \ref{eq:pk_est_text}, \ref{eq:bk_est_text}, \ref{eq:tk_est_text}) and the derived statistics obtained by subtracting the shot-noise estimated contributions (lower half, following equations \ref{eq:shot_noise_pkbk} and \ref{eq:shot_noise_tk}) for different randomly selected fractions of the total number of dark matter particles present in the simulations.
The results are shown for $z=1$, and the shaded areas are centred around the $100\%\,N_\mathrm{p}$ case, which is used also as denominator in each ratio sub-plot.
The green, red and blue points are respectively the cases considering $30\%$, $10\%$ and $5\%$ of the dark matter particles (randomly selected). While being significant in the cases of power spectrum and bispectrum especially at large $k$-values, for the i-trispectrum the shot-noise impact is not noticeable because of the large sample variance which is always dominant even  when lowering the tracers density (at least up to the $k_\mathrm{max}$ considered in this work).   
}
\end{figure}

%%%%%%%%%%%%%%%%%%%%%%%%%%%%%%%%%%%%%%%%%%%%%
%%%%%%%%%%%%%%%%%%%%%%%%%%%%%%%%%%%%%%%%%%%%%
%%%%%%%%%%%%%%%%%%%%%%%%%%%%%%%%%%%%%%%%%%%%%
\section{Shot-noise correction}
\label{app:shot_noise}
If the observed field is a Poisson sample of a (possibly hypothetical) continuous underlying field, the measured correlations would be affected by a shot-noise contribution which acts as a noise bias.  
In order to subtract the shot-noise contribution from the measured signal, we implement the  suitable estimators (combinations of power spectra, bispectra and average particles density number) necessary to reproduce the terms described in the appendix of \cite{Verde:2001pf} under the assumption of Poissonian shot-noise. The expressions for power spectrum and bispectrum are

\begin{eqnarray}
\label{eq:shot_noise_pkbk}
\langle\delta_i\delta_j\rangle^\mathrm{d}_\mathrm{c}
&=&
(2\pi)^3\delta^{D}\left(\mathbf{k}_{ij}\right)\left[P(k_i)+\dfrac{1}{\bar{n}}\right]
\notag \\
\langle\delta_i\delta_j\delta_l\rangle^\mathrm{d}_\mathrm{c}
&=&
(2\pi)^3\delta^{D}\left(\mathbf{k}_{ijl}\right)
\left[B(k_i,k_j,k_l)+\dfrac{1}{\bar{n}}\left(P(k_i)+P(k_j)+P(k_l)\right)
+\dfrac{1}{\bar{n}^2}
\right]\,,
\notag \\
\end{eqnarray}

\noindent where the index "d" and subscript "c" stand for "discrete" and "connected", respectively and $\bar{n}$ denotes the number density of objects. Notice that in this section we use the word "discrete" to indicate that the considered statistics are measured from a set of objects and not a continuous field. For the i-trispectrum, given the fact that for each set of $(k_i,k_j,k_l,k_m)$ there is an infinite number of possible quadrilateral shapes, the shot-noise expression from \cite{Verde:2001pf} needs to be modified. In particular for a given $\Delta k $ used to bin the $k$-values in the measurement there is a finite number of possible values of the diagonal $D$ ranging between a $D_\mathrm{min}$ and a $D_\mathrm{max}$ as defined in equation \ref{eq:diag_min_max}, $N_D=(D_\mathrm{max}-D_\mathrm{min})/\Delta k$. This of course is true in the case of the shot-noise correction being estimated directly from the data. Therefore, for each of the i-trispectrum configurations defined by a set of $k$-values $(k_i,k_j,k_l,k_m)$, the shot-noise expression becomes,
\begin{eqnarray}
\label{eq:shot_noise_tk}
&&\langle\delta_i\delta_j\delta_l\delta_m\rangle^\mathrm{d}_\mathrm{c}
\notag \\
&=&
(2\pi)^3\delta^{D}\left(\mathbf{k}_{ijlm}\right)
\Bigg\{T(k_i,k_j,k_l,k_m)
\notag \\
&+&
\dfrac{1}{3}\sum_{\substack{k_1,k_2,k_3,k_4 \\ k_1,k_3,k_2,k_4 \\ k_1,k_2,k_4,k_3}}
\Bigg[
\dfrac{1}{\bar{n}}
\left(B(|\mathbf{k}_i+\mathbf{k}_j|,k_l,k_m )\bigg|_{D_\mathrm{av.}}\,+\,5\,\mathrm{perm.}\right)
\notag \\
&+&
\dfrac{1}{{\bar n}^2}\bigg(P(|\mathbf{k}_i+\mathbf{k}_j+\mathbf{k}_l|)\bigg|_{D_\mathrm{av.}} \,+\,3\,\mathrm{perm.} 
\;+\;2P(|\mathbf{k}_i+\mathbf{k}_j|)\bigg|_{D_\mathrm{av.}} \,+\,2\,\mathrm{perm.} \bigg)
\Bigg]+\dfrac{1}{{\bar n}^3}\Bigg\}\,,
\notag \\
\end{eqnarray}

\noindent where we used the diagonal intermediate value of each configuration's possible range $D_\mathrm{av.}=(D_\mathrm{min}+D_\mathrm{max})/2$ to compute the vectors sums appearing the bispectra and power spectra arguments.
The terms proportional to powers of $1/{\bar n}$ constitute the shot-noise "bias" which should be subtracted from the measured signal.
Therefore these terms are first measured from the simulations and directly subtracted afterwards from the total signal measured using the estimators in equations \ref{eq:pk_est_text}, \ref{eq:bk_est_text} and \ref{eq:tk_total_est}. 
Equation \ref{eq:shot_noise_tk} is clearly an approximation, a more rigorous treatment would involve averaging the above terms over all the possible values of the diagonal $D$. We tested this and found no significant change. Most likely this is due to the fact the i-trispectrum measurement is dominated by statistical noise as it can be seen from figure \ref{fig:shot_noise}. Notice that the term $P(|\mathbf{k}_i+\mathbf{k}_j+\mathbf{k}_l|) = P(k_m)$ for the closure condition, and therefore does not need to be averaged.

Figure \ref{fig:shot_noise} shows the increase in relevance of the shot-noise term as the number of tracers $N_\mathrm{p}$ decreases.
For one realisation at $z=1$, the  dark matter particles present in the cubic box have been randomly sampled to include $5\%$, $10\%$, $30\%$ and $100\%$ of the total.
In the upper half of the figure shows the corresponding  discrete  power spectrum, bispectrum and connected part of the i-trispectrum, measured  using  the estimators in equations \ref{eq:pk_est_text}, \ref{eq:bk_est_text}, \ref{eq:tk_est_text}. In the lower half the same cases have been shown for the statistics of interest obtained subtracting the shot-noise terms as described in equations \ref{eq:shot_noise_pkbk} and \ref{eq:shot_noise_tk}.

Without shot-noise correction a significant bias would be induced in both power spectrum and bispectrum measurements, especially at large $k$-values. The relevance of this bias is given by its comparison with the error bars derived from the set of measurements on the simulations. For the i-trispectrum, even when decreasing the number of samples down to $5\%$ of the original set, the induced shift in the statistics is still smaller than the error bars. The shot-noise correction does not  have an impact as significative as in the case of power spectrum and bispectrum because of the dominant sample variance error. In other words, the large statistical error associated with measuring the i-trispectrum from a less dense catalogue of tracers dominates over the shot-noise effect, at least up to the $k_\mathrm{max}$ considered in this work.

%%%%%%%%%%%%%%%%%%%%%%%%%%%%%%%%%%%%%%%%%%%
%%%%%%%%%%%%%%%%%%%%%%%%%%%%%%%%%%%%%%%%%%%
%%%%%%%%%%%%%%%%%%%%%%%%%%%%%%%%%%%%%%%%%%%
\section{I-trispectrum covariance: Gaussian term in redshift space}
\label{app:tk_cov_rsd}
 Using the estimator defined in equation \ref{eq:tk_est_text}, the covariance matrix for the tracer field (halo or galaxy) i-trispectrum in redshift space is given by: 

\begin{eqnarray}
\label{eq:t0_gg_cov}
&&\mathrm{C}^{\mathcal{T}^{(0)}\mathcal{T}^{(0)}}_{\mathrm{G}}\left(k_1,k_2,k_3,k_4;k_a,k_b,k_c,k_d\right) =
\notag\\
&=&
\dfrac{V^4}{(2\pi)^{18}N^{\mathrm{q}}_{1234}N^{\mathrm{q}}_{abcd}}\prod^{4,d}_{i=1,a}\int_{V_{\mathbf{q}_i}}
\delta^D\left(\mathbf{q}_{1234}\right)
\delta^D\left(\mathbf{q}_{abcd}\right)
\notag \\
&\times&
\langle
\delta_{h}^{\mathrm{s}}\left(\mathbf{q}_1\right)\delta_{h}^{\mathrm{s}}\left(\mathbf{q}_2\right)\delta_{h}^{\mathrm{s}}\left(\mathbf{q}_3\right)\delta_{h}^{\mathrm{s}}\left(\mathbf{q}_4\right)\delta_{h}^{\mathrm{s}}\left(\mathbf{q}_a\right)\delta_{h}^{\mathrm{s}}\left(\mathbf{q}_b\right)\delta_{h}^{\mathrm{s}}\left(\mathbf{q}_c\right)\delta_{h}^{\mathrm{s}}\left(\mathbf{h}_d\right)
\rangle
\notag \\
&=&
\dfrac{k_\mathrm{f}^{-12}}{(2\pi)^{6}k_\mathrm{f}^{-18}V^{\mathrm{q}}_{1234}V^{\mathrm{q}}_{abcd}}\prod^{4,d}_{i=1,a}\int_{V_{\mathbf{q}_i}}
\delta^D\left(\mathbf{q}_{1234}\right)
\delta^D\left(\mathbf{q}_{abcd}\right)
\notag \\
&\times&
(2\pi)^{12}
\delta^D\left(\mathbf{q}_{1a}\right)
\delta^D\left(\mathbf{q}_{2b}\right)
\delta^D\left(\mathbf{q}_{c3}\right)
\delta^D\left(\mathbf{q}_{4d}\right)
\mathrm{P}^{s}_{\mathrm{h}}\left(\mathbf{q}_1\right)
\mathrm{P}^{s}_{\mathrm{h}}\left(\mathbf{q}_2\right)
\mathrm{P}^{s}_{\mathrm{h}}\left(\mathbf{q}_3\right)
\mathrm{P}^{s}_{\mathrm{h}}\left(\mathbf{q}_4\right)
\; + \,23\; \mathrm{perm.}
\notag \\
&=&
\dfrac{k_\mathrm{f}^{3}(2\pi)^{6}}{V^{\mathrm{q},2}_{1234}}
\delta^{\mathrm{K}}_{1a}\delta^{\mathrm{K}}_{2b}\delta^{\mathrm{K}}_{3c}\delta^{\mathrm{K}}_{4d}
\prod^{4}_{i=1}\int_{V_{\mathbf{q}_i}}
\delta^D\left(\mathbf{q}_{1234}\right)
\mathrm{P}^{s}_{\mathrm{h}}\left(\mathbf{q}_1\right)
\mathrm{P}^{s}_{\mathrm{h}}\left(\mathbf{q}_2\right)
\mathrm{P}^{s}_{\mathrm{h}}\left(\mathbf{q}_3\right)
\mathrm{P}^{s}_{\mathrm{h}}\left(\mathbf{q}_4\right)
+\,23\; \mathrm{perm.}
\notag \\
&\approx&
\dfrac{(2\pi)^{9}}{V_\mathrm{s}V^{\mathrm{q}}_{1234}}
\dfrac{\mathrm{R_{1234}}}{3}
\sum_{\substack{k_1,k_2,k_3,k_4 \\ k_1,k_3,k_2,k_4 \\ k_1,k_2,k_4,k_3}}
\dfrac{1}{8\pi^2\Delta D}
\int^{D_\mathrm{max}}_{D_\mathrm{min}}dD\int^{+1}_{-1}\,d\mu_D\int^{2\pi}_0\,d\phi_{12}\,\int^{2\pi}_0d\psi
\notag\\
&\times&
\mathrm{P}^{s}_{\mathrm{h}}\left(k_1,\mu_1\right)
\mathrm{P}^{s}_{\mathrm{h}}\left(k_2,\mu_2\right)
\mathrm{P}^{s}_{\mathrm{h}}\left(k_3,\mu_3\right)
\mathrm{P}^{s}_{\mathrm{h}}\left(k_4,\mu_4\right),
\end{eqnarray}{}

\noindent where we used the approximation made in \cite{Joachimi:2009zj} that ${\delta^{{D}}}^2\approx\dfrac{V_{\mathrm{s}}}{(2\pi)^3}\delta^{{D}} = k_\mathrm{f}^{-3}\delta^{{D}}$; $P^{s}_{\rm {\rm h}}$ denotes the redshift space tracer (halo or galaxy) power spectrum, $V_\mathrm{s}$ is the survey volume and $k_{\mathrm{f}}$ the fundamental frequency; 
$\mathrm{R}_{1234}$ is a symmetry factor that counts the number of possible combinations of equal sides between the two identical quadrilaterals sets. Depending on the number of equal sides, $\mathrm{R}_{1234}$ assumes the values:

\begin{align}
\label{eq:R_sym_fac}
    \mathrm{R}_{1234}=
    \begin{cases}
     1  &\quad \mathrm{for} \quad \left(k_a,k_b,k_c,k_d\right)\\
     2  &\quad \mathrm{for} \quad \left(k_a,k_a,k_c,k_d\right)\\
     4  &\quad \mathrm{for} \quad \left(k_a,k_a,k_c,k_c\right)\\
     6  &\quad \mathrm{for} \quad \left(k_a,k_a,k_a,k_d\right)\\
     24 &\quad \mathrm{for} \quad \left(k_a,k_a,k_a,k_a\right) \,. \\
    \end{cases}
\end{align}

%%%%%%%%%%%%%%%%%%%%%%%%%%%%%%%%%%%%%%%%%%%%%
%%%%%%%%%%%%%%%%%%%%%%%%%%%%%%%%%%%%%%%%%%%%%
%%%%%%%%%%%%%%%%%%%%%%%%%%%%%%%%%%%%%%%%%%%%%
\section{Primordial non-Gaussianity}
\label{sec:app_png}

We recap here the formalism presented in the appendix of Ref. \cite{Gualdi:2019sfc} for both power spectrum and bispectrum and extend it to include the i-trispectrum for the matter field in real space (replace the SPT kernels with the redshift space ones in appendix \ref{app:theo_models_rsd} for the redshift space case).

\noindent For the three statistics (power spectrum, bispectrum and i-trispectrum) we compute  the contribution due to the presence of a primordial non-Gaussian component of the local type  in the potential field. The primordial (subscript p) potential   (which in our work has units of $[c^2]$ in order for equation \ref{eq:dk_to_phi} appearing later to be dimensionless) can be thus  parametrised as:

\begin{eqnarray}
\Phi_{\mathrm{p}}(\mathbf{x})=\phi(\mathbf{x})+\dfrac{f_{\mathrm{nl}}}{c^2}\left[\phi^2(\mathbf{x})-\langle\phi^2(\mathbf{x})\rangle\right]+\dfrac{g_{\mathrm{nl}}}{c^4}\left[\phi^3(\mathbf{x})-3\phi(\mathbf{x})\langle\phi^2(\mathbf{x})\rangle\right] \;+\;.\,.\,. 
\end{eqnarray}{}
\noindent where $\phi$ represents a Gaussian field. In Fourier space it translates to:
\begin{eqnarray}
\label{eq:earlyphi_togauphi}
\Phi_{\mathrm{p}}(\mathbf{k})=\phi_{k}+\dfrac{f_{\mathrm{nl}}}{c^2}\left[ I^k_{ab}\phi_a\phi_b-\delta^D(\mathbf{k})\langle\phi^2\rangle\right]
+\dfrac{g_{\mathrm{nl}}}{c^4}\left[ I^k_{abc}\phi_a\phi_b\phi_c-\dfrac{3}{(2\pi)^3}\phi_k\langle\phi^2\rangle\right]\,,
\end{eqnarray}{}
\noindent where in Fourier space $\langle\phi^2\rangle=\int d\mathbf{q}^3P^{\phi}(\mathbf{q}) = (2\pi)^3\sigma^2_{\phi}$, and $\phi_k$ denotes the Fourier transform of $\phi$ and $P^\phi$ denotes the power spectrum of the $\phi$ field.
In order to write in a compact way we write for different quantities the argument as a subscript (e.g., $P^\phi(k)\rightarrow P^\phi_k$ or $F^{(2)}\left[\mathbf{k}_a,\mathbf{k}_b\right]\rightarrow F^{(2)}_{ab}$). For the same reason to express many integrals appearing in the perturbation expansion we have introduced the short notation:
\begin{eqnarray}
I^{k}_{ab} = \int\dfrac{d\mathbf{q}_a^3d\mathbf{q}_b^3}{(2\pi)^3}\,\delta^D(\mathbf{k}-\mathbf{q}_a-\mathbf{q}_b)\quad \mathrm{and}\quad
I^{k}_{abc} = \int\dfrac{d\mathbf{q}_a^3d\mathbf{q}_b^3d\mathbf{q}_c^3}{(2\pi)^6}\,\delta^D(\mathbf{k}-\mathbf{q}_a-\mathbf{q}_b-\mathbf{q}_c)\,.
\notag \\
\end{eqnarray}{}

\noindent Late-time  and primordial potentials are related by:

\begin{eqnarray}
\label{eq:lin_ev}
\Phi_{\mathrm{l.t.}}(k,a) ={\cal O}\left(\Phi_{\rm p}(k))\right)\simeq \dfrac{9}{10}\dfrac{D_+}{a}\mathbb{T}(k)\Phi_{\mathrm{p}}\,, 
\end{eqnarray}{}
where the operator ${\cal O}$ describing the full non-linear evolution of the primordial field has been linearised in the last step, $D_+(a)$  is the linear growth factor  as a function of the scale factor $a$ and  $\mathbb{T}(k)$  is the transfer function. 
At late-times the potential field is related to the density perturbation variable by the Poisson equation:
\begin{eqnarray}
\nabla^2\Phi_{\mathrm{l.t.}}(\mathbf{x},a)=\dfrac{3}{2}\dfrac{\Omega_\mathrm{m}H_0^2}{a}\delta(\mathbf{x},a)\,. 
\end{eqnarray}
\noindent This allows us to link the primordial potential with the late-time matter density perturbation assuming, as normally done in the literature, that  only the linearly evolved component of the primordial (potentially including non-Gaussianities) field $\Phi_\mathrm{p}$  sources  the late time gravitational potential $\Phi_\mathrm{l.t.}$.:

\begin{eqnarray}
\label{eq:dk_to_phi}
\delta_k = \dfrac{3}{5}\dfrac{D_+}{\Omega_\mathrm{m}H_0^2
}k^2\mathbb{T}_k\Phi_{\mathrm{p}}=
w\mathcal{M}_k\Phi_{\mathrm{p}}
\quad \mathrm{where}\quad \mathcal{M}_k=\dfrac{3}{5}\dfrac{D_+}{\Omega_\mathrm{m}H_0^2
}\,k^2\mathbb{T}_k
\,.
\end{eqnarray}{}

In the rest of this appendix, to use a more compact notation, the  wave-vector or wave-number in the argument is reported as a subscript index. 

\subsection{Power spectrum}

As described in \cite{Gualdi:2019sfc} up to order $\phi^4$ for the matter power spectrum we have the terms:

\begin{eqnarray}
&&\langle\delta\delta\rangle \,=\,
\langle\delta^{(1)}\delta^{(1)}\rangle + 
2\langle\delta^{(1)}\delta^{(2)}\rangle + 
\langle\delta^{(2)}\delta^{(2)}\rangle + 
2\langle\delta^{(1)}\delta^{(3)}\rangle \notag \\
&&\equiv\,P_{(11)}+P_{(12)}+P_{(22)}+P_{(13)}\,.
\end{eqnarray}{}

\noindent The first term $P_{(11)}$ is given by

\begin{eqnarray}
\label{eq:png_p11}
&&\langle\delta^{(1)}(\mathbf{k})\delta^{(1)}(\mathbf{q})\rangle
=
F^{(1)}_kF^{(1)}_q
\mathcal{M}_k\mathcal{M}_q
(2\pi)^3\delta^D({\bf k}+{\bf q}) 
\Bigg\{P^{\phi}_k
+\dfrac{2f_{\mathrm{nl}}^2}{c^4}\int \dfrac{d\mathbf{p}_a^3}{(2\pi)^3}P^{\phi}_aP^{\phi}_{|\mathbf{k}-\mathbf{p}_a|}
\Bigg\}\,.
\notag \\
\end{eqnarray}{}

\noindent where $F^{(1)}=1$ and we include it to simplify the transformation to the matter redshift space expressions, where $F^{(1)}\rightarrow Z^{(1)}$. Notice that the integral in the second term proportional to $f_\mathrm{nl}^2$ is divergent for $k\rightarrow0$. However these integrals should be performed for a $k_\mathrm{min}$ corresponding to the size of the causally connected patch of the Universe and therefore it is effectively a finite quantity.

The second term, $P_{(12)}$, reads

\begin{eqnarray}
2\langle\delta^{(1)}(\mathbf{k})\delta^{(2)}(\mathbf{q})\rangle
&&=
(2\pi)^3\delta^D({\bf k}+ {\bf q})\dfrac{4f_{\mathrm{nl}}}{c^2}F^{(1)}_k 
\mathcal{M}_k
\notag \\
&&\times\,
\int\dfrac{d\mathbf{p}_a^3}{(2\pi)^3}
\mathcal{M}_{p_a}\mathcal{M}_{|-\mathbf{k}-\mathbf{p}_a|}
F^{(2)}_{a,-\mathbf{k}-\mathbf{p}_a}P^{\phi}_{|-\mathbf{k}-\mathbf{p}_a|}\left[P^{\phi}_a+2P^{\phi}_k\right]\,.
\notag \\
\end{eqnarray}{}

\noindent The other terms in the power spectrum expansion, $P_{(22)}$ and $P_{(13)}$, are at first order already proportional to $\phi^4$ and therefore in our case they just return the standard terms for the Gaussian initial conditions.

\subsection{Bispectrum}
Limiting the expansion in terms of the Gaussian primordial potential up to order $\phi^4$, as described in \cite{Gualdi:2019sfc}'s appendix, for the bispectrum there is only a primordial term $B_{(111)}$ together with the standard one $B_{(112)}$ due to gravitational collapse. The expression for $B_{(111)}$ is given by

\begin{eqnarray}
\langle\delta^{(1)}\delta^{(1)}\delta^{(1)}\rangle
&&=\,
(2\pi)^3F^{(1)}_{k_1}F^{(1)}_{k_2}F^{(1)}_{k_3}
\mathcal{M}_{k_1}\mathcal{M}_{k_2}\mathcal{M}_{k_3}
\dfrac{2f_{\mathrm{nl}}}{c^2}
\delta^D(\mathbf{k}_1+\mathbf{k}_2+\mathbf{k}_3)P^{\phi}_{k_2}P^{\phi}_{k_3} \;+\; \mathrm{cyc.}
\notag\\
&&=\,
(2\pi)^3\delta^D(\mathbf{k}_1+\mathbf{k}_2+\mathbf{k}_3)
F^{(1)}_{k_1}F^{(1)}_{k_2}F^{(1)}_{k_3}
\dfrac{\mathcal{M}_{k_1}}{\mathcal{M}_{k_2}\mathcal{M}_{k_3}}
\dfrac{2f_{\mathrm{nl}}}{c^2}
P^{\mathrm{m}}_{k_2}P^{\mathrm{m}}_{k_3} \;+\; \mathrm{cyc.}\,,
\notag \\
\end{eqnarray}
\noindent where in the last line the primordial power spectrum was converted into the late-time matter power spectrum.

\subsection{Trispectrum}
\label{sec:subsec_tkpng}

The terms we report here were presented (in a less padagogical way) in  \cite{Jeong:2009vd} and \cite{Sefusatti:2009qh}. In this subsection, for clarity and  self-consistency, we also report them using a notation consistent with the rest of the paper.
In the case of i-trispectrum, primordial corrections appear at order $\phi^6$ at the lowest order in $\phi$. By converting $\phi$ into the late time matter perturbation density variable $\delta$, it can be noticed that this is the same order of the standard gravitational term for Gaussian initial conditions. It is indeed the lowest order at which the connected part of the four-point correlation function in Fourier Space appears. Therefore we will consider only terms of order $\phi^6$. Proceeding as  before:

\begin{eqnarray}
&&\langle\delta\delta\delta\delta\rangle = \langle(\delta^{(1)}+\delta^{(2)}+\delta^{(3)}+O(\delta^{(4)}))(\delta^{(1)}+\delta^{(2)}+\delta^{(3)}+O(\delta^{(4)}))
\notag \\
&&\times\, 
(\delta^{(1)}+\delta^{(2)}+\delta^{(3)}+O(\delta^{(4)}))(\delta^{(1)}+\delta^{(2)}+\delta^{(3)}+O(\delta^{(4)}))\rangle 
\notag \\
&&=\, 
\langle\delta^{(1)}\delta^{(1)}\delta^{(1)}\delta^{(1)}\rangle
+ 
4\langle\delta^{(1)}\delta^{(1)}\delta^{(1)}\delta^{(2)}\rangle 
+6\langle\delta^{(1)}\delta^{(1)}\delta^{(2)}\delta^{(2)}\rangle 
+ 
4\langle\delta^{(1)}\delta^{(1)}\delta^{(1)}\delta^{(3)}\rangle
\notag \\
&&\equiv \,
T_{(1111)}+T_{(1112)}+T_{(1122)}+T_{(1113)}\,. 
\end{eqnarray}{}

\noindent 
The i-trispectrum corresponding to Gaussian initial conditions is given by terms $T_{(1122)}$ and $T_{(1113)}$.
Without PNG, $T_{(1111)}$ would represent the unconnected part of the four-point correlator in Fourier space. However expanding in terms of PNG results in non-trivial terms proportional to both $f_{\mathrm{nl}}^2$ and $g_{\mathrm{nl}}$.
Finally $T_{(1112)}$ would be zero for Gaussian initial conditions since it is an odd moment. However with PNG, up to order $\phi^6$ it will produce at least one term proportional to $f_{\mathrm{nl}}$. 

We will follow the approach of first expanding each expression in terms of late-time density perturbation variables $\delta_k$ and then express each of these in terms of early time potential field $\Phi_\mathrm{p}$ using equation \ref{eq:dk_to_phi}. Finally we will convert $\Phi_\mathrm{p}$ to its expansion in terms of the early-time Gaussian potential $\phi$ using equation \ref{eq:earlyphi_togauphi}.

\subsubsection{\texorpdfstring{$T_{(1111)}$}{}}

\begin{eqnarray}
&&\langle\delta^{(1)}\delta^{(1)}\delta^{(1)}\delta^{(1)}\rangle 
\notag \\
&&\propto 
\langle
\left\{ \phi_{k_1}+\dfrac{f_{\mathrm{nl}}}{c^2}\left[ I^{k_1}_{ab}\phi_a\phi_b-\delta^D(\mathbf{k}_1)\langle\phi^2\rangle\right]
+\dfrac{g_{\mathrm{nl}}}{c^4}\left[ I^{k_1}_{cde}\phi_c\phi_d\phi_e-\dfrac{3}{(2\pi)^3}\phi_{k_1}\langle\phi^2\rangle\right] \right\}
\notag \\[2mm]
&&\times\,\left\{ \phi_{k_2}+\dfrac{f_{\mathrm{nl}}}{c^2}\left[ I^{k_2}_{fg}\phi_f\phi_g-\delta^D(\mathbf{k}_2)\langle\phi^2\rangle\right]
+\dfrac{g_{\mathrm{nl}}}{c^4}\left[ I^{k_2}_{hil}\phi_h\phi_i\phi_l-\dfrac{3}{(2\pi)^3}\phi_{k_2}\langle\phi^2\rangle\right] \right\}
\notag \\
&&\times\,\left\{\phi_{k_3}+\dfrac{f_{\mathrm{nl}}}{c^2}\left[ I^{k_3}_{mn}\phi_m\phi_n-\delta^D(\mathbf{k}_3)\langle\phi^2\rangle\right]
+\dfrac{g_{\mathrm{nl}}}{c^4}\left[ I^{k_3}_{oqr}\phi_o\phi_q\phi_r-\dfrac{3}{(2\pi)^3}\phi_{k_3}\langle\phi^2\rangle\right]\right\}
\notag \\
&&\times\,\left\{\phi_{k_4}+\dfrac{f_{\mathrm{nl}}}{c^2}\left[ I^{k_4}_{st}\phi_s\phi_t-\delta^D(\mathbf{k}_4)\langle\phi^2\rangle\right]
+\dfrac{g_{\mathrm{nl}}}{c^4}\left[ I^{k_4}_{uvz}\phi_u\phi_v\phi_z-\dfrac{3}{(2\pi)^3}\phi_{k_4}\langle\phi^2\rangle\right]\right\}
\rangle
\notag \\
\,.
\end{eqnarray}{}

\noindent We then have two kind of PNG-dependent terms,  one proportional to $f_{\mathrm{nl}}^2$ and one to $g_{\mathrm{nl}}$:

\begin{eqnarray}
\label{eq:tri_png_1111}
&&\langle\delta^{(1)}\delta^{(1)}\delta^{(1)}\delta^{(1)}\rangle =  F^{(1)}_{k_1}F^{(1)}_{k_2}F^{(1)}_{k_3}F^{(1)}_{k_4}
\mathcal{M}_{k_1}\mathcal{M}_{k_2}\mathcal{M}_{k_3}\mathcal{M}_{k_4}
\notag \\[2mm]
&&\times\,
\Bigg\{
\dfrac{f_{\mathrm{nl}}^2}{c^4}
\Bigg[
I^{k_3}_{mn}I^{k_4}_{st}\langle\phi_{k_1}\phi_{k_2}\phi_{m}\phi_{n}\phi_{s}\phi_{t}\rangle 
-I^{k_3}_{mn}\langle\phi_{k_1}\phi_{k_2}\phi_{m}\phi_{n}\rangle\delta^D(\mathbf{k}_4)\langle\phi^2\rangle
\notag \\
&&-\,I^{k_4}_ {st}\langle\phi_{k_1}\phi_{k_2}\phi_{s}\phi_{t}\rangle\delta^D(\mathbf{k}_3)\langle\phi^2\rangle
+\langle\phi_{k_1}\phi_{k_2}\rangle\delta^D(\mathbf{k}_3)\delta^D(\mathbf{k}_4)\langle\phi^2\rangle^2
\Bigg]\;+\;5\,\mathrm{p.}
\notag \\
&&+\,\dfrac{g_{\mathrm{nl}}}{c^4}
\Bigg[
I^{k_4}_{uvz}\langle\phi_{k_1}\phi_{k_2}\phi_{k_3}\phi_u\phi_v\phi_z\rangle - 3\langle\phi_{k_1}\phi_{k_2}\phi_{k_3}\phi_{k_4}\rangle\langle\phi^2\rangle
\Bigg]\;+\;3\,\mathrm{p.}\Bigg\}\,.
\notag \\
\end{eqnarray}{}

\noindent Given the several permutations let us split this computation in two terms proportional to the two PNG parameters. Proceeding with the first, $T_{(1111)}^{f_{\mathrm{nl}}^2}$, we can see that the six-point correlator can be decomposed in the following terms:

\begin{eqnarray}
T_{(1111)}^{f_{\mathrm{nl}}^2}\propto
\begin{cases}
\langle\phi_{k_1}\phi_{k_2}\rangle\langle\phi_m\phi_n\rangle\langle\phi_s\phi_t\rangle \\
\langle\phi_{k_1}\phi_{k_2}\rangle\langle\phi_m\phi_s\rangle\langle\phi_n\phi_t\rangle \times 2\\
\left(\langle\phi_{k_1}\phi_{k_m}\rangle\langle\phi_2\phi_n\rangle\langle\phi_s\phi_t\rangle+\langle\phi_{k_1}\phi_{k_s}\rangle\langle\phi_2\phi_t\rangle\langle\phi_m\phi_n\rangle\right) \times 2\\
\langle\phi_{k_1}\phi_{k_m}\rangle\langle\phi_2\phi_s\rangle\langle\phi_n\phi_t\rangle \times 4\; +\;1\,\mathrm{p.} \;(k_1\longleftrightarrow k_2)\\
\end{cases}
\end{eqnarray}{}

\noindent The symmetry factors are due to permutations over the auxiliary variables. Only the last term is fully connected, therefore we expect the others to be cancelled out by the other three terms appearing in the expression for $T_{(1111)}^{f_{\mathrm{nl}}^2}$. Let us find out:

\begin{eqnarray}
\label{eq:tri_png_111_fnl}
&&T_{(1111)}^{f_{\mathrm{nl}}^2}\propto
\Bigg\{
I^{k_3}_{mn}I^{k_4}_{st}\langle\phi_{k_1}\phi_{k_2}\rangle\langle\phi_m\phi_n\rangle\langle\phi_s\phi_t\rangle
+2 I^{k_3}_{mn}I^{k_4}_{st}\langle\phi_{k_1}\phi_{k_2}\rangle\langle\phi_m\phi_s\rangle\langle\phi_n\phi_t\rangle
\notag \\[2mm]
&&+\,2I^{k_3}_{mn}I^{k_4}_{st}\left[\langle\phi_{k_1}\phi_{k_m}\rangle\langle\phi_{k_2}\phi_n\rangle\langle\phi_s\phi_t\rangle
+
\langle\phi_{k_1}\phi_{k_s}\rangle\langle\phi_{k_2}\phi_t\rangle\langle\phi_m\phi_n\rangle\right]
\notag \\[2mm]
&&+\,
4I^{k_3}_{mn}I^{k_4}_{st}
\left[
\langle\phi_{k_1}\phi_{k_m}\rangle\langle\phi_{k_2}\phi_s\rangle\langle\phi_n\phi_t\rangle
+
\langle\phi_{k_1}\phi_{k_s}\rangle\langle\phi_{k_2}\phi_m\rangle\langle\phi_n\phi_t\rangle
\right]
\notag \\[2mm]
&&-\,I^{k_3}_{mn}\delta^D(\mathbf{k}_4)\left[\langle\phi_{k_1}\phi_{k_2}\rangle\langle\phi_m\phi_n\rangle\langle\phi^2\rangle+2\langle\phi_{k_1}\phi_m\rangle\langle\phi_{k_2}\phi_n\rangle\langle\phi^2\rangle\right]
\notag \\[2mm]
&&-\,
I^{k_4}_{st}\delta^D(\mathbf{k}_3)\left[\langle\phi_{k_1}\phi_{k_2}\rangle\langle\phi_s\phi_t\rangle\langle\phi^2\rangle+2\langle\phi_{k_1}\phi_s\rangle\langle\phi_{k_2}\phi_t\rangle\langle\phi^2\rangle\right]
+\delta^D(\mathbf{k}_3)\delta^D(\mathbf{k}_4)\langle\phi_{k_1}\phi_{k_2}\rangle\langle\phi^2\rangle^2\Bigg\}
\notag\\[2mm]
&&=\,
\Ccancel[red]{(2\pi)^3\delta^D(\mathbf{k}_{12})P^{\phi}_{k_1}\delta^D(\mathbf{k}_3)\delta^D(\mathbf{k}_4)\langle\phi^2\rangle^2}
+ 2(2\pi)^3\delta^D{(\mathbf{k}_{12})}\delta^D{(\mathbf{k}_{34})}P^{\phi}_{k_1}\int d\mathbf{p}_m^3P^{\phi}_mP^{\phi}_{|\mathbf{k}_4+\mathbf{p}_m|}
\notag \\[2mm]
&&+\,
2(2\pi)^3P^{\phi}_{k_1}P^{\phi}_{k_2}\langle\phi^2\rangle\left[\Ccancel[blue]{\delta^D(\mathbf{k}_{123})\delta^D(\mathbf{k}_4)+
\delta^D(\mathbf{k}_{124})\delta^D(\mathbf{k}_3)}\right]
\notag \\[2mm]
&&+\,
4(2\pi)^3\delta^D(\mathbf{k}_{1234})P^{\phi}_{k_1}P^{\phi}_{k_2}\left[P^{\phi}_{|\mathbf{k}_1+\mathbf{k}_3|}+P^{\phi}_{|\mathbf{k}_1+\mathbf{k}_4|}\right]
\notag \\[2mm]
&&-\,
(2\pi)^3\left[\Ccancel[red]{\delta^D(\mathbf{k}_{12})\delta^D(\mathbf{k}_3)\delta^D(\mathbf{k}_4)P^{\phi}_{k_1}\langle\phi^2\rangle^2} 
+
\Ccancel[blue]{2\delta^D(\mathbf{k}_{123})\delta^D(\mathbf{k}_4)P^{\phi}_{k_1}P^{\phi}_{k_2}\langle\phi^2\rangle}
\right]
\notag \\[2mm]
&&-\,
(2\pi)^3\left[\Ccancel[red]{\delta^D(\mathbf{k}_{12})\delta^D(\mathbf{k}_4)\delta^D(\mathbf{k}_3)P^{\phi}_{k_1}\langle\phi^2\rangle^2 }
+
\Ccancel[blue]{2\delta^D(\mathbf{k}_{124})\delta^D(\mathbf{k}_3)P^{\phi}_{k_1}P^{\phi}_{k_2}\langle\phi^2\rangle}
\right]
\notag \\[2mm]
&&+\,
\Ccancel[red]{(2\pi)^3\delta^D(\mathbf{k}_{12})\delta^D(\mathbf{k}_3)\delta^D(\mathbf{k}_4)P^{\phi}_{k_1}\langle\phi^2\rangle^2}
\notag\\[2mm]
&&=\,
2(2\pi)^3\delta^D(\mathbf{k}_{12})\delta^D(\mathbf{k}_{34})P^{\phi}_{k_1}\int d\mathbf{p}_m^3P^{\phi}_mP^{\phi}_{|\mathbf{k}_4+\mathbf{p}_m|}
+
4(2\pi)^3\delta^D(\mathbf{k}_{1234})P^{\phi}_{k_1}P^{\phi}_{k_2}\left[P^{\phi}_{|\mathbf{k}_1+\mathbf{k}_3|}+P^{\phi}_{|\mathbf{k}_1+\mathbf{k}_4|}\right]
\notag\\
\end{eqnarray}{}

\noindent From the last line we can see that one term did not cancel out (first one). However this represents (as it can be seen from the Dirac's deltas) an unconnected part of the four-point correlation function. It is indeed the primordial contribution to the power spectrum reported in equation \ref{eq:png_p11}.

From the Dirac deltas appearing in the above computation we also notice that for first part of the connected term $\mathbf{k}_m=-\mathbf{k}_1$, $\mathbf{k}_s=-\mathbf{k}_2$, $\mathbf{k}_n=\mathbf{k}_1 + \mathbf{k}_3$ and $\mathbf{k}_t=\mathbf{k}_2 + \mathbf{k}_4$. For the second one we have the relations $\mathbf{k}_s=-\mathbf{k}_1$, $\mathbf{k}_m=-\mathbf{k}_2$, $\mathbf{k}_n=\mathbf{k}_1 + \mathbf{k}_4$ and $\mathbf{k}_t=\mathbf{k}_2 + \mathbf{k}_3$.

We can now focus on the other term, $T_{(1111)}^{g_{\mathrm{nl}}}$:

\begin{eqnarray}
&&T_{(1111)}^{g_{\mathrm{nl}}}\propto
\begin{cases}
\left(\langle\phi_{k_1}\phi_{k_2}\rangle\langle\phi_{k_3}\phi_u\rangle\langle\phi_v\phi_z\rangle\;+\;2\,\mathrm{p.}\right)\times3\\
\langle\phi_{k_1}\phi_u\rangle\langle\phi_{k_2}\phi_v\rangle\langle\phi_{k_3}\phi_z\rangle\times6
\end{cases}
\end{eqnarray}{}

\noindent The first term can be easily spotted to be unconnected and should cancel out:

\begin{eqnarray}
\label{eq:tri_png_111_gnl}
&&T_{(1111)}^{g_{\mathrm{nl}}}\propto
\Bigg\{I^{k_4}_{uvz}\left[3\left(\langle\phi_{k_1}\phi_{k_2}\rangle\langle\phi_{k_3}\phi_u\rangle\langle\phi_v\phi_z\rangle\;+\;2\,\mathrm{p.}\right)
+6\langle\phi_{k_1}\phi_u\rangle\langle\phi_{k_2}\phi_v\rangle\langle\phi_{k_3}\phi_z\rangle\right]
\notag \\[2mm]
&&-\,
\dfrac{3}{(2\pi)^3}\left[\langle\phi_{k_1}\phi_{k_2}\rangle\langle\phi_{k_3}\phi_{k_4}\rangle\langle\phi^2\rangle\;+\;2\,\mathrm{p.}\right]\Bigg\}
\notag \\[2mm]
&&=\,3\left[\Ccancel[red]{\delta^D{(\mathbf{k}_{12})}\delta^D{(\mathbf{k}_{34})}(2\pi)^3P^{\phi}_{k_1}P^{\phi}_{k_3}\int d\mathbf{p}_v^3P^{\phi}_v\;+\;2\,\mathrm{p.}}\right]
+6(2\pi)^3\delta^D(\mathbf{k}_{1234})P^{\phi}_{k_1}P^{\phi}_{k_2}P^{\phi}_{k_3}
\notag \\
&&-\, \dfrac{3}{(2\pi)^3}\left[\Ccancel[red]{(2\pi)^6
\delta^D{(\mathbf{k}_{12})}\delta^D{(\mathbf{k}_{34})}
(2\pi)^3P^{\phi}_{k_1}P^{\phi}_{k_3}\langle\phi^2\rangle\;+\;2\,\mathrm{p.}}\right]
\notag \\[2mm]
&&=\,6(2\pi)^3\delta^D(\mathbf{k}_{1234})P^{\phi}_{k_1}P^{\phi}_{k_2}P^{\phi}_{k_3} \,.
\end{eqnarray}

\noindent From the Dirac deltas appearing in the above computation we also notice that for the connected term $\mathbf{k}_u=-\mathbf{k}_1$, $\mathbf{k}_v=-\mathbf{k}_2$ and $\mathbf{k}_z=-\mathbf{k}_3$. 

We can plug the results of equations \ref{eq:tri_png_111_fnl} and \ref{eq:tri_png_111_gnl} into the full expression of equation \ref{eq:tri_png_1111}:

\begin{eqnarray}
&&T_{(1111)}= F^{(1)}_{k_1}F^{(1)}_{k_2}F^{(1)}_{k_3}F^{(1)}_{k_4}
\mathcal{M}_{k_1}\mathcal{M}_{k_2}\mathcal{M}_{k_3}\mathcal{M}_{k_4}
\notag \\[2mm]
&&\times\,
\Bigg\{\dfrac{f^2_{\mathrm{nl}}}{c^4}4\,
P^{\phi}_{k_1}P^{\phi}_{k_2}
\left[
P^{\phi}_{|\mathbf{k}_1+\mathbf{k}_3|}
+
P^{\phi}_{|\mathbf{k}_1+\mathbf{k}_4|}
\right]\;+\;5\,\mathrm{p.}
+
\dfrac{g_{\mathrm{nl}}}{c^4}
\left[6P^{\phi}_{k_1}P^{\phi}_{k_2}P^{\phi}_{k_3}\;+\;3\,\mathrm{p.}\right]\Bigg\}\,.
\notag \\
\end{eqnarray}{}

\noindent
Basically the PNG terms $T^{f_{\mathrm{nl}}}_{(1111)}$ and $T^{g_{\mathrm{nl}}}_{(1111)}$ are respectively the equivalent of the terms $T_{(1122)}$ and $T_{(1113)}$ for the Gaussian initial condition due to gravitational non-linear evolution. 

As last step we express the above result in terms of the late-time matter power spectrum using equation \ref{eq:dk_to_phi}.

\begin{eqnarray}
&&T_{(1111)}= F^{(1)}_{k_1}F^{(1)}_{k_2}F^{(1)}_{k_3}F^{(1)}_{k_4}
\mathcal{M}_{k_1}\mathcal{M}_{k_2}\mathcal{M}_{k_3}\mathcal{M}_{k_4}
\notag \\[2mm]
&&\times\,
\Bigg\{\dfrac{f^2_{\mathrm{nl}}}{c^4}4\,
P^{\phi}_{k_1}P^{\phi}_{k_2}
\left[ 
P^{\phi}_{|\mathbf{k}_1+\mathbf{k}_3|}
+
P^{\phi}_{|\mathbf{k}_1+\mathbf{k}_4|}
\right]\;+\;5\,\mathrm{p.}
+
\dfrac{g_{\mathrm{nl}}}{c^4}
\left[6P^{\phi}_{k_1}P^{\phi}_{k_2}P^{\phi}_{k_3}\;+\;3\,\mathrm{p.}\right]\Bigg\}\,
\notag \\[2mm]
&&=\,
F^{(1)}_{k_1}F^{(1)}_{k_2}F^{(1)}_{k_3}F^{(1)}_{k_4}
\mathcal{M}_{k_1}\mathcal{M}_{k_2}\mathcal{M}_{k_3}\mathcal{M}_{k_4}
\notag \\[2mm]
&&\times\,
\Bigg\{\dfrac{f^2_{\mathrm{nl}}}{c^4}4\,
\dfrac{P^{\mathrm{m}}_{k_1}P^{\mathrm{m}}_{k_2}}
{\mathcal{M}_{k_1}^2\mathcal{M}_{k_2}^2}
\left[ 
\dfrac{P^\mathrm{m}_{|\mathbf{k}_1+\mathbf{k}_3|}}
{\mathcal{M}_{|\mathbf{k}_1+\mathbf{k}_3|}^2}
+
\dfrac{P^\mathrm{m}_{|\mathbf{k}_1+\mathbf{k}_4|}}
{{\mathcal{M}_{|\mathbf{k}_1+\mathbf{k}_4|}^2}}
\right]\;+\;5\,\mathrm{p.}
+
\dfrac{g_{\mathrm{nl}}}{c^4}
\left[6\dfrac{P^{\mathrm{m}}_{k_1}P^{\mathrm{m}}_{k_2}P^{\mathrm{m}}_{k_3}}
{\mathcal{M}_{k_1}^2\mathcal{M}_{k_2}^2\mathcal{M}_{k_3}^2}
\;+\;3\,\mathrm{p.}\right]\Bigg\}\,
\notag \\[2mm]
&&=\,
F^{(1)}_{k_1}F^{(1)}_{k_2}F^{(1)}_{k_3}F^{(1)}_{k_4}
\notag \\[2mm]
&&\times\,
\Bigg\{\dfrac{f^2_{\mathrm{nl}}}{c^4}4\,
\dfrac{\mathcal{M}_{k_3}\mathcal{M}_{k_4}}{\mathcal{M}_{k_1}\mathcal{M}_{k_2}}
P^{\mathrm{m}}_{k_1}P^{\mathrm{m}}_{k_2}
\left[ 
\dfrac{P^\mathrm{m}_{|\mathbf{k}_1+\mathbf{k}_3|}}
{{\mathcal{M}_{|\mathbf{k}_1+\mathbf{k}_3|}^2}}
+
\dfrac{P^\mathrm{m}_{|\mathbf{k}_1+\mathbf{k}_4|}}
{{\mathcal{M}_{|\mathbf{k}_1+\mathbf{k}_4|}^2}}
\right]\;+\;5\,\mathrm{p.}
+
\dfrac{g_{\mathrm{nl}}}{c^4}
\left[6
\mathcal{M}_{k_4}
\dfrac{P^{\mathrm{m}}_{k_1}P^{\mathrm{m}}_{k_2}P^{\mathrm{m}}_{k_3}}
{\mathcal{M}_{k_1}\mathcal{M}_{k_2}\mathcal{M}_{k_3}}
\;+\;3\,\mathrm{p.}\right]\Bigg\}\,.
 \notag \\[2mm]
\end{eqnarray}{}

%%%%%%%%%%%%%%%%%%%%%%%%%%%%%%%%%%%%%%%%%%%%%%%%%%%%%%%%%%%%%%%%%%%%%%%%%%%
%%%%%%%%%%%%%%%%%%%%%%%%%%%%%%%%%%%%%%%%%%%%%%%%%%%%%%%%%%%%%%%%%%%%%%%%%%%
%%%%%%%%%%%%%%%%%%%%%%%%%%%%%%%%%%%%%%%%%%%%%%%%%%%%%%%%%%%%%%%%%%%%%%%%%%%
\subsubsection{\texorpdfstring{$T_{(1112)}$}{}}
This is perhaps the most interesting PNG term of the i-trispectrum, since. as $B_{(111)}$ for the bispectrum, it comes from a correlator that would be null for Gaussian initial conditions.

\begin{eqnarray}
\label{eq:t1112_app}
&&\langle\delta^{(1)}\delta^{(1)}\delta^{(1)}\delta^{(2)}\rangle= \langle F^{(1)}_{k_1}F^{(1)}_{k_2}F^{(1)}_{k_3}
\mathcal{M}_{k_1}\mathcal{M}_{k_2}\mathcal{M}_{k_3}\mathcal{M}_{k_x}\mathcal{M}_{k_y}
I^{k_4}_{xy}F^{(2)}_{xy}
\notag \\[2mm]
&&\times\,\left\{ \phi_{k_1}+\dfrac{f_{\mathrm{nl}}}{c^2}\left[ I^{k_1}_{ab}\phi_a\phi_b-\delta^D(\mathbf{k}_1)\langle\phi^2\rangle\right]
+\dfrac{g_{\mathrm{nl}}}{c^4}\left[ I^{k_1}_{cde}\phi_c\phi_d\phi_e-\dfrac{3}{(2\pi)^3}\phi_{k_1}\langle\phi^2\rangle\right] \right\}
\notag \\[2mm]
&&\times\,\left\{ \phi_{k_2}+\dfrac{f_{\mathrm{nl}}}{c^2}\left[ I^{k_2}_{fg}\phi_f\phi_g-\delta^D(\mathbf{k}_2)\langle\phi^2\rangle\right]
+\dfrac{g_{\mathrm{nl}}}{c^4}\left[ I^{k_2}_{hil}\phi_h\phi_i\phi_l-\dfrac{3}{(2\pi)^3}\phi_{k_2}\langle\phi^2\rangle\right] \right\}
\notag \\
&&\times\,\left\{\phi_{k_3}+\dfrac{f_{\mathrm{nl}}}{c^2}\left[ I^{k_3}_{mn}\phi_m\phi_n-\delta^D(\mathbf{k}_3)\langle\phi^2\rangle\right]
+\dfrac{g_{\mathrm{nl}}}{c^4}\left[ I^{k_3}_{oqr}\phi_o\phi_q\phi_r-\dfrac{3}{(2\pi)^3}\phi_{k_3}\langle\phi^2\rangle\right]\right\}
\notag \\
&&\times\,\left\{\phi_{x}+\dfrac{f_{\mathrm{nl}}}{c^2}\left[ I^{x}_{st}\phi_s\phi_t-\delta^D(\mathbf{p}_x)\langle\phi^2\rangle\right]
+\dfrac{g_{\mathrm{nl}}}{c^4}\left[ I^{x}_{uvz}\phi_u\phi_v\phi_z-\dfrac{3}{(2\pi)^3}\phi_{x}\langle\phi^2\rangle\right]\right\}
\notag \\
&&\times\,
\left\{\phi_y+\dfrac{f_{\mathrm{nl}}}{c^2}\left[ I^{y}_{\epsilon\zeta}\phi_{\epsilon}\phi_{\zeta}-\delta^D(\mathbf{p}_y)\langle\phi^2\rangle\right]
+\dfrac{g_{\mathrm{nl}}}{c^4}\left[ I^{y}_{\eta\lambda\tau}\phi_{\eta}\phi_{\lambda}\phi_{\tau}-\dfrac{3}{(2\pi)^3}\phi_{y}\langle\phi^2\rangle\right]\right\}
\rangle\,
\notag \\
&&+\,
\;3\,\mathrm{p.}
\notag \\
\end{eqnarray}{}
 
\noindent There are two possible terms proportional to $f_{\mathrm{nl}}$ from the above expansion. The first corresponds to the $f_{\mathrm{nl}}$ associated to  a $\delta^{(1)}$, the second to that associated to $\delta^{(2)}$.

\begin{eqnarray}
\label{eq:t1112}
&&T_{(1112)} =\,  F^{(1)}_{k_1}F^{(1)}_{k_2}F^{(1)}_{k_3}
\mathcal{M}_{k_1}\mathcal{M}_{k_2}\mathcal{M}_{k_3}\mathcal{M}_{k_x}\mathcal{M}_{k_y}
I^{k_4}_{xy}F^{(2)}_{xy}\dfrac{f_{\mathrm{nl}}}{c^2}
\notag \\[2mm]
&&\times\,
\Bigg\{
\left[I^{k_1}_{ab}\langle\phi_a\phi_b\phi_{k_2}\phi_{k_3}\phi_x\phi_y\rangle-\delta^D(\mathbf{k}_1)\langle\phi_{k_2}\phi_{k_3}\phi_x\phi_y\rangle\langle\phi^2\rangle \;+\;2\,\mathrm{p.}\right]
\notag \\[2mm]
&&+\,
2\left[I^x_{st}\langle\phi_{k_1}\phi_{k_2}\phi_{k_3}\phi_{s}\phi_{t}\phi_y\rangle-\delta^D(\mathbf{p}_x)\langle\phi_{k_1}\phi_{k_2}\phi_{k_3}\phi_y\rangle\langle\phi^2\rangle\right]
\Bigg\}\,,
\end{eqnarray}{}

\noindent where the two permutations in the second line correspond to expanding $\mathbf{k}_2$ or $\mathbf{k}_3$ instead of $\mathbf{k}_1$. Proceed step by step  we now analyse the two terms separately, namely $T^{a}_{(1112)}$ and $T^{b}_{(1112)}$. Starting from the first:
\begin{eqnarray}
T^{a}_{(1112)}\propto
\begin{cases}
\langle\phi_a\phi_b\rangle\langle\phi_{k_2}\phi_{k_3}\rangle \langle\phi_x\phi_y\rangle \\
\langle\phi_{a}\phi_{k_2}\rangle\langle\phi_{b}\phi_{k_3}\rangle\langle\phi_x\phi_y\rangle \times 2 \\
\langle\phi_a\phi_b\rangle\langle\phi_{k_2}\phi_x\rangle\langle\phi_{k_3}\phi_y\rangle\times2 \\
\langle\phi_a\phi_x\rangle\langle\phi_{k_2}\phi_{k_3}\rangle\langle\phi_b\phi_y\rangle\times2 \\
\langle\phi_a\phi_{k_2}\rangle\langle\phi_{x}\phi_{k_3}\rangle\langle\phi_b\phi_y\rangle\times4 \; +\;1\,\mathrm{p.} \;(k_2\longleftrightarrow k_3)
\end{cases}
\end{eqnarray}{}

\noindent In order to proceed as generally as possible, we need to separately treat the combinations arising from the mixing between original wave vectors $\mathbf{k}_i$ with auxiliary ones $\mathbf{p}_j$ originating from different sources (equation \ref{eq:t1112_app}).

\begin{eqnarray}
&&T^{a}_{(1112)}\propto
\Bigg\{
I^{k_4}_{xy}I^{k_1}_{ab}F^{(2)}_{xy}
\langle\phi_a\phi_b\rangle\langle\phi_{k_2}\phi_{k_3}\rangle \langle\phi_x\phi_y\rangle
+
2I^{k_4}_{xy}I^{k_1}_{ab}F^{(2)}_{xy}
\langle\phi_{a}\phi_{k_2}\rangle\langle\phi_{b}\phi_{k_3}\rangle\langle\phi_x\phi_y\rangle
\notag \\[2mm]
&&+\,
2I^{k_4}_{xy}I^{k_1}_{ab}F^{(2)}_{xy}
\langle\phi_a\phi_b\rangle\langle\phi_{k_2}\phi_x\rangle\langle\phi_{k_3}\phi_y\rangle
+
2I^{k_4}_{xy}I^{k_1}_{ab}F^{(2)}_{xy}
\langle\phi_a\phi_x\rangle\langle\phi_{k_2}\phi_{k_3}\rangle\langle\phi_b\phi_y\rangle
\notag \\[2mm]
&&+\,
\left[4I^{k_4}_{xy}I^{k_1}_{ab}F^{(2)}_{xy}
\langle\phi_a\phi_{k_2}\rangle\langle\phi_{x}\phi_{k_3}\rangle\langle\phi_b\phi_y\rangle +k_2\longleftrightarrow k_3\right]
\notag \\[2mm]
&&-\,
\delta^D(\mathbf{k}_1)\langle\phi^2\rangle I^{k_4}_{xy}F^{(2)}_{xy}\left[\langle\phi_{k_2}\phi_{k_3}\rangle\langle\phi_x\phi_y\rangle + 2\langle\phi_{k_2}\phi_x\rangle\langle\phi_{k_3}\phi_y\rangle\right]
\Bigg\}\; + \;2\,\mathrm{p.}
\notag \\[2mm]
&&=\,
\Ccancel[red]{(2\pi)^3\delta^D({\bf k}_{23})\delta^D({\bf k}_1)\delta^D({\bf k}_4)P^{\phi}_{k_2}\langle\phi^2\rangle^2C_{\mathrm{b}}}
+
2(2\pi)^3\delta^D({\bf k}_{123})\delta^D({\bf k}_{4})P^{\phi}_{k_2}P^{\phi}_{k_3}\langle\phi^2\rangle C_{\mathrm{b}}
\notag \\[2mm]
&&+\,
\Ccancel[blue]{2(2\pi)^3\delta^D({\bf k}_{234})\delta^D({\bf k}_{1})P^{\phi}_{k_2}P^{\phi}_{k_3}F^{(2)}_{k_2k_3}\langle\phi^2\rangle}
\notag \\[2mm]
&&+\,
2(2\pi)^3\delta^D({\bf k}_{23})\delta^D({\bf k}_{14})P^{\phi}_{k_2}\int d\mathbf{p}_x^3 P^{\phi}_{x}P^{\phi}_{|\mathbf{k}_4-\mathbf{p}_x|}F^{(2)}\left[\mathbf{p}_x,\mathbf{k}_4-\mathbf{p}_x\right]
\notag \\[2mm]
&&+\,
4(2\pi)^3\delta^D({\bf k}_{1234})P^{\phi}_{k_2}P^{\phi}_{k_3}P^{\phi}_{|\mathbf{k}_3+\mathbf{k}_4|}F^{(2)}\left[-\mathbf{k}_3,\mathbf{k}_3+\mathbf{k}_4\right] +k_2\longleftrightarrow k_3
\notag \\[2mm]
&&-\,
\delta^D({\bf k}_{1})\langle\phi^2\rangle(2\pi)^3 \left[\Ccancel[red]{\delta^D({\bf k}_{23})P^{\phi}_{k_2}\delta^D({\bf k}_{4})\langle\phi^2\rangle C_{\mathrm{b}}}
+
\Ccancel[blue]{2(2\pi)^3\delta^D({\bf k}_{234})F^{(2)}_{k_2k_3}P^{\phi}_{k_2}P^{\phi}_{k_3}\langle\phi^2\rangle}\right]\; + \;2\,\mathrm{p.}
\notag \\[2mm]
&&=\,
\Bigg\{2(2\pi)^3\left[\delta^D({\bf k}_{123})\delta^D({\bf k}_{4})P^{\phi}_{k_2}P^{\phi}_{k_3}\langle\phi^2\rangle C_{\mathrm{b}}
+
\delta^D({\bf k}_{23})\delta^D({\bf k}_{14})P^{\phi}_{k_2}\int d\mathbf{p}_x^3 P^{\phi}_{x}P^{\phi}_{|\mathbf{k}_4-\mathbf{p}_x|}F^{(2)}\left[\mathbf{p}_x,\mathbf{k}_4-\mathbf{p}_x\right]\right]
\notag \\[2mm]
&&+\,
4(2\pi)^3\delta^D({\bf k}_{1234})P^{\phi}_{k_2}P^{\phi}_{k_3}P^{\phi}_{|\mathbf{k}_3+\mathbf{k}_4|}F^{(2)}\left[-\mathbf{k}_3,\mathbf{k}_3+\mathbf{k}_4\right] +k_2\longleftrightarrow k_3\Bigg\}\; + \;2\,\mathrm{p.}
\notag \\
\end{eqnarray}{}

\noindent The constant $C_{\mathrm{b}}=F^{(2)}\left[\mathbf{q},-\mathbf{q}\right] = b_2/2 + b_{\mathrm{s}^2}/3$ derives from the second order kernel including redshift space distortions in the case of opposite $k$-vectors. Only the term in the second line represents a connected part of the initial four-point correlator. From the Dirac deltas resulting in the connected part we can notice that $\mathbf{k}_x = -\mathbf{k}_3$ and $\mathbf{k}_y = \mathbf{k}_3+\mathbf{k}_4$. This is a clear example of mode coupling between non-linear evolution and primordial non-Gaussian initial conditions. Moving to the  second term $T_{(1112)}^{b}$:

\begin{eqnarray}
T_{(1112)}^{b}\propto
\begin{cases}
\langle\phi_{k_1}\phi_{k_2}\rangle\langle\phi_{k_3}\phi_y\rangle \langle\phi_s\phi_t\rangle\;+\;2\,\mathrm{p.} \\
\langle\phi_{k_1}\phi_{k_2}\rangle\langle\phi_t\phi_{k_3}\rangle\langle\phi_s\phi_y\rangle \times 2\;+\;2\,\mathrm{p.} \\
\langle\phi_{k_1}\phi_s\rangle\langle\phi_{k_2}\phi_t\rangle\langle\phi_{k_3}\phi_y\rangle\times2\;+\;2\,\mathrm{p.} \\
\end{cases}
\end{eqnarray}{}

\noindent Which then expanded becomes:

\begin{eqnarray}
&&T_{(1112)}^{b}\propto
\Bigg\{\left[I^{k_4}_{xy}I^x_{st}F^{(2)}_{xy}\langle\phi_{k_1}\phi_{k_2}\rangle\langle\phi_{k_3}\phi_y\rangle \langle\phi_s\phi_t\rangle\;+\;2\,\mathrm{p.}\right]
\notag \\[2mm]
&&+\,
\left[2I^{k_4}_{xy}I^x_{st}F^{(2)}_{xy}\langle\phi_{k_1}\phi_{k_2}\rangle\langle\phi_t\phi_{k_3}\rangle\langle\phi_s\phi_y\rangle\;+\;2\,\mathrm{p.}\right]
% \notag \\[2mm]
% &&+\,
+\left[2I^{k_4}_{xy}I^x_{st}F^{(2)}_{xy}\langle\phi_{k_1}\phi_s\rangle\langle\phi_{k_2}\phi_t\rangle\langle\phi_{k_3}\phi_y\rangle\;+\;2\,\mathrm{p.}\right]
\notag \\[2mm]
&&-\,
\left[\delta^D(\mathbf{p}_x)I^{k_4}_{xy}F^{(2)}_{xy}\langle\phi_{k_1}\phi_{k_2}\rangle\langle\phi_{k_3}\phi_y\rangle\langle\phi^2\rangle\;+\;2\,\mathrm{p.}\right]
\Bigg\}
\notag \\[2mm]
&&=\,
\Ccancel[red]{\left[(2\pi)^3\delta^D({\bf k}_{12})\delta^D({\bf k}_{34})P^{\phi}_{k_2}P^{\phi}_{k_3}F^{(2)}\left[\mathbf{k}_4+\mathbf{k}_3,-\mathbf{k}_3\right]\langle\phi^2\rangle\;+\;2\,\mathrm{p.}\right]}
\notag \\[2mm]
&&+\,
\left[(2\pi)^3\delta^D({\bf k}_{12})\delta^D ({\bf k}_{34})P^{\phi}_{k_2}P^{\phi}_{k_3}\int d\mathbf{p}_y^3F^{(2)}\left[\mathbf{k}_4-\mathbf{p}_y,\mathbf{p}_y\right]P^{\phi}_y\;+\;2\,\mathrm{p.}\right]\times2
\notag \\[2mm]
&&+\,
\left[(2\pi)^3P^{\phi}_{k_1}P^{\phi}_{k_2}P^{\phi}_{k_3}\delta^D(\mathbf{k}_{1234})F^{(2)}\left[\mathbf{k}_3+\mathbf{k}_4,-\mathbf{k}_3\right]\;+\;2\,\mathrm{p.}\right]\times2
\notag\\[2mm]
&&-\,
\Ccancel[red]{\left[(2\pi)^3P^{\phi}_{k_2}P^{\phi}_{k_3}\delta^D({\bf k}_{12})\delta^D({\bf k}_{34})F^{(2)}\left[\mathbf{k}_3+\mathbf{k}_4,-\mathbf{k}_3\right]\langle\phi^2\rangle\;+\;2\,\mathrm{p.}
\right]}
\notag \\[2mm]
&&=\,
\left[(2\pi)^3\delta^D({\bf k}_{12})\delta^D({\bf k}_{34}) P^{\phi}_{k_2}P^{\phi}_{k_3}\int d\mathbf{p}_y^3F^{(2)}\left[\mathbf{k}_4-\mathbf{p}_y,\mathbf{p}_y\right]P^{\phi}_y\;+\;2\,\mathrm{p.}\right]\times2
\notag \\[2mm]
&&+\,
\left[(2\pi)^3P^{\phi}_{k_1}P^{\phi}_{k_2}P^{\phi}_{k_3}\delta^D(\mathbf{k}_{1234})F^{(2)}\left[\mathbf{k}_3+\mathbf{k}_4,-\mathbf{k}_3\right]\;+\;2\,\mathrm{p.}\right]\times2\,,
\notag \\
\end{eqnarray}{}

\noindent where all the permutations above consist in changing $\mathbf{k}_3$ with either $\mathbf{k}_1$ or $\mathbf{k}_2$. Only the second term represents a connected part of the initial four-point correlator.
From the Dirac deltas resulting in the connected part we can notice that $\mathbf{k}_y = -\mathbf{k}_3$ and $\mathbf{k}_x = \mathbf{k}_3+\mathbf{k}_4$.
We can now proceed with equation \ref{eq:t1112}, using only the connected terms from $T^{a}_{(1112)}$ and $T^{b}_{(1112)}$:

\begin{eqnarray}
&&T_{(1112)}=  
2F^{(1)}_{k_1}F^{(1)}_{k_2}F^{(1)}_{k_3}
\mathcal{M}_{k_1}\mathcal{M}_{k_2}\mathcal{M}_{k_3}
\dfrac{f_{\mathrm{nl}}}{c^2}
\notag \\[2mm]
&&\times\,
\Bigg\{
\left[
2
\mathcal{M}_{k_3}\mathcal{M}_{|\mathbf{k}_3+\mathbf{k}_4|}
\,
P^{\phi}_{k_2}P^{\phi}_{k_3}P^{\phi}_{|\mathbf{k}_3+\mathbf{k}_4|}F^{(2)}\left[-\mathbf{k}_3,\mathbf{k}_3+\mathbf{k}_4\right] +k_2\longleftrightarrow k_3\right]\; + \;2\,\mathrm{p.}
\notag\\[2mm]
&&+\,
\left[
\mathcal{M}_{k_3}\mathcal{M}_{|\mathbf{k}_3+\mathbf{k}_4|}
\,
P^{\phi}_{k_1}P^{\phi}_{k_2}P^{\phi}_{k_3}F^{(2)}\left[\mathbf{k}_3+\mathbf{k}_4,-\mathbf{k}_3\right]\;+\;2\,\mathrm{p.}
\right]
\Bigg\} \;+\;3\,\mathrm{p.}
\notag\\[2mm]
&&=\,
2F^{(1)}_{k_1}F^{(1)}_{k_2}F^{(1)}_{k_3}
\mathcal{M}_{k_1}\mathcal{M}_{k_2}\mathcal{M}_{k_3}
\dfrac{f_{\mathrm{nl}}}{c^2}
\notag \\[2mm]
&&\times\,
\Bigg\{
\left[
2
\mathcal{M}_{k_3}\mathcal{M}_{|\mathbf{k}_3+\mathbf{k}_4|}
\,
P^{\phi}_{k_2}P^{\phi}_{k_3}P^{\phi}_{|\mathbf{k}_3+\mathbf{k}_4|}F^{(2)}\left[-\mathbf{k}_3,\mathbf{k}_3+\mathbf{k}_4\right]\; + \;5\,\mathrm{p.}\right]
\notag\\[2mm]
&&+\,
\left[
\mathcal{M}_{k_3}\mathcal{M}_{|\mathbf{k}_3+\mathbf{k}_4|}
\,
P^{\phi}_{k_1}P^{\phi}_{k_2}P^{\phi}_{k_3}F^{(2)}\left[\mathbf{k}_3+\mathbf{k}_4,-\mathbf{k}_3\right]\;+\;2\,\mathrm{p.}
\right]
\Bigg\} \;+\;3\,\mathrm{p.}\,.
\notag \\
\end{eqnarray}{}

\noindent As a final step we can convert the above result into an expression in terms of the late-time matter power spectrum using equation \ref{eq:dk_to_phi}

\begin{eqnarray}
&&T_{(1112)}= 
\dfrac{f_{\mathrm{nl}}}{c^2}
F^{(1)}_{k_1}F^{(1)}_{k_2}F^{(1)}_{k_3}
\notag \\[2mm]
&&\times\,
\Bigg\{
\left[
\mathcal{M}_{k_1}\mathcal{M}_{k_2}\mathcal{M}_{k_3}^2\mathcal{M}_{|\mathbf{k}_3+\mathbf{k}_4|}
P^{\phi}_{k_2}P^{\phi}_{k_3}P^{\phi}_{|\mathbf{k}_3+\mathbf{k}_4|}F^{(2)}\left[-\mathbf{k}_3,\mathbf{k}_3+\mathbf{k}_4\right]\; + \;5\,\mathrm{p.}\right]
\notag\\[2mm]
&&+\,
\left[
2
\mathcal{M}_{k_1}\mathcal{M}_{k_2}\mathcal{M}_{k_3}^2\mathcal{M}_{|\mathbf{k}_3+\mathbf{k}_4|}
P^{\phi}_{k_1}P^{\phi}_{k_2}P^{\phi}_{k_3}F^{(2)}\left[\mathbf{k}_3+\mathbf{k}_4,-\mathbf{k}_3\right]\;+\;2\,\mathrm{p.}
\right]
\Bigg\} \;+\;3\,\mathrm{p.}
\notag\\[2mm]
&&=\,
\dfrac{f_{\mathrm{nl}}}{c^2}
F^{(1)}_{k_1}F^{(1)}_{k_2}F^{(1)}_{k_3}
\notag \\[2mm]
&&\times\,
\Bigg\{
\left[
4 
\dfrac{\mathcal{M}_{k_1}}{\mathcal{M}_{k_2}}
P^{\mathrm{m}}_{k_2}P^{\mathrm{m}}_{k_3}
\dfrac{P^{\mathrm{m}}_{|\mathbf{k}_3+\mathbf{k}_4|}}
{\mathcal{M}_{|\mathbf{k}_3+\mathbf{k}_4|}}
F^{(2)}\left[-\mathbf{k}_3,\mathbf{k}_3+\mathbf{k}_4\right]\; + \;5\,\mathrm{p.}\right]
\notag\\[2mm]
&&+\,
\left[
2
\dfrac{\mathcal{M}_{|\mathbf{k}_3+\mathbf{k}_4|}}{\mathcal{M}_{k_1}\mathcal{M}_{k_2}}
P^\mathrm{m}_{k_1}P^\mathrm{m}_{k_2}P^\mathrm{m}_{k_3}F^{(2)}\left[\mathbf{k}_3+\mathbf{k}_4,-\mathbf{k}_3\right]\;+\;2\,\mathrm{p.}
\right]
\Bigg\} \;+\;3\,\mathrm{p.}
\,.
\notag \\
\end{eqnarray}

\noindent It is better to not further regroup together the two terms since the number of possible permutations is different. 

We conclude this i-trispectrum calculation by recalling that up to order $\phi^6$ the two remaining terms $T_{(1122)}$ and $T_{(1113)}$ return only the standard expression.

\subsection{Recap: all PNG terms}
\paragraph{Power Spectrum}

\begin{eqnarray}
\label{eq:png_pk}
&&P^{\mathrm{PNG}}(\mathbf{k}) = P_{11} + P_{12}
\notag\\[2mm]
&&=\,
F^{(1)}_kF^{(1)}_q
\mathcal{M}_k^2
\dfrac{2f_{\mathrm{nl}}^2}{c^4}\int \dfrac{d\mathbf{p}_a^3}{(2\pi)^3}P^{\phi}_aP^{\phi}_{|\mathbf{k}-\mathbf{p}_a|}
\notag\\[2mm]
&&+\,
\dfrac{4f_{\mathrm{nl}}}{c^2}F^{(1)}_k 
\mathcal{M}_k
\int\dfrac{d\mathbf{p}_a^3}{(2\pi)^3}
\mathcal{M}_{p_a}\mathcal{M}_{|-\mathbf{k}-\mathbf{p}_a|}
F^{(2)}_{a,-\mathbf{k}-\mathbf{p}_a}P^{\phi}_{|-\mathbf{k}-\mathbf{p}_a|}\left[P^{\phi}_a+2P^{\phi}_k\right]\,.
\notag \\
\end{eqnarray}{}

\paragraph{Bispectrum}
\begin{eqnarray}
\label{eq:png_bk}
&&B^{\mathrm{PNG}}(\mathbf{k}_1,\mathbf{k}_2,\mathbf{k}_3) = B_{(111)}
=
F^{(1)}_{k_1}F^{(1)}_{k_2}F^{(1)}_{k_3}
\dfrac{\mathcal{M}_{k_1}}{\mathcal{M}_{k_2}\mathcal{M}_{k_3}}
\dfrac{2f_{\mathrm{nl}}}{c^2}
P^{\mathrm{m}}_{k_2}P^{\mathrm{m}}_{k_3} \;+\; \mathrm{cyc.}\,.
\end{eqnarray}{}

\paragraph{Trispectrum}
\begin{eqnarray}
\label{eq:png_tk}
&&T^{\mathrm{PNG}}(\mathbf{k}_1,\mathbf{k}_2,\mathbf{k}_3,\mathbf{k}_4) =T_{(1111)}+ T_{(1112)}
\notag\\[2mm]
&&=\,
F^{(1)}_{k_1}F^{(1)}_{k_2}F^{(1)}_{k_3}F^{(1)}_{k_4}
\notag \\[2mm]
&&\times\,
\Bigg\{\dfrac{f^2_{\mathrm{nl}}}{c^4}4\,
\dfrac{\mathcal{M}_{k_3}\mathcal{M}_{k_4}}{\mathcal{M}_{k_1}\mathcal{M}_{k_2}}
P^{\mathrm{m}}_{k_1}P^{\mathrm{m}}_{k_2}
\left[ 
\dfrac{P^\mathrm{m}_{|\mathbf{k}_1+\mathbf{k}_3|}}
{{\mathcal{M}_{|\mathbf{k}_1+\mathbf{k}_3|}^2}}
+
\dfrac{P^\mathrm{m}_{|\mathbf{k}_1+\mathbf{k}_4|}}
{{\mathcal{M}_{|\mathbf{k}_1+\mathbf{k}_4|}^2}}
\right]\;+\;5\,\mathrm{p.}
+
\dfrac{g_{\mathrm{nl}}}{c^4}
\left[6
\mathcal{M}_{k_4}
\dfrac{P^{\mathrm{m}}_{k_1}P^{\mathrm{m}}_{k_2}P^{\mathrm{m}}_{k_3}}
{\mathcal{M}_{k_1}\mathcal{M}_{k_2}\mathcal{M}_{k_3}}
\;+\;3\,\mathrm{p.}\right]\Bigg\}\,
\notag \\[2mm]
&&+\,
\dfrac{f_{\mathrm{nl}}}{c^2}
F^{(1)}_{k_1}F^{(1)}_{k_2}F^{(1)}_{k_3}
\notag \\[2mm]
&&\times\,
\Bigg\{
\left[
4 
\dfrac{\mathcal{M}_{k_1}}{\mathcal{M}_{k_2}}
P^{\mathrm{m}}_{k_2}P^{\mathrm{m}}_{k_3}
\dfrac{P^{\mathrm{m}}_{|\mathbf{k}_3+\mathbf{k}_4|}}
{\mathcal{M}_{|\mathbf{k}_3+\mathbf{k}_4|}}
F^{(2)}\left[-\mathbf{k}_3,\mathbf{k}_3+\mathbf{k}_4\right]\; + \;5\,\mathrm{p.}\right]
\notag\\[2mm]
&&+\,
\left[
2
\dfrac{\mathcal{M}_{|\mathbf{k}_3+\mathbf{k}_4|}}{\mathcal{M}_{k_1}\mathcal{M}_{k_2}}
P^\mathrm{m}_{k_1}P^\mathrm{m}_{k_2}P^\mathrm{m}_{k_3}F^{(2)}\left[\mathbf{k}_3+\mathbf{k}_4,-\mathbf{k}_3\right]\;+\;2\,\mathrm{p.}
\right]
\Bigg\} \;+\;3\,\mathrm{p.}
\,.
\notag \\
\end{eqnarray}{}

\begin{figure}[tbp]
\centering 
\includegraphics[width=\textwidth]%,trim=0 380 0 200,clip]
{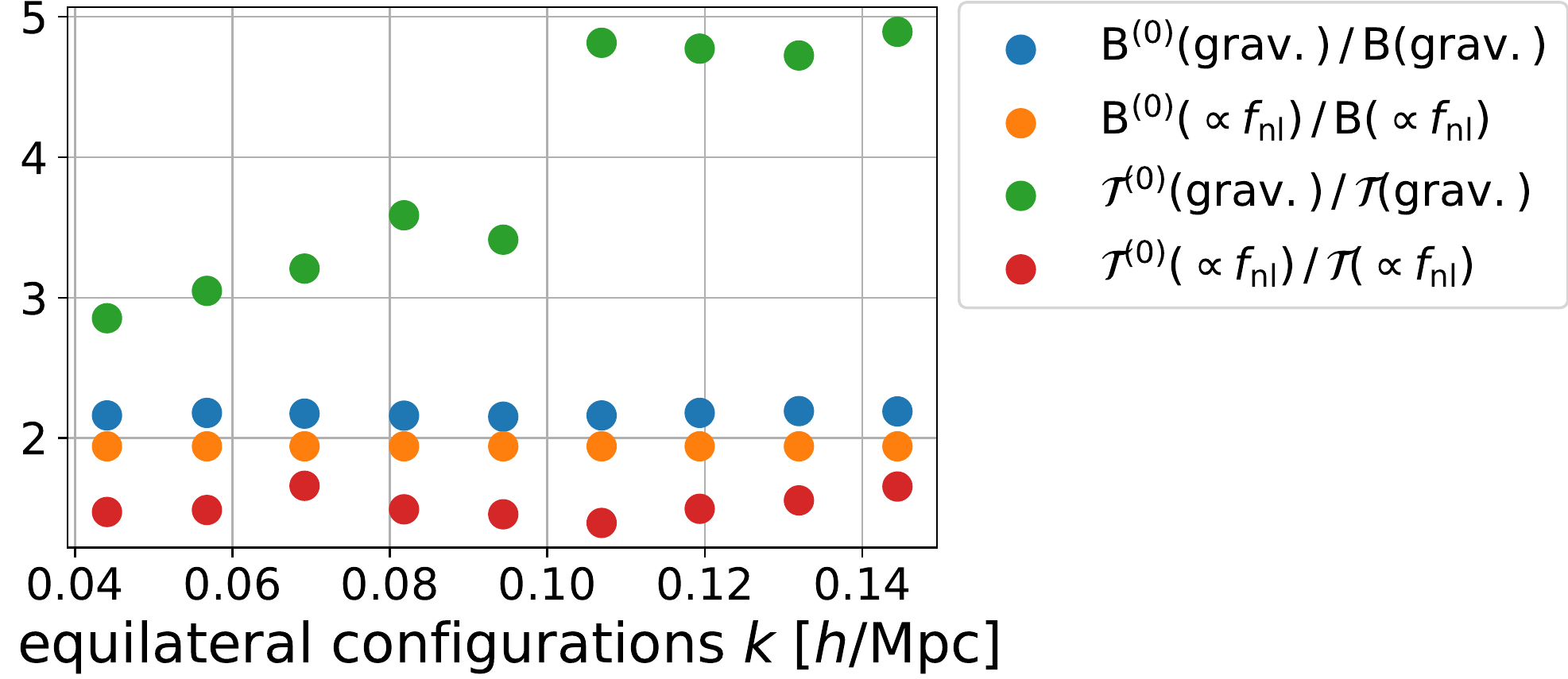}
\caption{\label{fig:rsd_vs_real_kaiserboost} 
Kaiser boost effect on gravitational and PNG components of the bispectrum and i-trispectrum signals at $z=0.5$. For clarity only the signal relative to equilateral configurations has been displayed.}
\end{figure}

%%%%%%%%%%%%%%%%%%%%%%%%%%%%%%%%%%%%%%%%%%%%%
%%%%%%%%%%%%%%%%%%%%%%%%%%%%%%%%%%%%%%%%%%%%%
%%%%%%%%%%%%%%%%%%%%%%%%%%%%%%%%%%%%%%%%%%%%%
\acknowledgments
D.G. and L.V. thank Francisco Villaescusa-Navarro for invaluable help with the extensive use of the \textsc{Quijote} simulations. D.G. is grateful to Nicola Bellomo and Alex Mead for useful discussions. All authors wish to thank the anonymous referee for the very constructive referee report which improved the paper's presentation.

\noindent The measurements from the simulations were performed in the TigerCPU cluster at Princeton.
L.V.  and D.G. acknowledge support of European Unions Horizon 2020 research and innovation programme ERC (BePreSySe, grant agreement 725327). Funding for this work was partially provided by the Spanish MINECO under projects  PGC2018-098866-B-I00 FEDER-EU.

\noindent HGM  acknowledges  the  support  from  `la  Caixa' Foundation  (ID100010434)  with  code

\noindent LCF/BQ/PI18/11630024.

\bibliographystyle{plain}
\bibliography{references}


% \begin{thebibliography}{99}

% \bibitem{a}
% Author, \emph{Title}, \emph{J. Abbrev.} {\bf vol} (year) pg.

% \bibitem{b}
% Author, \emph{Title},
% arxiv:1234.5678.

% \bibitem{c}
% Author, \emph{Title},
% Publisher (year).

% % Please avoid comments such as "For a review'', "For some examples",
% % "and Ref. s therein" or move them in the text. In general,
% % please leave only Ref. s in the bibliography and move all
% % accessory text in footnotes.

% % Also, please have only one work for each \bibitem.

% \end{thebibliography}
\end{document}